\documentclass[12pt]{article} 
\usepackage{amsmath,amssymb}
\usepackage{amsthm}
\usepackage{amsfonts}
\usepackage{natbib}
\usepackage{bigints}
\usepackage{comment}
\usepackage{graphicx}
\usepackage{caption} 
\usepackage{subcaption} 
\usepackage{relsize}
\usepackage{catchfilebetweentags} 
\usepackage{layout} 
\usepackage{bm} 

\usepackage{xcolor}  
\definecolor{darkgreen}{rgb}{0,0.5,0}

\usepackage[pdftex, pdfusetitle, plainpages=false, 
				letterpaper, bookmarks, bookmarksnumbered,
				colorlinks, linkcolor=black, citecolor=black,
	         filecolor=black, urlcolor=black]{hyperref}

\usepackage[shortlabels]{enumitem} 

\usepackage[normalem]{ulem} 
\newcommand{\robustout}[1]{\bgroup\markoverwith{\rule[0.5ex]{2pt}{0.4pt}}\ULon{#1}}

\usepackage{float}
\usepackage{multirow} 
\usepackage{multicol} 

\usepackage{bigints} 

\usepackage[margin=1in]{geometry}
\allowdisplaybreaks  

\usepackage[onehalfspacing]{setspace}

\usepackage{mathtools}

\usepackage{threeparttable}

\usepackage{pdflscape}

\newtheorem{theorem}{Theorem}
\newtheorem{lemma}{Lemma}
\newtheorem{definition}{Definition}
\newtheorem{proposition}{Proposition}
\newtheorem*{example*}{Example} 
\newtheorem{corollary}{Corollary}

\newtheorem{assumption}{Assumption}

\newcommand{\m}[1]{\mathcal{#1}}

\newcommand{\boup}[1]{\boldsymbol{\mathrm{#1}}}
\newcommand{\up}[1]{\mathrm{#1}}
\newcommand{\ha}[1]{\widehat{#1}}
\newcommand{\ti}[1]{\widetilde{#1}}
\newcommand{\lbar}[1]{\underline{#1}}
\newcommand{\ubar}[1]{\overline{#1}}

\newcommand{\mme}{\mathbb{E}}
\newcommand{\mmi}{\mathbb{I}}
\newcommand{\mmp}{\mathbb{P}}
\newcommand{\mmq}{\mathbb{Q}}
\newcommand{\mmr}{\mathbb{R}}

\newcommand{\mJ}{\m{J}}

\newcommand{\bQ}{\boup{Q}}
\newcommand{\bR}{\boup{R}}
\newcommand{\bS}{\boup{S}}
\newcommand{\mS}{\m{S}}

\newcommand{\bLambda}{\boup{\Lambda}}

\newcommand{\eps}{\varepsilon} 

\makeatletter
\def\old@comma{,}
\catcode`\,=13
\def,{%
  \ifmmode%
    \old@comma\discretionary{}{}{}%
  \else%
    \old@comma%
  \fi%
}
\makeatother

\begin{document}

\title{
Causal Effects in Matching Mechanisms with Strategically Reported Preferences\thanks{\scriptsize This research was conducted in part when Luflade and Mourifi\'e were visiting the
Becker Friedman Institute (BFI), and Bertanha was visiting the Kenneth C. Griffin Department of Economics, both at the University of Chicago. The authors thank their respective hosts for their hospitality and support.
We are grateful to 
Nikhil Agarwal,
Sebastian Gallegos, 
Guillaume Haeringer,
YingHua He,
Sam Hwang,
Yuichi Kitamura,
Kory Kroft, 
Maciej Kotowski,
Chris Neilson,
Marcin Peski,
Kevin Song,
and
Bumin Yemnez
for valuable comments and discussions.  
We thank Maxi Machado for excellent research assistance.
The paper has also benefited from feedback received during presentations at
Stanford, 
Notre Dame,
Northwestern,
Toronto, 
FGV Rio,
Michigan,
UC Santa Barbara,
Boston College,
Harvard-MIT,
Texas A\&M, 
UIUC,
UC Santa Cruz,
Emory,
Tennessee,
Duke, 
Wash U,
Rutgers, 
Sciences Po,
Wisconsin,
Georgetown,
Pittsburgh, 
UC Santa Barbara
and various conferences. 
}
}

\author{Marinho Bertanha\footnote{ \scriptsize Department of Economics, University of Notre Dame.
  Address: 3060 Jenkins Nanovic Halls, Notre Dame, IN 46556.
  Email: mbertanha@nd.edu.
  }
  \and
  Margaux Luflade\footnote{ \scriptsize Department of Economics, University of Pennsylvania. Address: 133 South 36th Street, 
Philadelphia, PA 19104. Email: mluflade@sas.upenn.edu.
   }
  \and
  Ismael Mourifi\'e\footnote{\scriptsize Department of Economics,
  Washington University in St. Louis \& NBER. Address: One Brookings Drive
St. Louis, MO 63130-4899, USA. Email: ismaelm@wustl.edu.
}
}

\date{
First version: July 26th, 2023
\\
This version: March 9th, 2026
}

\maketitle

\vspace{-1cm}

\footnotesize

\begin{center}
\textbf{Abstract}
\end{center}

A growing number of authorities use mechanisms to allocate students to schools in a way that reflects student preferences and school priorities.
However, most real-world mechanisms incentivize students to strategically misreport their preferences.
Misreporting complicates the identification of causal parameters that depend on true preferences, which are necessary inputs for a broad class of counterfactual analyses.  
We provide an identification approach robust to misreporting and derive sharp bounds on causal effects of school assignment.
Our approach applies to allocation rules characterized by placement scores and cutoffs.
We use data from a deferred acceptance mechanism that assigns students to university programs in Chile.
Matching theory predicts and empirical evidence shows that students behave strategically in Chile because they face constraints on preference submission and have good prior information about school accessibility.
Our bounds are informative enough to reveal significant heterogeneity in graduation success with respect to preferences and school assignment.


\vspace{0cm}

\vspace{0.5cm}



\normalsize

\newpage

\section{Introduction}
\indent
 
One of the most important decisions students make is their choice of field and institution of education. 
Identification of the impact of such choices on future outcomes is a critical step in the study of this decision process and of public education policies. 
The causal effects of schooling may vary
widely across economic agents because of heterogeneous skills and preferences. 
Moreover, individuals' expectations about potential returns may prompt them to choose schools strategically. 
Heterogeneity and selection make identification of causal effects very challenging, 
especially when students face a large number of unordered options. 
On the positive side, a growing number of schools use centralized assignment mechanisms, which produce credible instruments based on the discontinuities generated by the assignment of comparable students to different schools (\cite{kirkeboen2016} and \cite{abdul_ecta2022}).

Many centralized school assignment mechanisms effectively amount to a quasi\--ex\-per\-i\-men\-tal design where two groups of individuals who share similar scores are assigned to different schools based on how their scores relate to admission cutoffs.
The assignment in a matching characterized by such cutoffs depends on the student's preferences over feasible schools.
Unlike in a typical regression discontinuity (RD) design, in this setting, students on the same side of the cutoff do not necessarily receive the same assignment.
For example, individuals with similar scores just above a certain cutoff may all prefer to go to the same ``school $j$'' but could have very different second-best options if they fall on the other side of the cutoff.
In such a context, \cite{kirkeboen2016} construct comparable groups of individuals near
a cutoff by conditioning on local preferences---that is, by selecting students whose preferences yield identical first- and second-best options if they fall, respectively, above and below that cutoff.
Controlling for local preferences that equal a pair of schools, e.g., $(j,k)$, allows the RD to identify the causal effect of a change in the school assignment from $k$ to $j$, averaged over individuals who prefer $j$ over $k$.

Most real-world school assignment mechanisms create incentives for students to misreport their true preferences.
\cite{agarwal2018} and \cite{fack2019} provide thorough discussions with several real-world examples of this.
Subsequent empirical work have employed their methods to estimate true preferences or test for strategic behavior, for example,  \cite{campos_impact_2024}, \cite{andersson_beyond_2026}, and many others.
The empirical literature on school choice typically engages in counterfactual policy evaluations and an important class of such counterfactuals requires identification of causal parameters that are conditional on true 
preferences (\cite{Artemov2023}, Section V.B).
The identification of such causal parameters is the primary goal of our paper and we propose a novel RD strategy that controls for true local preferences. The challenge resides on the fact that true preferences are unobserved but may be partially identified under general forms of misreporting behavior.

This paper derives sharp bounds for causal effects of school assignment on future outcomes in mechanisms with cutoff characterization and strategic student behavior. 
We devise a two-step identification approach that is robust to strategic reporting of preferences. 
In the first step, the researcher partially identifies local preferences and constructs local preference sets for each student.
We provide several tools for constructing these sets in the context of 
a student-proposing deferred acceptance (DA) mechanism where constraints on the submitted preferences lead to strategic behavior.
Outside such a context, researchers may employ alternative tools to partially identify local preferences, as the second step of our procedure does not require a particular method to be used in the first step.
For example, the identification method of \cite{agarwal2018} applies to a general class of mechanisms with cutoff representation that includes variants of the DA mechanism, Boston, First Preferences First, Chinese Parallel, etc.
Finally, in the second step, the researcher employs an RD identification strategy that controls for local preference sets and partially identifies the causal effects of school assignment. 
Our bounds can be substantially tightened—and may even collapse to a single point—depending on the data and on how strong the first-stage assumptions are.
The more restrictions we impose on strategic behavior, the more we learn about true local preferences, which then leads to narrower bounds.

Strategic behavior among students depends on the characteristics of the assignment mechanism. 
A mechanism is said to be strategy proof if submitting true preferences is a weakly dominant strategy for all students. 
For example, \cite{dubins1981} demonstrate that the DA mechanism is strategy proof. 
However, this result breaks down when the mechanism imposes constraints on the preferences that students can submit.
In many real-world school assignment mechanisms, the number of schools is too large for students to feasibly rank all schools. 
The central authorities running these systems may either limit the number of schools that students may rank or impose costs on the basis of the number of schools submitted (see Table 1 Panel B by \cite{fack2019} for examples).

Our first-step tools for partial identification of local preferences naturally require assumptions on students' strategic behavior. 
We motivate our assumptions following the important contributions of \cite{haeringer2009}.
They study a game where students submit constrained preference rankings and a central mechanism allocates the students to schools.
One of their important findings is that it is rational for students to submit partial orders of their true preferences in some mechanisms.
Specifically, suppose that a mechanism is strategy proof when students are free to rank any number of schools, as under, e.g., the unconstrained DA or Top Trading Cycles (TTC) mechanisms.
Then, if the preference rankings are constrained to having at most $K$ schools, a student can do no better than selecting $K$ schools among her acceptable schools and ranking them according to her true preferences.

The key assumption for our first-step tools is that students submit only partial orders of their preferences.
This feature is in addition to the cutoff characterization of the matching, which we assume throughout the paper.  
Cutoff characterization means that a student is matched to her best feasible school, where ``best'' is defined according to  
her true preferences and a school is feasible if the student's placement score clears that school's admission cutoff. 
Our assumption about cutoff characterization is satisfied when the matching outcome is stable
\citep{azevedo2016}. 
Stability means that each student is matched to an acceptable school and all slots in preferred schools have been filled with people who have better placement scores. 
In constrained DA and TTC mechanisms, 
stability 
occurs in Nash equilibrium of the preference revelation game under appropriate conditions on the placement scores (Theorems 6.3 and 6.4, \cite{haeringer2009}). 
Stability occurs in Nash equilibrium without restrictions on the scores in the constrained serial dictatorship (SD) mechanism, which is a particular case of DA.
We characterize sharp local preference sets for every individual that are compatible with the observed data and these model assumptions. 
For an individual drawn at random, this becomes a random set that contains the true local preference with probability one.
We show how to shrink these sets by proposing a menu of behavioral assumptions on strategic behavior that researchers may find more or less appropriate to their empirical contexts. The smaller the sets are in the first step, the narrower the bounds will be in the second step, even collapsing to point identification in some cases.

The second step of our approach relies on the local preference sets constructed in the first step with either our method or an alternative method. 
Given interest in a pair of schools $(j,k)$, 
we select all individuals whose local preference sets contain $(j,k)$ and whose placement scores are close to the cutoff for admission at school $j$.
This subpopulation of individuals contains all individuals whose true local preferences equal $(j,k)$, but also other individuals.
The average outcome in the subpopulation equals a weighted average of two averages: 
first, the average outcome for individuals with true local preferences $(j,k)$, which is interesting for the identification of causal effects,
and second,
the average outcome for individuals with true local preferences that differ from $(j,k)$. 
We do not know which individuals have preferences $(j,k)$, 
but we do characterize sharp bounds on the proportion of such individuals in the subpopulation
using the random sets constructed in the first step.
Thus, our setting fits the identification problem with corrupted data studied by \cite{horowitz1995}. 
This method allows us to derive closed-form bounds on the first out of the two average outcomes above, which then leads to bounds on the average causal effects.
This closed-form approach offers some intuition on when we can expect the bounds to be informative about or equal to the actual average causal effect (i.e., point identification). 
Although practical and intuitive, these closed-form bounds may not be sharp. 
Thus, building on \cite{molinari2020} and using random set theory, we characterize sharp bounds that are numerically computable when outcomes take finitely many values.

Methods combining RD identification with school matching data have been popular among applied and theoretical researchers in economics for at least 15 years \citep{kirabo_nber2009,popurquiola2011,bertanha_jmp2014}.
To the best of our knowledge, our paper is the first to prove RD identification of returns of school assignment in matching mechanisms with strategically reported preferences.
We note that our two-step approach differs from the usual control function approach because our first step partially identifies the control variable instead of point-identifying it as in the usual approach. For this reason, we call it control mapping approach. Various other examples in econometrics feature this structure, see \cite{han_kaido_2025}.

Our paper unifies and complements two branches of the literature. 
One branch features the methods proposed by \cite{agarwal2018} and \cite{fack2019} that take into account strategic reporting and identify students' true preferences, but
they do not focus on causal effects of school matches on future outcomes. 
The other branch features the methods of \cite{kirkeboen2016} and \cite{abdul_ecta2022} that identify the causal effects of different assignments but control for reported instead of true preferences.\footnote{In the same branch, \cite{chen2026} provides a comprehensive set of causal parameters  that are identified when  
assignment is based both on lottery- and RD-driven variation.}
An RD strategy that controls for reported 
preferences 
identifies causal parameters that could be useful for a class of counterfactuals that differs from the one we consider in this paper.
It is important to note, however, that in some empirical contexts, agents may have precise \textit{ex ante} information about \textit{ex post} matching cutoffs at the time when they choose preferences to report. In some cases, this can even lead to a discontinuous change of submission behavior at the \textit{ex post} cutoffs.
We present evidence of such behavior in the empirical context of Chile (Section \ref{sec:empir:strategicbehavior}).
Such discontinuous behavior may confound identification in an RD that controls for reported preferences; 
we find that controlling for true preferences makes the RD strategy robust to this possibility.

Our work, and the kind of parameters we target, open avenues for studying identification in various policy settings while allowing for strategic reporting. 
Recent work in causal inference seeks to identify counterfactual policy effects  under interference from market equilibrium or centralized assignment \citep{Arkhangel2025,munro_causal_2025}.
This line of work relies on truthful reporting but acknowledges its limitations and the importance of new developments that allow for strategic reporting.

We apply our two-step identification strategy to matching data from Chile.
Chile has a centralized DA mechanism that assigns students to university--major pairs.
To have an idea of the figures, in 2010, 88,000 students were constrained to rank at most eight university-major pairs ($K=8$) out of a total of 1,092 options available, in which case the theory  predicts lack of truth telling.
Prior work documents strategic behavior in Chile \citep{larroucaurios2020,larroucaurios2021} and we present two additional pieces of evidence.
First, \textit{ex post} admission cutoffs are highly predictable \textit{ex ante} for many university-major pairs (Figure \ref{fig:cutoff}).
Second, preference submission behavior changes when students' \textit{ex ante} placement scores are close to \textit{ex post} cutoffs, strongly suggesting students act on information on prior cutoffs (Figure \ref{fig:appliproba_distcutoffmean}). 
For some major pairs, the submission behavior even changes discontinuously at the \textit{ex post} cutoffs (Figure \ref{fig:discontinuities}).
The methods proposed by \cite{kirkeboen2016} and \cite{abdul_ecta2022} are unable to provide the answers we seek with the Chilean data.
First, strategic behavior leads their methods to identify a parameter that differs from an average treatment effect conditional on true preferences, which is our interest.
Second, even if all agents report the truth, their methods are not directly applicable because of some specific aspects of the Chilean data.\footnote{
The methods of \cite{kirkeboen2016} apply to the SD mechanism, which is a particular case of the Chilean DA mechanism. In DA contexts, individual counterfactual sets must be defined in a more general way (see Definitions \ref{def:counter:budget} and \ref{def:nextbest} below).
While \cite{abdul_ecta2022} address this issue by creating a propensity-score control variable, their technique relies on placement scores with integer priorities plus continuous full-support lottery priorities. 
This particular structure does not correspond to the Chilean case, where program-specific placement scores are computed as functions of five primitive scores, and sometimes these functions are nonlinear. 
The Chilean setting requires again the counterfactual sets of schools to be carefully defined such that controlling for local preferences does not violate the continuity assumptions required by RD (see Assumption \ref{aspt:continuity} and Lemma \ref{lemma:contqj} below).
}

Our analysis of the Chilean data proceeds in two  parts.
First, we show evidence that  outcomes are heterogenous with respect to true local preferences.
We compute bounds for average graduation rates if assigned to program $k$ among students with  true local preference $(j,k)$.
We fix the next-best alternative $k$ to be the Bachillerato de Ingreso Común at the University of Chile---a selective entry program with a large applicant pool that allows students to explore multiple disciplines before settling on a major, and is often viewed as an alternative route into competitive programs.
We vary the first-best option $j$ and find non-overlapping bounds in several cases.
The results reveal substantial heterogeneity across $j$, indicating that students’ preferences matter for graduation outcomes and are likely correlated  with unobserved factors such as effort or ability not captured by test scores.
The second part of our analysis focuses on average effects on graduation outcomes from changing the initial assignment  from $k$ to $j$ among students with true local preference $(j,k)$.
We choose $j$ to be Medicine at PUC Santiago because it is a highly-selective and popular program with a large applicant pool. 
We fix $j$ and plot bounds for common next-best alternatives $k$.
In most cases, our bounds identify a positive effect of assignment into $j$ on graduating from $j$, with magnitudes that vary across alternatives $k$. 

More generally, we compute bounds for three hundred college-major pairs $(j,k)$ under assumptions of varying strength on strategic behavior, which provides a valuable sensitivity exercise for researchers to evaluate the empirical content of their assumptions.
We highlight three important takeaways from our bounding exercise. 
First, bounds are generally wide under the weakest of our assumptions, namely, that students simply select any subset with at most $K$ options from their true preference list while preserving the true relative ordering. 
We call this weak partial order (WPO). 
Wide WPO bounds essentially say that RD alone applied to matching data with strategic behavior does not reveal much unless researchers are willing to make stronger assumptions on strategic behavior.
Second, bounds are generally narrow and even collapse to point identification in some cases under the strongest of our assumptions: strong partial order (SPO). 
SPO says that agents who rank less than $K$ programs reveal their true preferences.  
Third, point-estimates produced by assuming truth-telling for all agents fall outside our SPO bounds in many cases.
This is relevant because the truth-telling assumption implies the SPO assumption, and we would expect point-estimates inside the bounds. 
Our finding suggests violation of the truth-telling assumption despite the fact that 80\% of students in our sample rank less than $K$ programs.
This runs counter to the common intuition that students will report truthfully whenever the $K$ constraint is non-binding for most of them.

The rest of this paper proceeds as follows. 
Section \ref{sec:model} lays out the matching model for a continuum population of students and a finite number of schools.
Section \ref{sec:id_ur} examines point identification of average treatment effects when students are truth-tellers. 
Section \ref{sec:id_r} examines partial identification when students strategically report their preferences, with two subsections:
Section \ref{sec:id_r:qj_id} provides tools for construction of local preference sets that apply to constrained DA mechanisms, and Section \ref{sec:id_r:partial_id_givenqj} discusses how to use local preference sets constructed in this or other ways to derive bounds on the average treatment effects. 
We illustrate our identification approach with the Chilean data in Section \ref{sec:app}.
The appendix presents all proofs for the paper plus additional results.

\section{Model}\label{sec:model}
\indent 

We consider a continuum population of students and a set of $J$ schools, 
$\m{J} := \{1,\ldots, J\}$, 
that have capacities $\{ q_1, \ldots, q_J \}$ defined in terms of shares of the student population \citep{azevedo2016}.
Denote by $\Omega$  the set of all students in the universe of interest and use $\omega$ to index an individual student type.
The student type consists of three objects.
First, $Q(\omega)$ denotes the true (strict) preference relation of student $\omega$ over the set of options  $\m{J}^0 := \mathcal{J} \cup\{0\}$, which includes schools $\m{J}$ and 
an outside option $0$.
For example, if $J=2$ and $Q(\omega) = \{1,2,0\}$, then  1 is preferred to 2 (i.e., $1 Q(\omega) 2 $), 1 is preferred to 0 (i.e., $1 Q(\omega) 0 $), and 2 is preferred to 0 (i.e., $2 Q(\omega) 0$).
Let $\mathcal{Q}$ be the set of all strict preference relations over $\mathcal{J}^0$ that admit at least one school that is acceptable.
A school $j \in \m{J}$ is ``acceptable'' for student $\omega$ if it is preferred to that student's  outside option, i.e., $jQ(\omega)0$.
We define $\bar{Q}$ as the weak preference relation induced by $Q$, i.e., $j \bar{Q} k \Leftrightarrow j Q k $ or $j=k$.
The second object of the student type is a vector of scores $\bR(\omega):=(R_1(\omega), \ldots, R_J(\omega)) \in \m{R} \subseteq \mmr^J$, where each school $j$ utilizes $R_j$ to rank students for admission. 
The third and last object, $Y(\omega,d)$, is the potential outcome of student $\omega$ if the student is assigned to option $d \in \mathcal{J}^0$.
Each student has a potential outcome function $Y(\omega,\cdot)$ that maps from 
$\mathcal{J}^0$ to $\m{Y} \subseteq \mathbb{R}$.
We call $\boup{\Gamma}$ the set of all possible potential outcome functions.
The set of all student types is $\Omega := \mathcal{Q} \times \m{R} \times \boup{\Gamma}$.
In a continuum economy, there is a probability measure $\mathbb P$ over $\Omega$ and the 
Borel $\sigma$-algebra of the product space $\Omega$.
We suppress the argument $\omega$ whenever it is unnecessary for ease of notation, e.g., $Y(d)$ vs. $Y(\omega,d)$ and $Q$ vs. $Q(\omega)$.

A ``matching'' is described by a measurable function $\mu:\Omega \to \mathcal{J}^0$ that satisfies two conditions: 
for every $j\in\mathcal{J}$, 
(i) the mass of students matched to $j$ is less than or equal to the capacity of school $j$, i.e., $\mathbb P \{\omega: \mu(\omega) = j \} \leq q_j$;
and
(ii) the set of students who weakly prefer option $j \in \m{J}^0$ over their matching, i.e., $\{\omega: j \bar Q(\omega) \mu(\omega) \}$, is an open set.\footnote{ 
\cite{azevedo2016} impose the same condition to rule out multiplicity of stable matchings that differ in a set of types with measure zero.}
For every student type $\omega$, $\mu(\omega)$ is either the school $j$ to which the student is matched or zero.
When $\mu(\omega)=0$, the student is unmatched and takes an outside option.
An important definition for this paper is that of stability.
\begin{definition}[Stability]\label{def:stable}
The matching $\mu: \Omega \to \m{J}^0$ is a stable matching if three conditions are satisfied for every $\omega \in \Omega$:
(i) $\mu(\omega) \bar Q(\omega) 0$ (individual rationality);
(ii) for any $j \in \m{J}$, if $j Q(\omega) \mu(\omega)$, then $j$ is full (no waste);
and
(iii) for any $j \in \m{J}$ that is full, if $\mu(\omega')=j$ and $ j Q(\omega) \mu(\omega)$, then $R_j(\omega') > R_j(\omega )$ (no justified envy).
\end{definition}

A mechanism $\varphi$ matches students to schools by mapping the students’ scores and submitted preference lists to schools.
Student $\omega$ submits a preference list $P(\omega) \subseteq \m{J}$, which is an ordered list of her acceptable schools.
For example, for $J=3$, if $Q(\omega)=\{1,2,0,3\}$, then $P(\omega)=\{1,2\}$, as long as the student submits her true list of acceptable schools.
The number of schools in $P$, denoted $|P|$, is at least one because everyone participating in the match has at least one acceptable school. 
As with $\bar Q$, we also define $\bar P$ as the weak preference relation induced by $P$.
A mechanism takes as inputs everyone's submitted preferences (i.e., $P$ which is a mapping from $\Omega$ to an ordered subset of $\m{J}$) and everyone's scores (i.e., a vector-valued function $\bR:\Omega \to \m{R}$) 
and gives rise to a matching function.
Formally, $\varphi(P,\bR):\Omega \to \m{J}^0$.
We say a mechanism $\varphi$ is strategy proof if, for every student, submitting the true ranking of acceptable schools is a weakly dominant strategy---in other words, 
any student $\omega$'s misreporting of $P$ never leads to a better option and sometimes leads to a worse option, depending on what other students submit.
We say a student is a truth-teller if her $P$ equals her true ranking of acceptable schools. 
Otherwise, we say she is strategic or not a truth-teller.

The ability to characterize a matching allocation on the basis of cutoffs is fundamental for this paper.
\begin{definition}[Cutoff Characterization]
\label{def:cutoff}
For placement scores 
$\bS : \Omega \to \m{S} \subseteq \mmr^J$,
$\bS(\omega):=(S_1(\omega), \ldots, S_J(\omega))$ 
and admission cutoffs
$\boup{c} \in \m{S}$, $\boup{c}:=(c_1,\ldots, c_J)$,
the set of feasible options of a student $\omega$ equals all schools for which her placement scores clear the admission cutoffs plus the outside option: 
$\{0\} \cup \{j \in \m{J} : S_j(\omega) \geq c_j \}$;
student $\omega$'s best feasible option is the option that ranks first according to $Q(\omega)$ among her feasible options.
We say the matching $\mu: \Omega \to \m{J}^0$
has cutoff characterization if there exist placement scores 
$\bS : \Omega \to \m{S}$
and admission cutoffs
$\boup{c} \in \mathcal S$
such that, for every $\omega \in \Omega$,
the matching $\mu(\omega)$ equals student $\omega$'s best feasible option according to $Q(\omega)$.
\end{definition}

This paper considers mechanisms that produce matching functions with a cutoff characterization according to Definition \ref{def:cutoff}.
Placement scores $\bS$ may or may not equal school priority scores $\bR$.
The definition gives researchers the freedom to construct special placement scores $\bS$ if the mechanism that they consider does not admit cutoff characterization by means of priority scores $\bR$.
The idea behind this definition stems from the logic of the general class of mechanisms of \cite{agarwal2018}. 
Moreover, we assume throughout the paper that, for any $j\in \m{J}$, 
$S_j$ has a distribution that is absolutely continuous with respect to (wrt) the Lebesgue measure and support $\mS_j$ that contains a closed interval around $c_j$.
The support of $\bS$ is $\mS$.
For any two scores $S_j$ and $S_l$, we assume that either $\mmp[S_j = S_l]=1$
or
$\mmp[f(S_j) = S_l]<1$ for any measurable function $f$.
This says that the only deterministic function relating any two scores may be the identity function.

\cite{azevedo2016} demonstrate that, if the matching function is stable, 
the matching has cutoff characterization 
with $\bS = \bR$ and admission cutoffs constructed as follows.
For each $j \in \m{J}$, $c_j := \inf \{ S_j(\omega) : \text{ for } \omega \text{ with } \mu(\omega)=j \}$ 
if some individuals are matched to $j$ or 
$c_j:=\inf \mS_j$ if nobody is matched to $j$.
In fact, most empirical work applying RD to matching data in the last 15 years rely on this stability-based method for constructing cutoffs.
Many mechanisms produce stable matchings.
For example, SD and DA are strategy proof and lead to stable matchings if agents are truth-tellers.

Regarding settings where agents are not truth-tellers, e.g., because they face constraints in the submission of $P$,
\cite{haeringer2009} and \cite{fack2019} show how stability arises in equilibrium with strategic agents.
In particular, \cite{fack2019} formally test for stability in their empirical context of Parisian schools and do not find evidence against it. 
One possible threat to stability is agents that make mistakes during submission of their preferences. 
\cite{artemovCheHe2017,Artemov2023} explicitly study stable matching with mistaken agents. 
They argue that most of the mistakes observed in equilibrium are likely to be payoff-irrelevant in large markets, and thus not hurting stability of the match.
Despite the strong support for stability in the literature, we emphasize that Definition \ref{def:cutoff} allows for the general cutoff characterization of
\cite{agarwal2018} that does not require stability.

\subsubsection*{Parameter of Interest}

\indent

Having laid out the physical environment of our matching economy, we now motivate the type of causal parameter that concerns this paper. 
Causal parameters are often used as policy tools for their \textit{internal validity}—evaluating the impact of historical interventions on outcomes—or for their \textit{external validity}—assessing the impact of new interventions based on knowledge of historical interventions, that is, constructing counterfactual analyses; 
see \cite{heckman2005scientific} for a detailed discussion. 

In this paper, we aim to provide strategies for identifying a set of parameters that are not only internally valid but also possess substantial external validity, thereby enabling the evaluation of counterfactual policies.
We seek to identify conditional moments of treatment effects $Y(d')-Y(d)$ when
the econometrician has access to an infinite amount of data and observes
the joint distribution of the following random objects: 
$P(\omega)$, $\bS(\omega)$, $\mu(\omega)$, and $Y(\omega) := Y(\omega, \mu(\omega))$.\footnote{
We abuse the notation and employ the letter $Y$ for both the observed outcome, $Y(\omega)$, 
and the potential outcome of being assigned to school $d$, $Y(\omega,d)$.}

Section \ref{sec:app:parinteres} in the appendix provides a detailed discussion of a class of counterfactual analyses that is advocated by \cite{Artemov2023}. 
Analyzing the welfare effects of a policy change in that class relies on identification of 
the average structural functions (ASF) 
\begin{equation}
s \mapsto \mathbb{E}[Y(j) \mid Q = q, \; S_j = s]
\label{eq:asf}
\end{equation}
for $j \in \{0,\ldots,J\}$ and $q \in \mathcal{Q}$. 
 Nonparametric identification of these functions is extremely challenging.
However, this paper shows that combining an RD identification strategy with data from assignment mechanisms that satisfy our assumptions yields set identification of \textit{differences} of the ASFs in \eqref{eq:asf}; in other words, our proposed methods set-identify  
\begin{equation}
\mathbb{E}[ Y(j) - Y(k) \mid Q = q, \; S_j = c_j]
\label{eq:param0}
\end{equation}
at finitely many values of $j$, $k$, and $q$ (see Propositions \ref{result:truth:identif} and \ref{result:bounds_RD}).

The parameters in \eqref{eq:param0} are the object of interest of this paper.
They carry the simple intuition of the massively popular RD designs while providing useful identifying information for a large class of counterfactual policy evaluations. 
Section \ref{sec:app:parinteres} in the appendix describes 
how researchers may combine all information identified by RD with smoothness assumptions on the ASFs to construct bounds on a variety of counterfactual policy effects.

\section{Identification with Truthful Reports}
\label{sec:id_ur}
\indent 

In this section, we consider identification of causal effects when all students are truth-tellers, that is, when they submit their true ranking of acceptable schools.
We start with truth-telling in order to introduce our approach and notation in a simple behavioral setting before moving to strategic reports in Section \ref{sec:id_r}.

\begin{assumption}[Truth-telling]\label{aspt:truth:telling}
Students submit their true list of acceptable schools.  
\end{assumption}

The identification strategy of this paper resembles a sharp RD design. 
Our goal is to identify the effects of the school of assignment on future outcomes. 
Another interesting question is the effect of the school of \textit{graduation} on future outcomes---to answer it we would require a strategy resembling a fuzzy RD because some students do not graduate from the same school they are assigned to.
We defer this identification problem to future work as several issues beyond the scope of this paper (e.g., multiple compliance types with unordered treatments) arise in that case.

Unlike in a standard sharp RD, that $S_j(\omega)$ clears the cutoff $c_j$ does not automatically determine that student $\omega$ is allocated to school $j$.
This is the case only when $j$ is the most preferred school among the schools that are feasible to the student, 
that is, when $j$ is the favorite school in the set of schools for which the student clears the cutoff.

The first step in the RD is to correctly identify the marginal individuals for a given cutoff 
and a given change in schools.
For example, for any individual with score $S_j$ just to the right of $c_j$, we need to determine two things: that the individual is matched to school $j$, and that the individual would have been matched to school $k$ had her score been just to the left of $c_j$. 
The cutoff representation implies that these two things depend
on the counterfactual sets of available schools on either side of the cutoff and on the individual's preferences over these sets.

It is straightforward to obtain counterfactual sets of available schools
in the case where all schools rely on the same placement score, that is, $S_j=S_1$ for every $j$.
For example, this is the case under the SD mechanism. 
In this case, the set of feasible schools is all schools with a cutoff below or equal to score $S_1$.
Note that everyone just above (or just below) cutoff $c_j$ has exactly the same set of feasible schools. 
For someone with $S_1 \geq c_j$, the counterfactual scenario has the score crossing to the left of cutoff $c_j$, 
and school $j$ is dropped from the set of feasible schools.
In turn, for someone with $S_1 < c_j$, the counterfactual scenario adds school $j$ to the set of feasible schools.
Unlike under SD, agents near and on the same side of a cutoff in DA differ in their sets of feasible schools. 
It is not immediately obvious which schools appear in their counterfactual sets.
In DA, schools use different scores, and these scores may be functions (e.g., weighted averages) of a small set of primitive scores.
That is the case in our application with the Chilean data.
This makes the joint support of the distribution of scores highly dependent and complicates the counterfactual analysis.
Dealing with this complexity is empirically relevant since many real-world higher education assignment mechanisms use DA.
See Table 1 Panel B in \cite{fack2019} for a list of examples.\footnote{
A common practice in applied work consists of ``cleaning'' irrelevant schools from the submitted preference lists in cases where $S_j=S_1$ for every $j$.
For instance, say an individual submits $P=\{ 1,2,3 \}$ and $c_2 > S_1 > c_1 > c_3$.
Given the cutoff characterization and truth-telling, 
the matching assignment of this individual is school $1$; 
the counterfactual assignment when $c_2 >  c_1 > S_1 >  c_3$ is school $3$ even though $2P3$;
this is the case because school $2$ has a cutoff higher than the cutoff of school $1$. 
In this case, the irrelevant school to be cleaned from $P$ is school $2$.
The general idea is to remove all schools ranked below $1$ that have cutoffs higher than $c_1$. 
See the description of this practice by \cite{estrada2017}.
The practice cannot be used to identify counterfactual assignments in cases where different schools use 
different placement scores, as under, e.g., the DA mechanism. 
}

Our framework allows for a variety of joint distributions on the vector of scores $\bS$ and works with the definition of counterfactual budget sets below.
\begin{definition}[Counterfactual Budget Sets]
\label{def:counter:budget}
Consider a student with a vector of scores $\bS$.
The budget set for this student is her set of feasible options,
\begin{align}
B(\bS) := & 
\{ 0 \} \cup \{m \in \m{J} ~:~ S_m \geq c_m \}.
\notag
\end{align}
Fix a school $j \in \mJ$ with cutoff $c_j$.
The right-counterfactual budget set for this student at cutoff $c_j$
is $B^+_j(\bS) := B(\bS) \cup \{m : S_m=S_j \text{ and } c_m=c_j\}$;
the left-counterfactual is 
$B^-_j(\bS) := B(\bS)\setminus \{m : S_m=S_j \text{ and } c_m=c_j \}$,
where $C\setminus D$ equals the set $C$ minus the elements of set $D$.
\end{definition}

To fix ideas, we consider a simple example throughout the paper in the context of the SD mechanism.

\begin{example*}[SD Example, Part I] 
In SD, $S_j=S_1$ for every $j$, and the definition of the budget set above equals $B(\bS) = \{ 0 \} \cup \{m : S_1 \geq c_m \}$.
Suppose we have four schools with cutoffs $c_1 < c_2 < c_3 < c_4$.
Individuals in this economy have five possible budget sets: 
$\{0\}$, $\{0, 1\}$, $\{0,1,2\}$, $\{0,1,2,3\}$, and $\{0,1,2,3,4\}$.
For individuals near cutoff $c_4$, the counterfactual budget sets are $B^-_4(\bS) = \{0,1,2,3\}$ and $B^+_4(\bS) = \{0,1,2,3,4\}$.\footnote{ 
In SD or in DA with independent placement scores, the definitions of the counterfactual equal
$B^+_j(\bS) = B(\bS) \cup \{m : c_m=c_j\}$ and
$B^-_j(\bS) = B(\bS) \setminus \{m : c_m=c_j \}$;
moreover, if the cutoffs are unique, 
$B^+_j(\bS) = B(\bS) \cup \{j\}$
and
$B^-_j(\bS) = B(\bS) \setminus \{j \}$. 
}
\end{example*}

Next, we follow the intuition of \cite{kirkeboen2016} and define the concept of local preferences, that is, 
the first- and second-best choices for a marginal individual at any given cutoff. 
This will later become the control variable in our RD identification strategy with truthful agents.

\begin{definition}[Local Preferences]
\label{def:nextbest}
Fix a school $j \in \mJ$ with cutoff $c_j$.
Consider a student $\omega$ with preference $Q(\omega)$ and scores $\bS(\omega) $.
For any pair of options $(k,l)\in \m{J}^0 \times \m{J}^0$,
we say that $(k,l)$ is the local preference of 
student $\omega$ 
at cutoff $c_j$
if the favorite feasible option of student $\omega$ shifts from $l$ to $k$ as we exogenously increase $S_j(\omega)$ from being smaller than $c_j$ to being larger than $c_j$.
We define the true local preference of this student as the pair $Q_j(\omega) := (k,l)$.
Formally, for a set of options $B \subseteq \mJ^0$, define the best option in $B$ according to $Q$ as $Q(B)$.
We have that $Q(B) = m \Leftrightarrow m \in B \text{ and } m \bar Q(\omega) n ~ \forall n \in B.$
Finally, $Q_j(\omega) = (k,l)$ if, and only if,
$Q(B^+_j(\bS))=k$ 
and 
$Q(B^-_j(\bS))=l$.
The reported local preference $P_j(\omega)$ is defined in a similar fashion.
If $B \cap P \neq \emptyset$,
$P(B) = m \Leftrightarrow m \in B \text{ and } m \bar P(\omega) n ~ \forall n \in B;$
otherwise, if $B \cap P = \emptyset$, $P(B)=0$.
We have that $P_j(\omega) = (k,l)$ if, and only if,
$P(B^+_j(\bS))=k$ 
and
$P(B^-_j(\bS))=l$.
\end{definition}

\begin{example*}[SD Example, Part II] 
Consider four individuals whose submitted preferences are
$P^{(1)}=\{3,4,2,1\}$, $P^{(2)}=\{4,1,2,3\}$, $P^{(3)}=\{4,2,3,1\}$, and $P^{(4)}=\{4,3,1,2\}$.
Their corresponding local preferences at cutoff $c_4$ are
$P^{(1)}_4=(3,3)$, $P^{(2)}_4=(4,1)$, $P^{(3)}_4=(4,2)$, and $P^{(4)}_4=(4,3)$.
\end{example*}

There is no distinction between $P_j$ and $Q_j$ at this stage because of Assumption \ref{aspt:truth:telling}.
Students are truthful when they submit their list of acceptable schools, so $P$ equals the schools listed higher than $0$ in $Q$ and $P_j=Q_j$.
Section \ref{sec:id_r} considers the case of strategic misreporting. 
In that case, $P_j$ is observed but $Q_j$ is not.
Cutoff characterization implies that an individual with $Q_j=(k,l)$ is matched to school $k$ if $S_j$ is just above $c_j$ 
or to school $l$ if $S_j$ is just below $c_j$.
The same applies for individuals with $P_j=(k,l)$ under Assumption \ref{aspt:truth:telling}.

The local preference pair $(j,k)$ at a certain cutoff $c_j$ is useful for identification
only if there exists a positive fraction of individuals in the data near cutoff $c_j$ with those local preferences. 
We collect such useful pairs in the set $\m{P}$.

\begin{definition}[Comparable Pairs] \label{def:compairs} We say $(j,k) \in \m{J} \times \m{J}$, $j \neq k$,  is a comparable pair of alternatives if 
(i) $c_j$ is an interior point of the support $\m{S}_j$  
and
(ii) $\mmp [ Q_j =(j,k) | S_j =s]$ is bounded away from zero for 
$s$ in an open neighborhood of $c_j$.
Finally, we define $\mathcal P \subseteq \mathcal{J} \times \mathcal{J}$ as the set of all comparable pairs.\footnote{
We adopt the convention that comparable pairs do not involve the outside option $0$.
We do this because it may be hard to interpret the treatment effects of a change from the outside option to a school when the outside option varies across individuals.
We do not consider pairs with $j=k$ because the initial school assignment does not change for these individuals. 
We also exclude pairs $Q_j = (j',k)$ with $j'\neq j$ from $\m{P}$ to avoid redundancy.
We may find individuals with $Q_j = (j',k)$ whenever schools $j$ and $j'$ use the same score and have the same cutoff.
As the score $S_j = S_{j'} $ crosses the cutoff $c_j = c_{j'}$, access is granted to both schools $j$ and $j'$, and individuals may differ in their preferences for these schools.
Individuals who prefer $j'$ will not appear in $\m{P}$ as having $Q_j = (j',k)$, but they may appear in $\m{P}$ with $Q_{j'} = (j',k)$.
}
\end{definition}

The purpose of defining counterfactual sets and local preferences is to construct a 
variable for every student and use it as a control variable in the RD.
In this section, this variable is $P_j$, which equals to $Q_j$ because of Assumption \ref{aspt:truth:telling}.
When we focus on students with $P_j =(j,k)$, $k\neq j$, the marginal switch in the allocation around cutoff $c_j$ becomes a function of $S_j$.
Controlling for $P_j$ ensures that we apply the RD strategy to all individuals whose assignment switches from $k$ to $j$ at the cutoff;
this identifies the effect of the change in the assignment as long as the typical RD continuity assumptions are satisfied.

RD identification requires continuity assumptions on the distribution of individual types conditional on the relevant placement score.
In our case, we also need to verify continuity after we condition on $P_j=Q_j$.
After all, we do not want to condition on a variable that breaks the central argument for identification in RD:
that individuals to the right and the left of the cutoff are ``similar on average''. 
Below, we state an assumption on the continuity of types and prove that it implies the kind of smoothness required by RD.
Before we do so, we define the following set of events.
For every school $j\in \m{J}$, partition the placement scores $\bS$ as
the score of school $j$ and all other scores: $\bS \equiv (S_j,\bS_{-j})$. 
Define $\boup{A}_{-j}$ to be the collection of events on $\bS_{-j}$ that determine the availability
of all non-$j$ schools. 
There are $2^{J-1}$ such events in $\boup{A}_{-j}$.
For example, if $J=2$, $\boup{A}_{-1}=\{ \{ S_2\geq c_2\}, \{ S_2 < c_2\}\}$; 
if $J=3$,
$\boup{A}_{-1}=\{ 
\{ S_2 \geq c_2, S_3 \geq c_3 \},
\{ S_2 \geq c_2, S_3 < c_3 \},
\{ S_2 < c_2, S_3 \geq c_3 \},
\{ S_2 < c_2, S_3 < c_3 \} 
\}$; etc.

\begin{assumption}(Continuity of Types)\label{aspt:continuity}
Consider a school $j$ with cutoff $c_j$ in the interior of the support $\m{S}_j$.
Assume the following functions of $s$ are all continuous
at $s=c_j$: 
(i) $\mmp[\bS_{-j} \in A_0,  Q =Q_0|   S_j = s ]$ for any $A_0 \in \boup{A}_{-j}$ and $Q_0 \in \m{Q}$
and 
(ii) $\mme[ ~ g (Y(d)) ~  \mathbb{I}\{ \bS_{-j} \in A_0,  Q =Q_0 \} ~  | ~  S_j = s   ]$
	for any $A_0 \in \boup{A}_{-j}$, $Q_0 \in \m{Q}$, and $g\in \m{G}$, where $\m{G}$ is a set of measurable functions 
	$g:\mathbb{R} \to \mathbb{R}$ that includes the constant function $g(y)=1$ and the identity function $g(y)=y$.
\end{assumption}

\begin{lemma}
\label{lemma:contqj}
Suppose Assumption \ref{aspt:continuity}  holds.
Consider a school $j$ with cutoff $c_j$ in the interior of the support $\m{S}_j$,
and choose two schools $k,l  \in \m{J}^0$ such that
$\mmp[Q_j=(k,l)|S_j=c_j]>0$.
Then, for any function $g \in \m{G}$ and any 
$d \in \mathcal{J}^0$, we have that
$ \mme[ g (Y(d))   | Q_j=(k,l),  S_j = s ]$ 
and
$\mmp[ Q_j=(k,l) |  S_j = s ]$
are continuous functions of $s$ at $s=c_j$.
\end{lemma} 
\noindent The proof of this lemma and all other proofs appear in the appendix. 
Finally, Assumptions \ref{aspt:truth:telling} and \ref{aspt:continuity} give sufficient conditions for identification for comparable pairs of school changes.

\begin{proposition}
\label{result:truth:identif}
Suppose Assumptions \ref{aspt:truth:telling}--\ref{aspt:continuity} hold.
For any pair $(j,k)  \in \m{P}$, 
\begin{align*}
&\mme[g(Y(j)) - g(Y(k)) | Q_j=(j,k), S_j=c_j ] 
\\
& \hspace{2cm}= \mme[g(Y) | P_j =(j,k), S_j=c_j^+] -  \mme[g(Y) | P_j = (j,k), S_j=c_j^-],
\end{align*}
where 
the condition $S_j=c_j^+$ denotes the limit as $S_j \downarrow c_j$
and 
the condition $S_j=c_j^-$ denotes the limit as $S_j \uparrow c_j$.
\end{proposition}

Proposition \ref{result:truth:identif} shows that a standard RD is valid in the truth-telling case as long as we control for $P_j$.
The parameter of interest is the average treatment effect on $g(Y)$ from a change in the school of assignment from $j$ to $k$, 
averaged over individuals at the cutoff $c_j$ and with true local preferences $(j,k)$ ---fomally:
\begin{equation}\label{eq:param}
\mme\left[ g(Y(j)) - g(Y(k)) \left| Q_j=(j,k), S_j=c_j \right. \right], \; (j,k) \in \m{P}, \; g\in \m{G}.
\end{equation}

Note that our parameter of interest \eqref{eq:param} does not condition on the full preference profile ~$Q = q$ as in \eqref{eq:param0} but on local preferences $Q_j=(j,k)$.
We do so having in mind the data constraints inherent to the local nature of  RD estimation. 
It follows that one value of $Q_j$ maps to multiple values of $Q$ for individuals with $S_j=c_j$  and 
\eqref{eq:param} equals  weighted averages of 
\eqref{eq:param0} over values of $Q$ (see proof of Lemma \ref{lemma:contqj} in Section \ref{proof:lemma:contqj} of the appendix).

\section{Identification with Strategic Reports}
\label{sec:id_r} 
\indent 

This section studies identification of causal effects when students are strategic in reporting their rankings of acceptable schools.
Strategic reports make $P_j$ generally different from $Q_j$, and $Q_j$ is not observed.
In contrast to Proposition \ref{result:truth:identif}, controlling for $P_j$ likely identifies a weighted average of various average treatment effect parameters; 
however, the weights are unknown because they depend on the unknown mapping $\omega \mapsto Q_j(\omega)$.
In addition, there is the possibility of a more extreme form of strategic behavior, that is, agents may change their preference submission discontinuously at the cutoff. 
We provide empirical evidence of such behavior in Section \ref{sec:empir:strategicbehavior}.
In this case, controlling for $P_j$ may break internal validity of the RD, and the strategy of controlling for $P_j$ no longer identifies a weighted average of average treatment effect parameters.
Although our methods are also robust to this extreme possibility, we emphasize that it is not the only situation where researchers may find our methods useful. 
Our methods seek to identify \eqref{eq:param} but controlling for $P_j$ does not identify \eqref{eq:param} under general forms of strategic behavior, regardless if it is discontinuous or not.

We propose a two-step identification approach. 
In the first step, the researcher characterizes the set of true local preferences $Q_j$ that is compatible with the data and appropriate behavioral assumptions.
In the second step, the researcher controls for the constructed local preference sets and partially identifies the parameters in \eqref{eq:param}.
We discuss the first and second steps in Sections \ref{sec:id_r:qj_id} and \ref{sec:id_r:partial_id_givenqj}, respectively. 
Note that our two-step approach differs from the usual two-step control function approach in econometrics.
The usual approach is to point-identify the control variable in the first step, while
our approach involves partially identifying the control variable. 
Thus, we refer to our two-step procedure as a control mapping approach.

Section \ref{sec:id_r:qj_id} presents several tools for the identification of local preference sets. 
These tools rely on assumptions known to be appropriate in SD and DA contexts, although we do not rule out their applicability in contexts with other mechanisms;
e.g., the TTC mechanism satisfies one of our assumptions, such that some of the tools from Section \ref{sec:id_r:qj_id} are still useful.
More generally, researchers may utilize preference identification tools that work under alternative assumptions, for example, 
the methods of \cite{agarwal2018} and \cite{fack2019}.
Either way, the researcher must construct a set of local preferences for each individual in the first step.

Section \ref{sec:id_r:partial_id_givenqj} describes the second step of our procedure.
This step features high-level assumptions imposed on the local preference sets such that the researcher is not restricted to the methods of Section \ref{sec:id_r:qj_id}. 
In particular, depending on the data and the strength of assumptions in the first step, each individual local preference set may collapse to a unit set. 
This leads to  point identification in our second step.
For example, this will be the case if researchers utilize the parametric identification tools of \cite{agarwal2018} or \cite{fack2019} in the first step.

\subsection{Partial Identification of Local Preferences}\label{sec:id_r:qj_id}
\indent

This section provides tools for set identification of local preferences using assumptions on agents' behavior and the mechanism.
These assumptions are specific to this subsection, and we motivate them with reference to the context of the constrained DA mechanism studied by \cite{haeringer2009}.
\cite{haeringer2009} study a game where students submit constrained preference rankings and a mechanism matches students to schools as a function of $P$, $\bR$, and schools’ capacities.
Although the unconstrained DA mechanism is strategy proof, many real-world implementations of DA restrict the number of schools that students can submit in their rankings. 
In this case, there is a cap $K<J$ such that $1 \leq |P| \leq K$, and the submitted ranking $P$ is generally different from the list of acceptable schools in $Q$.
When implemented in this way, the DA mechanism is not strategy-proof, and there are no dominant strategies.
Strategyproofness also breaks down if, instead of facing a cap, students incur an application cost as a function of the number of schools submitted \citep{fack2019}.

Lemma 4.2 by \cite{haeringer2008} shows that, if a mechanism is strategy proof when $K=J$, then, in the game with $K<J$, any constrained ranking of schools is weakly dominated by the same set of schools ranked according to true preferences.
This result implies that a student cannot lose and may possibly gain by taking any arbitrary list with less than or equal to $K$ schools, dropping the unacceptable schools, and ranking the acceptable schools according to her true preferences.
A further implication is that if a student's number of acceptable schools is less than or equal to $K$, then her dominant strategy is to submit her true list of acceptable schools (Proposition 4.2 of \cite{haeringer2009}).
These implications give rise to a class of undominated strategies according to the following definitions of partial order.

\begin{definition}(Weak and Strong Partial Order)\label{def:partialorder}
We say $P$ is a weak partial order of $Q$ if $P$ is any selection of up to $K$ schools among the acceptable schools in $Q$ 
and that selection of schools is ranked according to $Q$.
Formally, 
(i) $1 \leq |P| \leq K$, $P \subseteq \{d\in Q: d Q 0 \} $; and
(ii) for every $d,d' \in P$, $d' P d \Leftrightarrow d' Q d$.
We say $P$ is a strong partial order of $Q$ when a third condition holds in addition to (i) and (ii).
Namely, (iii) $|P| = \min\{K, | \{d\in Q: d Q 0 \} | \}$. 
In other terms, if the number of acceptable schools in $Q$ is less than or equal to $K$ and $P$ is a strong partial order of $Q$, then $P$ equals the list of acceptable schools in $Q$;
otherwise, if the number of acceptable schools in $Q$ is greater than $K$, $P$ is a subset of $K$ schools among the acceptable schools in $Q$.
\end{definition}

Lemma \ref{result:partial_order} in Section \ref{sec:app:partial_order} of the appendix summarizes 
the implications of the result on partial orders from
\cite{haeringer2008,haeringer2009} in terms of our Definition \ref{def:partialorder}.
In short, for a student with true preferences $Q$, any $P$ is weakly dominated by a weak partial order $P^*$ of $Q$ that has the same acceptable schools as $P$; in turn, $P^*$ is weakly dominated by a strong partial order $P^{**}$ of $Q$ that contains the same set of acceptable schools as $P^*$.
Every strong partial order is a weak partial order, but the converse is not true.
Our definition of a weak partial order strategy is similar to the definition of the dropping strategy from \cite{kojima2009incentives}.

Assuming that agents always submit a strong partial order implies they reveal their true ordered list of acceptable schools whenever they submit $P$ with fewer schools than the cap $K$. 
This could be a strong behavioral assumption in some contexts where agents have more than $K$ acceptable schools but have a strong expectation that they will gain admission to a smaller-than-$K$ set of schools. 
In this case, they may submit $|P|< K$ not because it reflects their full list of acceptable schools but simply because they may not want to incur the costs of ranking all schools up to $K$.
In the rest of this subsection, we consider mechanisms that impose a cap $K$ on $P$ and assume that students submit a weak partial order of their true preferences.

\begin{assumption}[Submission of Weak Partial Order]\label{aspt:weakpo}
Students submit a weak partial order of their true preferences.
\end{assumption}

Assumption \ref{aspt:weakpo} replaces Assumption \ref{aspt:truth:telling} to accommodate mechanisms that are not strategy proof. 
Submitting a weak partial order is rational in DA mechanisms with cap constraints. 
That is true for any mechanism that becomes strategy proof once we remove the cap constraint, for example, the TTC mechanism.

An assumption maintained throughout this paper is that the cutoff characterization from Definition \ref{def:cutoff} applies to the mechanism. 
This assumption says that $\mu(\omega) = Q(B(\bS(\omega))$ for every $\omega \in \Omega$.
\cite{azevedo2016} show that stability is equivalent to cutoff characterization 
with $\bS=\bR$ and cutoffs that equal the minimum score of the admitted students in each school.
Thus, it is worth discussing the stability of the constrained DA mechanism. 
Theorem 6.3 from \cite{haeringer2009} demonstrates that any Nash equilibrium in constrained DA where $\bR$ satisfies Ergin acyclicity
leads to a stable matching in the finite economy.\footnote{ 
Ergin acyclicity ensures that no student can block a potential improvement for any two other students without affecting her own assignment. See \cite{ergin2002} for the formal definition.
} 
Even without Ergin acyclicity, some Nash equilibria still produce stability. 
SD always satisfies Ergin acyclicity, and thus every Nash equilibrium in SD produces a stable matching. 
\cite{fack2019} also study the constrained DA mechanism.
They extend Theorem 6.3 from \cite{haeringer2009} to (pure-strategy) Bayesian Nash equilibria in the continuum economy (Proposition A3, Online Appendix A.2.5, \cite{fack2019}).
\cite{fack2019} also provide primitive conditions for finite economies where students play partial orders to 
converge to a continuum economy with a stable equilibrium (Proposition 5, \cite{fack2019}).
They further provide a test for implications of stability and find no empirical or simulation evidence against it.
In the context of the constrained TTC mechanism, any Nash equilibrium leads to 
a stable matching as long as $\bR$ satisfies Kesten acyclicity 
(Theorem 6.4 from \cite{haeringer2009}).
Therefore, constrained SD, DA, and TTC all satisfy the assumption of cutoff characterization as in Definition \ref{def:cutoff}
with $\bS=\bR$ under the appropriate conditions.

There is another interesting feature of the cutoff characterization of DA mechanisms.
We know that DA produces a stable matching if agents are truth-tellers.
In case agents are not truth-tellers, the matching outcome continues to be ``stable'' if we replace $Q$ with $P$ in the definition of stability.

\begin{definition}[Stability wrt $P$]\label{def:stableP}
We say the matching $\mu: \Omega \to \m{J}^0$ is a stable matching wrt $P$ if three conditions are satisfied for every $\omega \in \Omega$:
(i) $\mu(\omega) \bar P(\omega) 0$  (individual rationality);
(ii) for any $j \in \m{J}$, if $j P(\omega) \mu(\omega)$, then $j$ is full (no waste);
and
(iii) for any $j \in \m{J}$ that is full,  if $\mu(\omega')=j$ and $ j P(\omega) \mu(\omega)$, then $R_j(\omega') > R_j(\omega )$ (no justified envy),
where we adopt the convention that $m P 0$ for every $m \in P$.
This is the same as Definition \ref{def:stable} except that $P$ appears in the place of $Q$.
\end{definition}

The DA mechanism, constrained or unconstrained, produces a matching that is stable wrt reported preferences $P$.
Stability wrt $P$ leads to a cutoff characterization wrt $P$ according to the work of \cite{azevedo2016}.
This cutoff characterization has scores $\bS=\bR$ and admission cutoffs that equal the smallest scores of admitted students in each school.
In other words, this is the same cutoff characterization from Definition \ref{def:cutoff}
except that $Q$ is replaced with $P$.
Cutoff characterization wrt $P$ is natural in DA but not necessarily in other mechanisms, so we state it in the following assumption.

\begin{assumption}[Cutoff Characterization wrt $P$]\label{aspt:cutoff2}
In addition to the maintained assumption of cutoff characterization as in  
Definition \ref{def:cutoff}, the matching function $\mu$ satisfies 
$\mu(\omega) = P(B(\bS(\omega))$ for every $\omega \in \Omega$.
\end{assumption}

Assumption \ref{aspt:cutoff2} essentially says that agents are matched to their best feasible options, where best is now defined according to $P$. 
Assumption \ref{aspt:cutoff2} is convenient because it allows us to write a simple expression for the identified set of local preferences in Proposition \ref{result:partial_id:nextbest} below; 
however, it is not a necessary assumption for the identification of those sets. 
The convenience comes from the fact that Assumption \ref{aspt:cutoff2} implies
$\mu = P(B(\bS)) = Q(B(\bS))$, where both $\mu$ and $P$ are observed and $P$ and $Q$ are related via the weak partial order assumption.
If we drop Assumption \ref{aspt:cutoff2}, we have only one equality $\mu = Q(B(\bS))$, which leads to larger sets of local preferences in Proposition \ref{result:partial_id:nextbest} below. 
This is useful to know for settings such as those with the TTC mechanism, which is not stable wrt $P$.

Next, we characterize all possible pairs of local preferences at a cutoff that are
compatible with the data and Assumptions \ref{aspt:weakpo} and \ref{aspt:cutoff2}.

\begin{proposition}[Identification of Local Preference Sets]\label{result:partial_id:nextbest}
Suppose Assumptions \ref{aspt:weakpo} and \ref{aspt:cutoff2} hold.
Select a school $j$ with cutoff $c_j$.
Consider a student with scores $\bS \equiv (S_j, \bS_{-j})$ 
and submitted preferences $P$.
Call $(a,b) = P_j$.
For this student, define 
 $N_j^+ = B^+_j( \bS ) \setminus \left\{ P \cup \{ 0 \} \right\}$
and
$N_j^- = B^-_j( \bS ) \setminus \left\{ P \cup \{ 0 \} \right\}$,
respectively, 
the sets of unlisted feasible schools in the counterfactual budget sets to the right and the left of the cutoff.
Then, the $Q_{j}$ of this student belongs to $\bQ_j$, where the set $\bQ_j$ is defined as follows:
\vspace{-.25cm}
\begin{align}
\bQ_j = 
\left\{ 
\begin{array}{ll}
\{ (a,b) \}, & \text{ if } S_j \geq c_j \text{ and } a=b, 
\\
\{ (a,b) \} \cup \left(\{ a \} \times N_{j}^-\right), & \text{ if } S_j \geq c_j \text{ and } a \neq b, 
\\
\{ (a,b) \} \cup \left( (N_{j}^+ \setminus N_{j}^-) \times \{ b \} \right) & \text{ if } S_j < c_j,
\end{array}
\right.
\label{eq:boldqj_weakpo}
\end{align}
where 
$\left(\{ a \} \times N_{j}^- \right)$ denotes the set formed by the Cartesian product of $a$ and elements in $N_{j}^-$
and
$\left(\{ a \} \times N_{j}^- \right) = \emptyset$ if $ N_{j}^- = \emptyset$.

Moreover, assume $P$ is a strong partial order of $Q$.
Then, $\bQ_j$ becomes:
\vspace{-.25cm}
\begin{align}
\bQ_j = 
\left\{ 
\begin{array}{ll}
\{ (a,b) \}, & \text{ if } |P|<K, \text{ or if } |P|=K,~ S_j \geq c_j, \text{ and } a=b, 
\\
\{ (a,b) \} \cup \left(\{ a \} \times N_{j}^-\right), & \text{ if } |P|=K,~ S_j \geq c_j, \text{ and } a \neq b, 
\\
\{ (a,b) \} \cup \left( (N_{j}^+ \setminus N_{j}^-) \times \{ b \} \right) & \text{ if } |P|=K \text{ and } S_j < c_j.
\end{array}
\right.
\label{eq:boldqj_po}
\end{align}

Finally, the characterization in \eqref{eq:boldqj_weakpo} is sharp if the distribution of $Q$ conditional on $P$ and $\bS$ has full support, that is, if every $Q \in \mathcal{Q}$ that satisfies Assumptions \ref{aspt:weakpo} and \ref{aspt:cutoff2} is in that support.
Likewise, \eqref{eq:boldqj_po} is sharp if the distribution of $Q$ conditional on $P$ and $\bS$ has full support under Assumptions \ref{aspt:weakpo} and \ref{aspt:cutoff2} and $P$ being a strong partial order.
\end{proposition} 

 We illustrate the proposition in terms of the SD Example. 

\begin{example*}[SD Example, Part III] 
Suppose the cap constraint is $K=3$ and the four schools are acceptable for everyone.
We consider all agents whose $P_4=(4,2)$.
For example, if agents submit strong partial orders, they submit either $P=\{4,2,1\}$ or $P=\{4,2,3\}$.
The assumption of cutoff characterization wrt $P$ (Assumption \ref{aspt:cutoff2}) says that these agents are matched to school $4$ if $S_1\geq c_4$ and to school $2$ otherwise. To keep things simple, consider five different types of true preferences:
$Q^{(1)}=\{4,2,3,1,0\}$, $Q^{(2)}=\{4,3,2,1,0\}$, $Q^{(3)}=\{4,1,3,2,0\}$, $Q^{(4)}=\{3,4,2,1,0\}$, and $Q^{(5)}=\{1,3,2,4,0\}$.
The weak partial order assumption rules out $Q^{(5)}$ because $2 Q^{(5)} 4$ contradicts $4$ being reported preferred to $2$.
The maintained assumption of cutoff characterization (Definition \ref{def:cutoff}) further rules out more types of $Q$, depending on whether
$S_1 \geq c_4$ or $S_1 < c_4$:
\begin{enumerate}
\item if $S_1\geq c_4$, $Q^{(4)}$ is not possible 
because the matching assignment is $4$ but the best feasible option according to $Q^{(4)}$ is $3$;
in this case, the possible true local preferences are: 
$Q^{(1)}_4=(4,2)$, $Q^{(2)}_4=(4,3)$, and $Q^{(3)}_4=(4,1)$;
for a student who submits $P=\{4,2,1\}$, $\bQ_4 = \{(4,2), (4,3)\} $;
otherwise, for someone who submits $P=\{4,2,3\}$, $\bQ_4 = \{(4,2), (4,1)\} $;

\item if $S_1 < c_4$, none of $Q^{(2)}$, $Q^{(3)}$, or $Q^{(4)}$ is possible
because the matching assignment is $2$ but the best feasible options according to these $Q$s differ from $2$;
in this case, the only possible true local preference is $Q^{(1)}_4=(4,2)$, so that $\bQ_4 = \{(4,2)\} $.
\end{enumerate}
\end{example*}

This example illustrates why an RD at $c_4$ that controls for $P_4=(4,2)$ might be problematic.
The range of possibilities for true preference types is different between individuals above and below the cutoff.
We see types $(4,1)$, $(4,2)$, and $(4,3)$ above the cutoff but only type $(4,2)$ below the cutoff. 
Suppose in an extreme case there is a nonzero fraction of individuals with types $(4,1)$ or $(4,3)$ right above  the cutoff.
True preferences arguably affect outcomes and this implies that the distribution of outcomes conditional on $P_4=(4,2)$ and $S_1=s$ changes as $s$ crosses the cutoff $c_4$ even without a treatment effect. 
This extreme case breaks internal validity of RD identification.
Even if the fraction of types $(4,1)$ or $(4,3)$ at the cutoff is zero and RD identification is valid, a continuous but sudden increase in the fraction of such agents on the right of the cutoff may bring severe bias issues and make inference impossible (e.g., \cite{bertanha_impossible}, in particular, Sections 5 and A.6).

For the issue to arise when we control for $P_4=(4,2)$, there must be at least a fraction of agents with true local preferences $(4,1)$ and $(4,3)$ who change their submission behavior discontinuously as a function of $S_1$ as $S_1$ crosses the value of $c_4$; i.e., they must change $P$ such that $P_4 \neq (4,2)$ below the cutoff and 
$P_4 = (4,2)$ above the cutoff. 
Intuitively, this requires these agents to have good \textit{ex-ante} knowledge about the \textit{ex-post} value of the cutoff $c_4$.
The more knowledge they have about the \textit{ex-post} cutoff values, the sharper will be their change in behavior around the cutoff, and the closer we are to the extreme case mentioned above. 
Section \ref{sec:app:control:P} in the appendix presents a numerical example of an economy with agents that maximize expected utility in face of cutoff uncertainty and exhibit such discontinuous behavior at the \textit{ex-post} cutoffs. 
An interesting feature of that example is that agents do not need to have perfect knowledge of the cutoffs to exhibit discontinuous behavior.
Finally, Section \ref{sec:empir:strategicbehavior} displays empirical evidence from Chile that supports agents having good 
\textit{a priori} knowledge of cutoffs; we also show that submission behavior changes near \textit{ex post} cutoffs and the change is discontinuous in some cases.

Proposition \ref{result:partial_id:nextbest} identifies all possible values of $Q_j$ for students near a cutoff $c_j$
as a function of their scores and submitted preferences.
In some contexts, students may have a large number of feasible but unlisted programs, resulting in sets $\bQ_j$ with many possible values. For instance, in the Chilean data, $K=8$ but there are over 1,000 programs; a student may have many feasible options but choose not to list most of them. We now introduce one approach to reducing the size of $\bQ_j$ by placing additional restrictions on the expectations students hold when submitting $P$. Other context-specific assumptions can also be employed—for example, in Section \ref{sec:app}, we use an assumption based on preference for fields of study when analyzing post-secondary education in Chile.

\cite{agarwal2018} propose a general framework to rationalize strategic reporting as the optimal solution to an expected utility maximization problem.
In this framework, agents have private information about their preferences and scores and form beliefs about the distribution of other people's preferences and scores. 
These beliefs plus knowledge of the mechanism lead the rational agent to derive probabilities of admission to the various schools as a function of the agent's private information and expectations about other agents. 
The agent then chooses the submission $P$ that maximizes her expected utility.

For our next proposition, we assume that agents are expected utility maximizers where the uncertainty about their match comes from uncertainty about what the admission cutoffs will be after the matching algorithm is run. As such, we assume each agent forms beliefs on admission cutoffs (see Section \ref{sec:app:uncertainty} in the appendix for a formal definition of the problem of the agent). 
Uncertainty about cutoffs is key in our continuum economy with cutoff characterization because cutoffs and scores fully characterize the agent's budget set.
Given the student's scores, a distribution of possible cutoffs translates into a distribution of possible budget sets.
Under Assumption \ref{aspt:cutoff2}, the student is admitted to the best school according to the submission $P$ among the available schools in the budget set.
Therefore, beliefs on cutoffs translate into probabilities of admission to various schools for any given $P$. 
We make an assumption on the distribution of cutoffs expected by agents
that has to do with the concept of uniformly more accessible schools.

\begin{definition}[Uniformly More Accessible Schools]\label{def:umas} 
For a pair of distinct schools $(d,e)$, we say $e$ is uniformly more accessible than $d$ if
two conditions are satisfied: first, if access to school $d$ implies access to school $e$,
\vspace{-.35cm}
\[
\{\omega: S_d(\omega) \geq c_d \} \subseteq \{\omega: S_e(\omega) \geq c_e \},
\]
and second, if replacing option $d$ with option $e$ in any submission $P$ alters the likelihood of admission for at least one school listed in $P$;
formally, for any two fixed (i.e., nonrandom) submissions $P$ and $\ti P$ 
such that $P$ has $d$ but does not have $e$ 
and
$\ti P$ equals $P$ except for $e$ in the place of $d$,
there exists $u \in \{0,1, \ldots, |P| \}$ for which
\[
\mmp\left[ P(B(\bS)) = P^u \right] \neq \mmp\left[ \ti P(B(\bS)) = \ti{P}^u \right],
\]
where $P(B)$ denotes the best choice in set $B$ according to $P$ (Definition \ref{def:nextbest})
and
$P^u$ denotes the school ranked in the $u$-th position in $P$.
In short, we say $(d,e) \in UMAS$, where $UMAS\subseteq \m{J} \times \m{J}$ is the set of all such pairs.
\end{definition}

Definition \ref{def:umas} says that $e$ is uniformly more accessible than $d$ if 
everyone who qualifies for school $d$ also qualifies for school $e$.
Schools $d$ and $e$ must also be relevant in the sense of the second condition: 
there is always a strictly positive fraction of individuals
for whom listing $e$ in the place of $d$ changes their best feasible options.
In the SD case, a sufficient condition for Assumption \ref{aspt:umas} is that $c_d>c_e$ and the cutoffs are distinct interior points in the support of the placement score. 
Uniformly more accessible schools do not always exist.
Whether they do depends on the mechanism in place and the joint distribution of the placement scores. 
We use this definition to impose a mild restriction on the expectations of agents regarding cutoffs.

\begin{assumption}\label{aspt:umas}
Consider a student with scores $\boup{s} \in \mS$ who views uncertain cutoffs as random variables $C_1, \ldots, C_J$
before the matching assignment.
Let $\ti{B} = \{ 0 \} \cup \{j \in \m{J} ~:~ s_j \geq C_j \}$ be the student’s corresponding random budget set.
For every pair $(d,e) \in UMAS$, 
the distribution of cutoffs for this student is such that two conditions are satisfied:
first,
\[
\{s_d \geq C_d \} \subseteq \{s_e \geq C_e \},
\]
and second, for any two fixed (i.e., nonrandom) submissions $P$ and $\ti P$ 
such that $P$ has $d$ but does not have $e$ 
and
$\ti P$ equals $P$ except for $e$ in the place of $d$,
there exists $u \in \{0,1, \ldots, |P| \}$ for which
\[
\mmp\left[ P(\ti{B}) = P^u \right] \neq \mmp\left[ \ti P(\ti{B}) = \ti{P}^u \right].
\]
This is true for every student in the economy. 
\end{assumption}

Assumption \ref{aspt:umas} says that students correctly anticipate which schools will be uniformly more accessible after the matching assignment.
For example, agents may learn this information by observing past realizations of the matching in the economy. 
If a school $e$ is well known to be accessible to everyone who has access to school $d$,
then it is natural for a student to expect to have access to $e$ if she ever has access to school $d$. 
Note that the assumption does not pin down the expected probability of admission or the set of schools to which the student will have access in the \textit{ex-post} economy.
It restricts only the expected hierarchy of school access according to $UMAS$.
This assumption has implications for the joint distribution of $(P,Q)$.

\begin{proposition}\label{result:umas}
Suppose Assumptions \ref{aspt:weakpo}--\ref{aspt:umas} hold.
Consider a student with reported preference ranking $P$.
Let $\left( P \times P^c \right)$ be the Cartesian product of listed and unlisted schools, respectively, $P$ and $P^c$.
If $(d,e) \in UMAS \cap \left( P \times P^c \right)$, 
then $d Q e$.
\end{proposition}

Proposition \ref{result:umas} says that if an agent lists school $d$ but does not list the uniformly more accessible school $e$, it must be that this agent prefers $d$ over $e$.
This result offers a refinement of Proposition \ref{result:partial_id:nextbest} above.

\begin{corollary}\label{result:umas_refine} 
Consider the setup of Proposition \ref{result:partial_order}, where $P_j=(a,b)$, and suppose Assumption \ref{aspt:umas} holds.
Define
$A_j^- = N^-_j \setminus \left\{ e: \exists d \in P \text{ with which } (d,e) \in UMAS \text{ and } b \bar{P} d \right\}$
and\\
$A_j^+ = \left(N^+_j \setminus N^-_j \right) \setminus \left\{ e: \exists d \in P \text{ with which } (d,e) \in UMAS \text{ and } a \bar{P} d \right\}$.
Then, under weak partial order,
\vspace{-.4cm}
\begin{align}
\bQ_j = 
\left\{ 
\begin{array}{ll}
\{ (a,b) \}, & \text{ if } S_j \geq c_j \text{ and } a=b, 
\\
\{ (a,b) \} \cup \left(\{ a\} \times A_{j}^-\right), & \text{ if } S_j \geq c_j \text{ and } a \neq b, 
\\
\{ (a,b) \} \cup \left( A_j^+ \times \{ b \} \right) & \text{ if } S_j < c_j.
\end{array}
\right.
\label{eq:boldqj_weakpo_uas}
\end{align}
Under strong partial order,
\begin{align}
\bQ_j = 
\left\{ 
\begin{array}{ll}
\{ (a,b) \}, & \text{ if } |P|<K, \text{ or if } |P|=K,~ S_j \geq c_j, \text{ and } a=b, 
\\
\{ (a,b) \} \cup \left(\{ a \} \times A_{j}^-\right), & \text{ if } |P|=K,~ S_j \geq c_j, \text{ and } a \neq b, 
\\
\{ (a,b) \} \cup \left( A_j^+ \times \{ b \} \right) & \text{ if } |P|=K \text{ and } S_j < c_j.
\end{array}
\right.
\label{eq:boldqj_po_uas}
\end{align}
These characterizations are sharp as long as \eqref{eq:boldqj_weakpo}--\eqref{eq:boldqj_po} are sharp in their respective contexts in Proposition
\ref{result:partial_order}
and
imposing Assumption \ref{aspt:umas} sets to zero only the following probabilities:
$\mmp\left[ e Q d | P, \bS \right]$ for every $(d,e) \in UMAS \cap \left( P \times P^c \right)$.

\end{corollary}

Corollary \ref{result:umas_refine} describes how to use Proposition \ref{result:umas} to potentially reduce the number of elements in the $\bQ_j$ constructed in Proposition \ref{result:partial_order}.
The intuition runs as follows. 
Suppose that a student submits $P$ and $P_j=(a,b)$.
If school $e$ is uniformly more accessible than school $d$ and $d$ is listed in $P$ but $e$ is not listed in $P$, 
then
we know the student truly prefers $d$ over $e$. 
This excludes some possibilities of $Q_j$ in the $\bQ_j$ defined by Proposition \ref{result:partial_order}.
For instance, this person cannot have $Q_j=(a,e)$ if $b \bar{P} d$ because that presupposes $e Q b \bar{Q} d $, which contradicts $d Q e$.
Likewise, this person cannot have $Q_j=(e,b)$ if $a \bar{P} d$.

\begin{example*}[SD Example, Part IV] 
The set of uniformly more accessible schools is
$UMAS = \{(2,1), (3,2), (3,1), (4,3), (4,2), (4,1) \}$.
Assumption \ref{aspt:umas} shrinks the set $\bQ_4$ of those agents 
with $S_1\geq c_4$ and $P=\{4,2,3\}$.
Applying the assumption changes $\bQ_4 =\{(4,2), (4,1)\}$
to $\bQ_4 = \{(4,2)\}$
because $1$ is uniformly more accessible than $2$, $2$ is listed, and $1$ is not listed, so Proposition \ref{result:umas} implies $2Q1$.
\end{example*}

\subsection{Partial Identification of Causal Effects}\label{sec:id_r:partial_id_givenqj} 

\indent

In this section, we lay out conditions and derive bounds on average treatment effects.
We assume that the researcher has already identified the set of local preferences at a cutoff of interest.
This means that the researcher has a set-valued variable $\bQ_j$ for all students in the
vicinity of a cutoff $j$ corresponding to a comparable pair $(j,k)$ in $\m{P}$. 
Researchers may construct $\bQ_j$ using the methods in Section \ref{sec:id_r:qj_id} if they find it reasonable to rely on at least some of the specific assumptions in that subsection; otherwise, they may use any other method to construct $\bQ_j$.
There is no restriction on the choice of the method for constructing $\bQ_j$ except for a couple of high-level conditions that we assume to hold in this section. 
We start by defining the conditional support of partially identified true local preferences.

\begin{definition}[Support of Local Preference Sets]
\label{def:supp_bQj}
Consider a pair $(j,k) \in \m{P}$ and corresponding cutoff $c_j$.
The support of partially identified true local preferences conditional on $S_j=s$ is defined as
\vspace{-.35cm}
\[
\bLambda_{j}(s) = \left\{ 
B\subseteq \m{J}^0 \times \m{J}^0 : 
\mmp\left[ \bQ_j = B | S_j=s \right] >0 
\right\}.
\]
The union set of this support is defined as the collection of all unions of sets in $\bLambda_{j}(s)$,
namely,
\[
\bLambda^{\cup}_{j}(s)
=
\left\{ 
B^{\cup} \subseteq \m{J}^0 \times \m{J}^0 :
\exists B_1, B_2, \ldots \in \bLambda_{j}(s) \text{ with }
B^{\cup} = \cup_{i} B_i
\right\}.
\]
\end{definition}

The set $\bLambda_{j}(s)$ collects all values of $\bQ_j$ that occur with positive probability conditional on $S_j=s$.
In the specific context of Section \ref{sec:id_r:qj_id}, $\bQ_j$ is constructed from the mapping of observables $(P,\bS)$ to a subset of $\m{J}^0 \times \m{J}^0$, i.e.,
$\bQ_j = \psi_j (P,\bS)$.
For example, Proposition \ref{result:partial_id:nextbest} and Corollary \ref{result:umas_refine} give examples of such mapping $\psi_j$.
A set $B$ of pairs $(a,b) \in \m{J}^0 \times \m{J}^0$ belongs to the support set $\bLambda_{j}(s)$ if there is a set of values in the support of the conditional distribution of $(P,\bS)$ given $S_j=s$ such that $\psi_j$ maps those values to the set $B$.
The union set $\bLambda^{\cup}_{j}(s)$ collects all possible unions of support points of $\bQ_j$ conditional on $S_j=s$.
These definitions are instrumental in the computation of the partially identified distribution of $Q_j$, 
as explained in Proposition \ref{result:partial_id:nextbest-dist} below.

Sharpness of identification of the distribution of $Q_j$ requires sharpness in the construction of the sets $\bQ_j$.
Proposition \ref{result:partial_id:nextbest} and Corollary \ref{result:umas_refine} gave the conditions for sharpness of $\bQ_j$ in the context of Section \ref{sec:id_r:qj_id}.
Outside that context, researchers may construct $\bQ_j$ in a different way, so we impose sharpness of $\bQ_j$ in the general form of the assumption below. 

\begin{assumption}[Sharp Local Preference Sets]\label{aspt:sharp_bQj}
Consider a pair $(j,k) \in \m{P}$ and corresponding cutoff $c_j$.
Assume that:

(i) the random variable $Q_{j}$ and the random set $\bQ_j$ are both measurable maps on the same probability space
and $\mmp\left[ Q_j \in \bQ_j ~|~ S_j \right]=1$ with probability 1; and

(ii) $\up{supp}\left[ Q_j ~|~ \bQ_j, S_j \right] = \bQ_j $ with probability $1$, 
where $\up{supp}\left[ Y ~|~ X \right]$ denotes the support set of the distribution of $Y$ conditional on $X$.
\end{assumption}

Assumption \ref{aspt:sharp_bQj}(i) says that $\bQ_j(\omega)$ of individual $\omega$ contains the true pair of local preferences $Q_j(\omega)$ of that individual (for almost all individuals), which is a minimum requirement for the construction of $\bQ_j(\omega)$.
This does not say anything about the sharpness of $\bQ_j$. 
For example, $\bQ_j = \m{J}^0 \times \m{J}^0$ is completely uninformative and trivially satisfies Assumption \ref{aspt:sharp_bQj}(i). 
The sharpness requirement is stated in Assumption \ref{aspt:sharp_bQj}(ii).
It says that
all possibilities of local preferences listed in $\bQ_j$ actually occur in the data with positive probability. 
This rules out unnecessarily large sets $\bQ_j$. 
Assumption \ref{aspt:sharp_bQj}(ii) may be dropped at the cost of lacking sharpness in the identified sets in the rest of this section.

Partial identification of true local preferences and treatment effects occurs at the limit, as $S_j$ approaches $c_j$,
and is conditional on $\bQ_j$.
For this to work, we impose regularity conditions on the distribution of potential outcomes and $\bQ_j$
conditional on $S_j$ at the limit $c_j$.

\begin{assumption}[Distribution of Local Preference Sets]\label{aspt:dist_bQj}
Consider a pair $(j,k) \in \m{P}$ and corresponding cutoff $c_j$.
Assume that:

(i) 
there exist a small $\eps>0$ and collections of subsets of $\m{J}^0 \times \m{J}^0$
denoted $\bLambda_{j}^{+}$ and $\bLambda_{j}^{-}$ such that 
$\bLambda_{j}^{+} = \bLambda_{j}(c_j+e)$ $\forall e \in [0,\eps)$
and
$\bLambda_{j}^{-} = \bLambda_{j}(c_j-e)$ $\forall e \in (0,\eps)$;
consistent with Definition \ref{def:supp_bQj}, we define $\bLambda_{j}^{\cup+}$ 
and
$\bLambda_{j}^{\cup-}$ 
as union sets of 
$\bLambda_{j}^{+}$
and
$\bLambda_{j}^{-}$, respectively;

(ii) for any $g \in \m{G}$ of Assumption \ref{aspt:continuity} and any $\lbar{\tau}, \ubar{\tau} \in \mmr \cup \{-\infty, +\infty \}$, $\lbar{\tau} < \ubar{\tau}$, 
the side limits of the following expectations are well defined: 
$\mme\left[ g(Y) \mmi\{ \bQ_j = A, \lbar{\tau} < g(Y) < \ubar{\tau} \} ~|~ S_j= c_j^+ \right] \; \; \forall A \in \bLambda_{j}^{+}$
and $\mme\left[ g(Y) \mmi\{ \bQ_j = A, \lbar{\tau} < g(Y) < \ubar{\tau} \} ~|~ S_j= c_j^- \right] \; \; \forall A \in \bLambda_{j}^{-}$.

\end{assumption}

Assumption \ref{aspt:dist_bQj}(i) concerns the distribution of $\bQ_j$ conditional on $S_j$: the support set of $\bQ_j$
is constant as $S_j=s$ approaches the cutoff $c_j$ from either side of it.
Part (ii) of the assumption concerns the joint distribution of potential outcomes and $\bQ_j$ conditional on $S_j$. 
For example, Assumption \ref{aspt:dist_bQj}(ii) implies that
$\mmp\left[ \bQ_j = A ~|~ S_j= c_j^+ \right]$ 
and
$\mme[ Y ~|~ \bQ_j = A, Y < \tau, S_j= c_j^+ ]$
are well-defined limits for any $A \in \bLambda_{j}^{+}$ and $\tau \in \mmr \cup +\infty$
provided that
$\mmp [ \bQ_j = A, Y < \tau ~|~ S_j= c_j^+ ]>0.$
The next result gives inequalities to construct bounds on $\mmp[Q_j=(a,b) | S_j=c_j]$ for any pair $(a,b)$.

\begin{proposition}[Sharp Set of Distributions of Local Preferences]\label{result:partial_id:nextbest-dist}
Consider a pair $(j,k) \in \m{P}$.
Suppose Assumptions \ref{aspt:continuity}, \ref{aspt:sharp_bQj}, and \ref{aspt:dist_bQj} hold. 
Then, 
the sharp set of all possible discrete probability distributions of $Q_{j}$ conditional on  $S_j=c_j$
is characterized as follows.
For every $A \in \bLambda^{\cup+}_{j} \cup \bLambda^{\cup-}_{j}$, 
each probability distribution in that set implies a value for
$\mmp\left[ Q_j \in A | S_j= c_j \right]$ 
that satisfies one of the three inequalities below:
\begin{enumerate}[(i)]
	\item if $A \in  \bLambda^{\cup+}_{j} \cap \bLambda^{\cup-}_{j}$, 
    \vspace{-.25cm}
		\[
			\mmp\left[ Q_j \in A | S_j= c_j \right] \geq 			
			\max \left\{~
				\mmp\left[ \bQ_j \subseteq A | S_j= c_j^+ \right]
				~;~
				\mmp\left[ \bQ_j \subseteq A | S_j= c_j^- \right]
				~\right\};
		\]
		
	\item if $A \in  \bLambda^{\cup+}_{j} \setminus \bLambda^{\cup-}_{j}$, 
    \vspace{-.5cm}
		\[
			\mmp\left[ Q_j \in A | S_j= c_j \right]
			\geq 
			\mmp\left[ \bQ_j \subseteq A | S_j= c_j^+ \right]; \text{ or }
		\]
	\item if $A \in  \bLambda^{\cup-}_{j} \setminus \bLambda^{\cup+}_{j}$,
    \vspace{-.5cm}
		\[
			\mmp\left[ Q_j \in A | S_j= c_j \right]
			\geq 
			\mmp\left[ \bQ_j \subseteq A | S_j= c_j^- \right].
		\]
\end{enumerate}

\end{proposition}

Proposition \ref{result:partial_id:nextbest-dist} provides a way to construct the sharp partially identified set of 
all possible distributions of $Q_{j}$ conditional on  $S_j=c_j$.
A distribution of $Q_{j}$ conditional on  $S_j=c_j$ consists of values $p_{a,b} \in [0,1]$ 
for every $(a,b) \in \m{J}^0 \times \m{J}^0$
such that $\sum_{(a,b) \in \m{J}^0 \times \m{J}^0} ~ p_{a,b} = 1$, where $p_{a,b} = \mmp\left[ Q_j =(a,b) | S_j= c_j \right].$
The sharp set is constructed by finding all values of $p_{a,b}$ 
where
$\sum_{(a,b) \in A} ~ p_{a,b}$ satisfies the inequalities of Proposition \ref{result:partial_id:nextbest-dist}
for every $A \in \bLambda^{\cup+}_{j} \cup \bLambda^{\cup-}_{j}$.

\begin{example*}[SD Example, Part V] 
Continue to assume that the four schools are acceptable for everyone. 
Suppose for a moment that all combinations of $(P,Q)$ that satisfy Assumptions \ref{aspt:weakpo}--\ref{aspt:umas} exist in the economy, for both
$S_1 \geq c_4$ and $S_1<c_4$.
Then, the list of all possible $\bQ_4$ is as follows:
\begin{enumerate}
\item if $S_1 \geq c_4$, 
$\{(1,1)\}$, 
$\{(2,2)\}$,
$\{(3,3)\}$,
$\{(4,1)\}$,
$\{(4,2)\}$,
$\{(4,3)\}$,
$\{(4,1),(4,2),(4,3)\}$,
$\{(4,1),(4,3)\}$, and
$\{(4,2),(4,3)\}$;
\item if $S_1 < c_4$,
$\{(1,1)\}$, 
$\{(2,2)\}$,
$\{(3,3)\}$,
$\{(4,1)\}$,
$\{(4,2)\}$,
$\{(4,3)\}$,
$\{(1,1),(4,1)\}$,
$\{(2,2),(4,2)\}$, and
$\{(3,3),(4,3)\}$.
\end{enumerate}

Let us focus on the case that $S_1\geq c_4$.
To keep things simple, suppose three types of $\bQ_4$ occur with positive probability conditional on $S_1=s$ for any $s \geq c_4$:
$\{(4,2)\}$ with probability $0.1$,
$\{(4,3)\}$ with probability $0.3$,
and
$\{(4,2),(4,3)\}$ with probability $0.6$.
It follows that $\bLambda_{4}(s)=\bLambda_{4}^+=\{ \{(4,2)\},\{(4,3)\}, \{(4,2),(4,3)\} \} $
and
$\bLambda_{4}^{\cup}(s)= \bLambda_{4}^{\cup+}= \{ \{(4,2)\},\{(4,3)\}, \{(4,2),(4,3)\} \} $.
The lower bounds $\mmp[\bQ_4 \subseteq A | S_1 = c_4^+]$ of Proposition \ref{result:partial_id:nextbest-dist} are as follows:
$0.1$ for $A=\{(4,2)\}$;
$0.3$ for $A=\{(4,3)\}$;
and
$1$ for $A=\{(4,2),(4,3)\}$.
Thus, $\mmp[Q_4 = (4,2) | S_1 = c_4^+]$ has lower bound
$0.1$, $\mmp[Q_4 = (4,3) | S_1 = c_4^+]$ has lower bound $0.3$, and the sum of the two equals $1$.
When we look at each individual probability, the bounds are $[0.1,0.7]$ on $\mmp[Q_4 = (4,2) | S_1 = c_4^+]$ 
and $[0.3,0.9]$ on $\mmp[Q_4 = (4,3) | S_1 = c_4^+]$.
\end{example*}

Recall that the construction of the random set $\bQ_j$ depends 
on assumptions regarding the behavior of agents when they submit $P$.
For example, Section \ref{sec:id_r:qj_id} characterizes $\bQ_j$ by assuming weak partial order and cutoff characterization wrt $P$.
Alternatively, the identification approach of \cite{agarwal2018} makes different types of assumptions on agents' expectations and requires data variation in the choice environment. 
The theoretical credibility of these types of assumptions depends on the mechanism faced by agents; in practice, the assumptions have testable implications for what we should observe in the data.
Section \ref{sec:app:falsification} in the appendix provides a testable implication of our assumptions.

Partial identification of the distribution of local preferences
allows us 
to bound the fraction of individuals near cutoff $c_j$ who have $Q_j=(j,k)$.
The average outcome near the cutoff is a weighted average of the average outcomes from two different groups:
first, individuals with $Q_j=(j,k)$, who interest us for the identification of treatment effects;
and
second, individuals with $Q_j \neq (j,k)$.
The overall average is identified, but the average in each of the groups is not. 
A strictly positive lower bound on the fraction of individuals in the $Q_j=(j,k)$ group allows us to construct
lower and upper bounds on the average outcome for that group.

Start with all individuals above and near cutoff $c_j$ whose $\bQ_j$ contain the comparable pair of interest, $(j,k) \in \m{P}$.
The fraction of those individuals who have $Q_j=(j,k)$ equals 
\[
\delta_{j,k}^+ = 
\frac{
\mmp\left[ Q_j=(j,k)|S_j=c_j \right]
}{
\mmp\left[ \bQ_j \cap \{ (j,k) \} \neq \emptyset |S_j=c_j^+ \right]
},
\]
where both numerator and denominator are strictly positive by virtue of $(j,k)$ being a comparable pair (Definition \ref{def:compairs}) and of the sharpness of $\bQ_j$ (Assumption \ref{aspt:sharp_bQj}).
The denominator of $\delta_{j,k}^+$ is identified from the data,
and Proposition \ref{result:partial_id:nextbest-dist} bounds the numerator.
All we need for identification of treatment effects is a lower bound on $\delta_{j,k}^+$,
which comes from a lower bound on its numerator.
Let $\lbar{p}_{j,k}$ denote the infimum over all probability values for $\mmp\left[ Q_j=(j,k)|S_j=c_j \right]$ 
that belong to the partially identified set of Proposition \ref{result:partial_id:nextbest-dist}.
The sharp lower bound on $\delta_{j,k}^+$ equals 
\[
\lbar{\delta}_{j,k}^+ = 
\frac{
\lbar{p}_{j,k}
}{
\mmp\left[ \bQ_j \cap \{ (j,k) \} \neq \emptyset |S_j=c_j^+ \right]
}.
\]
The denominator of $\lbar{\delta}_{j,k}^+$ is strictly positive, but $\lbar{p}_{j,k}$ may or may not be strictly positive.

The same idea applies for individuals just below the cutoff:
\begin{align*}
& \delta_{j,k}^- = 
\frac{
\mmp\left[ Q_j=(j,k)|S_j=c_j \right]
}{
\mmp\left[ \bQ_j \cap \{ (j,k) \} \neq \emptyset |S_j=c_j^- \right]
},
\hspace{1cm}
\lbar{\delta}_{j,k}^- = 
\frac{
\lbar{p}_{j,k}
}{
\mmp\left[ \bQ_j \cap \{ (j,k) \} \neq \emptyset |S_j=c_j^- \right]
}.
\end{align*}

\begin{example*}[SD Example, Part VI] 
We have that $\mmp\left[ \bQ_4 \cap \{ (4,2) \} \neq \emptyset |S_1=c_4^+ \right]=0.7$
and the bounds on $\mmp\left[ Q_4=(4,2)|S_1=c_4 \right]$ are $[0.1,0.7]$.
These imply $\lbar{p}_{4,2}=0.1$ and $\lbar{\delta}_{4,2}^+ = 1/7$.
\end{example*}

The following result utilizes the proportions $\lbar{\delta}_{j,k}^+$ and $\lbar{\delta}_{j,k}^-$ to partially identify
average outcomes for individuals with $Q_j=(j,k)$ on either side of the cutoff.
Taking differences of these bounds yield bounds for the averages of the treatment effects $Y(j)-Y(k)$.

\begin{proposition}\label{result:bounds_RD}
Suppose Assumptions \ref{aspt:continuity}, \ref{aspt:sharp_bQj}, and \ref{aspt:dist_bQj} hold.
Consider a pair $(j,k) \in \m{P}$ such that $\lbar{p}_{j,k}>0$, 
then we have the following bounds on $\mme[g(Y(j)) | Q_j=(j,k), S_j=c_j ]$ 
and $\mme[g(Y(k))  |  Q_j=(j,k), S_j=c_j ]$:
\begin{align*}
& \mme\left [F_{j,k+}^{-1}(U)  \left|  \bQ_j \cap \{ (j,k) \} \neq \emptyset,  U < \lbar{\delta}_{j,k}^+, S_j=c_j^+ \right. \right] 
\\
&  
\hspace{1cm} \leq 
~ 
\mme\left[g(Y(j)) \left|  Q_j=(j,k), S_j=c_j \right. \right] 
~
\leq  
\\ 
& \mme\left [F_{j,k+}^{-1}(U) \left|  \bQ_j \cap \{ (j,k) \} \neq \emptyset,  
U > 1-\lbar{\delta}_{j,k}^+, S_j=c_j^+ \right. \right], 
\end{align*}
\vspace{-.4cm}
and
\vspace{-.4cm}
\begin{align}\label{eq:bounds_ASF_k}
\begin{split}
& \mme \left[ F_{j,k-}^{-1}(U) \left|  \bQ_j \cap \{ (j,k) \} \neq \emptyset,  U < \lbar{\delta}_{j,k}^-, S_j=c_j^- \right. \right] 
\\
& \hspace{1cm} \leq ~ \mme\left[ g(Y(k)) \left| Q_j=(j,k), S_j=c_j \right. \right] ~ \leq  
\\ 
& \mme \left[ F_{j,k-}^{-1}(U) \left|  \bQ_j \cap \{ (j,k) \} \neq \emptyset,  
U > 1 - \lbar{\delta}_{j,k}^-, S_j=c_j^- \right. \right], 
\end{split}
\end{align}
where $U \sim \text{Uniform}[0,1]$ (independent of everything else), \\ $F_{j,k+}^{-1}(u):=\inf\left\{y: \mmp\left[ g(Y) \leq y \left|  \bQ_j \cap \{(j,k)\} \neq \emptyset, S_j=c_j^+ \right. \right] \geq u \right\}$ 
and \\ $F_{j,k-}^{-1}(u) :=\inf\left\{y: \mmp\left[ g(Y) \leq y \left|   \bQ_j \cap \{(j,k)\} \neq \emptyset, S_j=c_j^- \right. \right] \geq u \right\}$.
\end{proposition}
The expressions for the bounds simplify significantly in the cases where $Y$ is binary or continuous. For the sake of brevity, we relegate the detailed formulas to the Appendix (Section \ref{proof:result:bounds_RD}).
The bounds in Proposition \ref{result:bounds_RD} build on the work by \cite{horowitz1995}.
To see the intuition, consider the case where $g(Y)=Y$ is a continuous random variable, and focus on individuals just above the cutoff. 
Among all individuals in the subpopulation with $\bQ_j \cap \{(j,k) \} \neq \emptyset$ and $S_j=c_j^+$,
a fraction ${\delta}^+_{j,k}$ of them has $Q_j=(j,k)$ and $Y=Y(j)$ by the cutoff characterization.
We do not know who these individuals are among those in the subpopulation.
However, the lowest possible value for $\mme[ Y(j) | Q_j=(j,k), S_j=c_j ]$
occurs if all such individuals are located at the lower tail of the distribution of outcomes in the subpopulation. 
Likewise, the highest possible value for $\mme[ Y(j) | Q_j=(j,k), S_j=c_j ]$
occurs if all of that same fraction of individuals are located in the upper tail of the distribution of outcomes. 
We do not know ${\delta}^+_{j,k}$, but we do know that it is no smaller than $\lbar{\delta}^+_{j,k}>0$.
The bounds only grow wider as the fraction ${\delta}^+_{j,k}$ decreases, so the bounds evaluated at
${\delta}^+_{j,k} = \lbar{\delta}^+_{j,k}$ take into account all possible values for ${\delta}^+_{j,k}$.

Although intuitive and analytically simple, these bounds are not necessarily sharp because $\lbar{\delta}^+_{j,k}$ is not exogenously given as considered by \cite{horowitz1995};
$\lbar{\delta}^+_{j,k}$ is constructed from the marginal distribution of $Q_j$ but there could be additional identifying information in the joint distribution of $Q_j$ and $Y$.
Providing a complete characterization of the sharp bounds is complex because it involves deriving bounds on the joint distribution of potential outcomes and $Q_j$,
which may not be practical when the potential outcomes are continuous. 
For the sake of simplicity, we relegate the sharp characterization to Section \ref{sec:app:sharp_bounds} in the appendix.
The bounds of Proposition \ref{result:bounds_RD} 
contain the sharp bounds of Section \ref{sec:app:sharp_bounds} as long as our model assumptions are true. 
The lack of sharpness may not matter in practice when
Proposition \ref{result:bounds_RD}
yields tight bounds for a given dataset. 
However, if our assumptions are not true, the bounds of Proposition \ref{result:bounds_RD} may not contain the sharp bounds of 
Section \ref{sec:app:sharp_bounds}.
That said, we may obtain tight bounds from the data using Proposition \ref{result:bounds_RD}, but this does not mean they contain the true parameter \citep{kedagni2020}.
Therefore, it is advisable to assess the testable implications of Corollary \ref{result:fals-test} as a matter of routine.

\section{Assignment to College Programs and Graduation in Chile}\label{sec:app}

\indent

In this section, we illustrate our method using data from college applications in Chile. Before estimating bounds for the effect of assignment to a program on graduation outcomes, we document the presence of strategic behavior in this setting.

\subsection{Institutional Setting and Evidence of Strategic Behavior}\label{sec:empir:strategicbehavior} 

\paragraph{Centralized college application system in Chile.} We use publicly available data on Chile's centralized college application and assignment system from 2004 to 2010 and on graduation from 2007 to 2020. The institutional setting has been described in detail by \cite{hastingsneilsonzimmerman2013} and \cite{larroucaurios2020,larroucaurios2021}, among others. College choice in Chile is organized as a semicentralized system---a subset of universities participate in a centralized market in which a clearinghouse collects rank-order lists from applicants and determines assignments using a variant of the DA algorithm. Students can submit rank-order lists of up to eight major--university pairs (``programs'') out of more than 1,000.\footnote{The cap was increased to ten in 2012.} Priorities are program-specific and determined by a weighted average of scores obtained in a national standardized test (the PSU, for \textit{prueba de selecci\'on universitaria}) and of high-school GPA.\footnote{In 2014, students' relative rank within their high school was added as one of the ``primary'' scores to be averaged to construct priorities.}
Descriptive statistics about the sample of students and programs are shown in Tables \ref{apptable:ds_programs} and  \ref{apptable:ds_students}.

\paragraph{Strategic behavior.} \cite{larroucaurios2020,larroucaurios2021} thoroughly document that Chilean college applicants behave strategically. Using a 2014 survey linked to administrative data on applications, they show that listed programs often do not coincide with the truly preferred programs as elicited by the survey. 
Focusing on medicine programs, they show that as application scores decline, students become more likely to exclude medicine from the top of their submitted list—even when ranking it first in the survey—and application rates drop sharply in the 700–750 range, where most medicine cutoffs lie. This pattern indicates that students forgo medicine as their admission prospects weaken, despite preferring it over other programs. 

\begin{figure}[t]  
\caption{Cutoffs of Selective Programs}
\label{fig:cutoff}

\begin{center}
    \begin{subfigure}[b]{0.48\textwidth}  
        \centering
        \includegraphics[width=\textwidth]{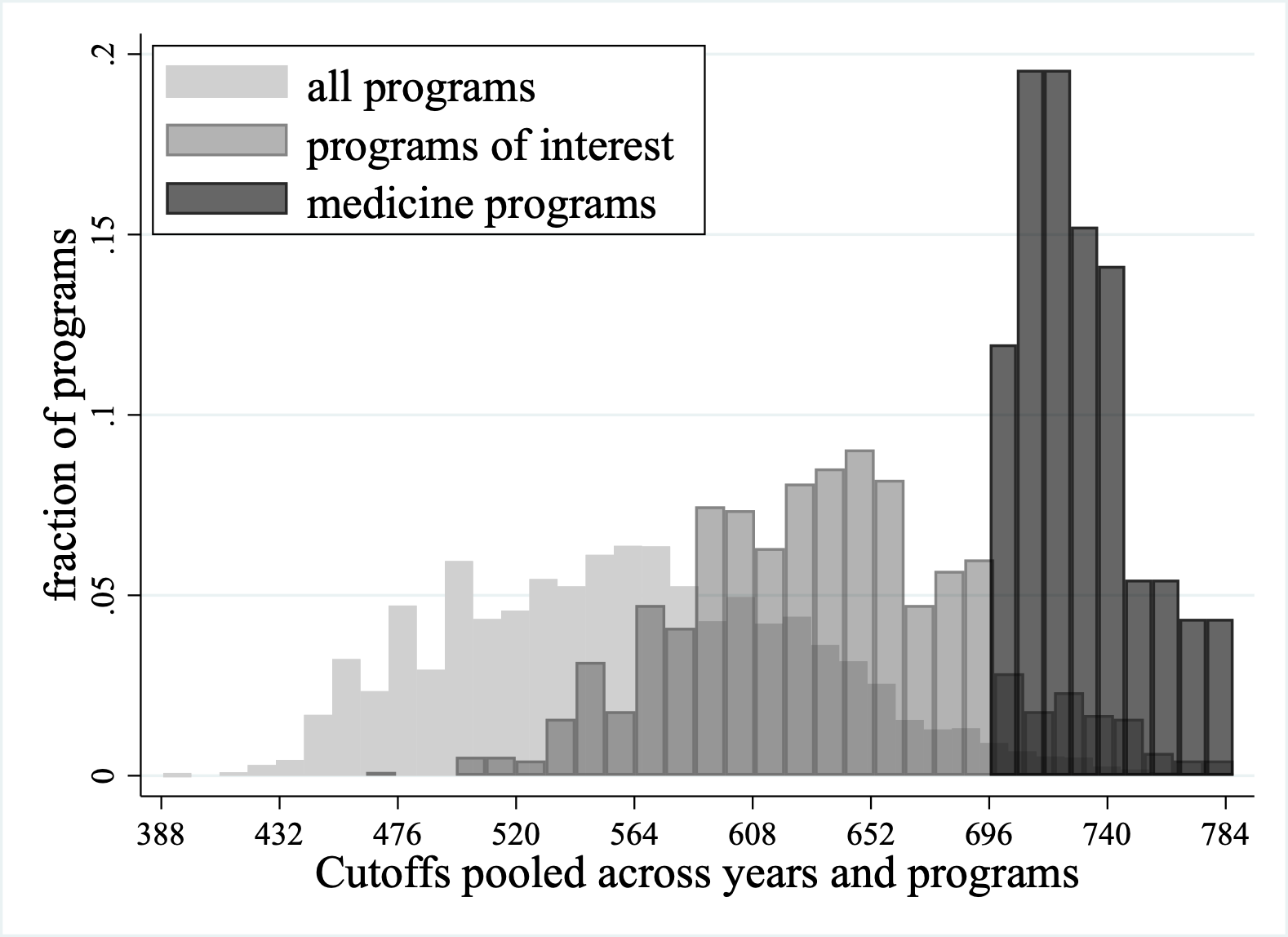}  
        \caption{}
        \label{fig:cutoff:pool}  
    \end{subfigure}
    \hfill  
    \begin{subfigure}[b]{0.48\textwidth}  
        \centering
        \includegraphics[width=\textwidth]{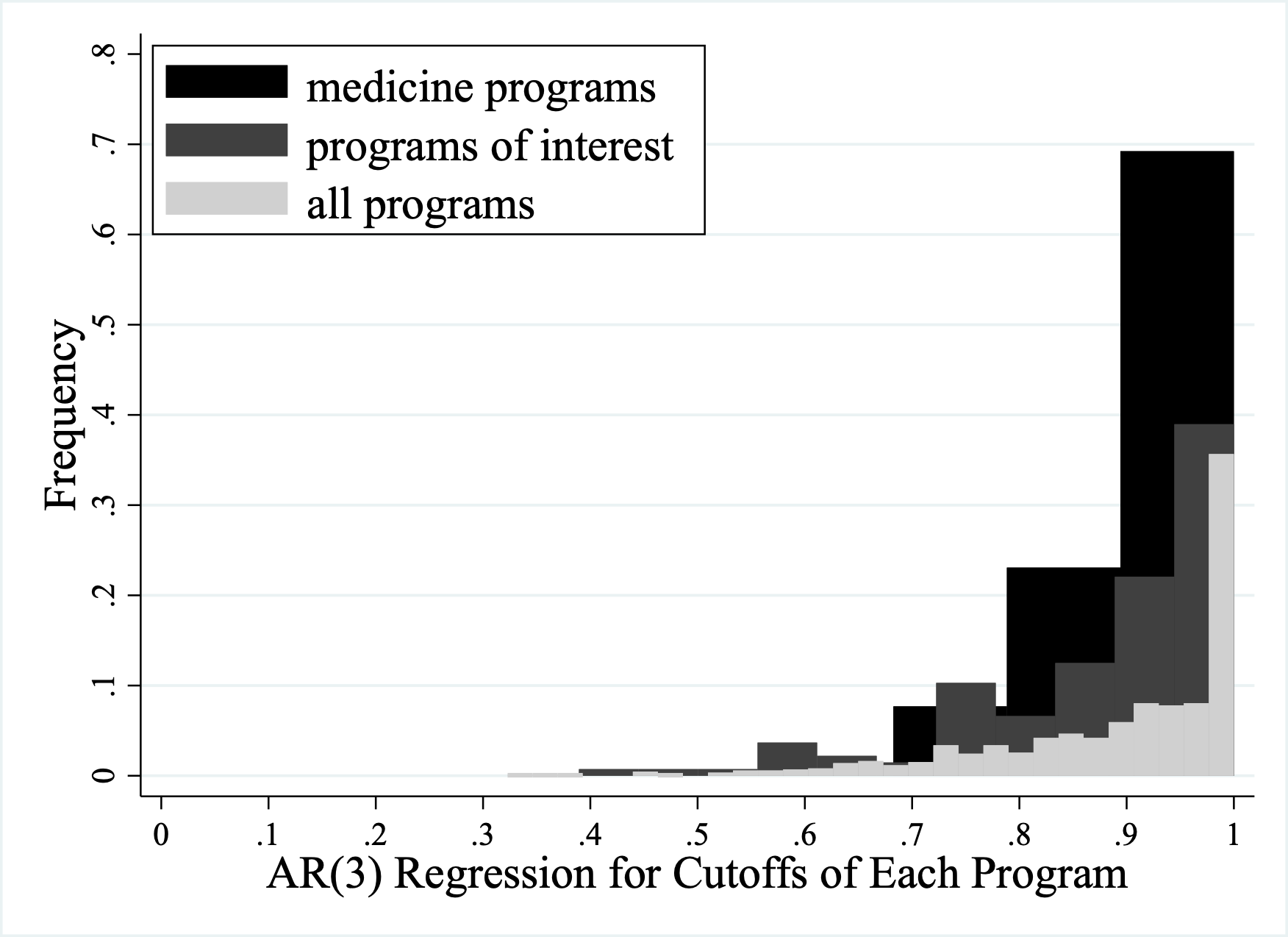}  
        \caption{}
        \label{fig:cutoff:r2}  
    \end{subfigure}

\end{center}
\vspace{-.7cm}
\footnotesize{This figure displays distributions of cutoffs of college-program pairs in Chile for the years of 2004--2010. 
Histograms in both panels show distributions for three sets of programs: all programs (light gray), the subset of programs selected for the analysis of Section \ref{sec:empir:results}, i.e., programs of interest (medium gray), and medicine programs (dark gray). 
Panel (a) plots the cutoff values pooled over all years.
Panel (b) displays the distribution of program-specific $R^2$'s. 
We construct an $R^2$ for each program by regressing cutoff at time $t$ on three lagged cutoff values after de-meaning the seven years of cutoff data.}
    
\end{figure}

Expectations about cutoffs play a central role in program choice. In Chile, students can readily infer these from historical cutoff data, and many programs exhibit stable cutoffs over time (see, e.g., \cite{kapor2024}, Figure 1). Because students observe their priority scores before submitting $P$, cutoff stability implies considerable certainty about which programs are feasible.
Figure \ref{fig:cutoff} shows that more selective programs have not only higher cutoffs but also more \ predictable cutoffs over time.
Panel (a) plots the histogram of cutoff values pooled across years and then compares that with two subsets of programs:
first, with programs that we select to illustrate our methods in Section \ref{sec:empir:results} below; 
and second, with medicine programs.
Section \ref{sec:empir:results} focuses on programs with a sufficiently large number of applications because we need adequate sample sizes to implement our estimation procedures.
We refer to them as programs of interest.
Programs of interest are more selective than most, and Medicine programs stand out as exceptionally selective. 
We see in Panel (a) that cutoffs generally increase with selectivity and tend to be more compressed at the top of the distribution.
In order to assess how predictable program-specific cutoffs are, we estimate an auto-regressive (AR) model for cutoff value at time $t$ for each program as a function of past cutoff values. 
Our sample has seven years of data which allows us to estimate an AR with three lags after de-meaning the data.
Panel (b) displays the distribution of these program-specific $R^2$'s. 
That distribution is again compared with the programs of interest of Section \ref{sec:empir:results} and with medicine programs. 
We find that selectivity is generally associated with predictability of cutoffs.
For most programs of interest, the $R^2$ is larger than 0.70, while 70\% of medicine programs have an $R^2$ greater than 0.90. 
The survey data evidence on strategic behavior regarding Medicine is consistent with agents having strong \textit{a priori} knowledge about \textit{ex post} cutoffs for Medicine.

\begin{figure}[t!]
\begin{center}
\caption{Applications to a program peak among applicants with scores close to the cutoff}\label{fig:appliproba_distcutoffmean}
\includegraphics[width=.6\textwidth]{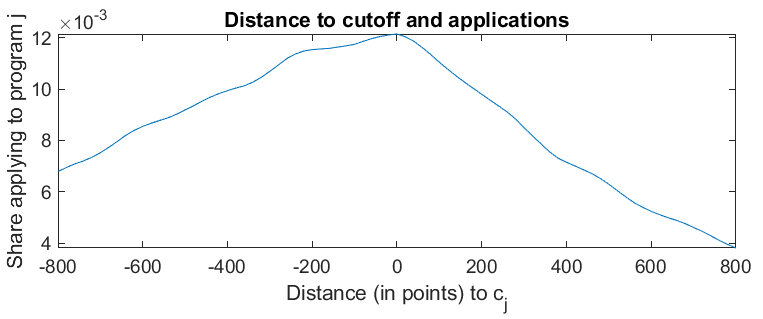}
\begin{tabular}{p{16cm}}
\footnotesize{This figure provides evidence of strategic behavior in student applications. Pooling all programs $j$ together, the figure shows the fraction of students listing program $j$ in their rank-order lists (ROLs or $P$ in terms of our notation), as a function of the distance between their priority score for program $j$ and the cutoff $c_j$. }
\end{tabular} 
\end{center}
\end{figure}

Figure \ref{fig:appliproba_distcutoffmean} provides additional evidence of strategic behavior arising from a priori knowledge of cutoffs. Consider a program $j$ with cutoff $c_j$. Suppose that students tend to prefer programs of higher quality, consistent with what is found in the literature. If applicants behave strategically, one would expect applications to program $j$ to peak among students with application scores close to $c_j$. If cutoffs tend to remain in the same neighborhood across years, students with application scores much higher than $c_j$ can expect to be admissible to more selective, higher-quality programs than $j$, which they prefer over $j$. Hence, we expect very few of these students to include $j$ on their list. As application scores drop and are closer to $c_j$, students' chances of admission to the most selective programs decrease, and program $j$ becomes one of the most selective (desirable) programs among those for which they still have a high admission probability. Hence, we expect applications to $c_j$ to increase as application scores decrease and draw closer to $c_j$. As application scores decrease below $c_j$, students realize that their probability of admission to program $j$ is lower, and while program $j$ remains a relatively desirable (selective) alternative, we expect these expectations to drive applications down. This application pattern, expected if students behave strategically, is exactly what we observe in the top panel of Figure \ref{fig:appliproba_distcutoffmean}.
Pooling all programs $j$ together, Figure \ref{fig:appliproba_distcutoffmean} shows the fraction of students listing program $j$ in their rank-order lists (ROLs or $P$ in terms of our notation), as a function of the distance between their priority score for program $j$ and the cutoff $c_j$.

\begin{figure}[t!]
    \begin{center}
\caption{Evidence of strategic behavior of the discontinuous type }\label{fig:discontinuities}
\includegraphics[width=0.49\linewidth]{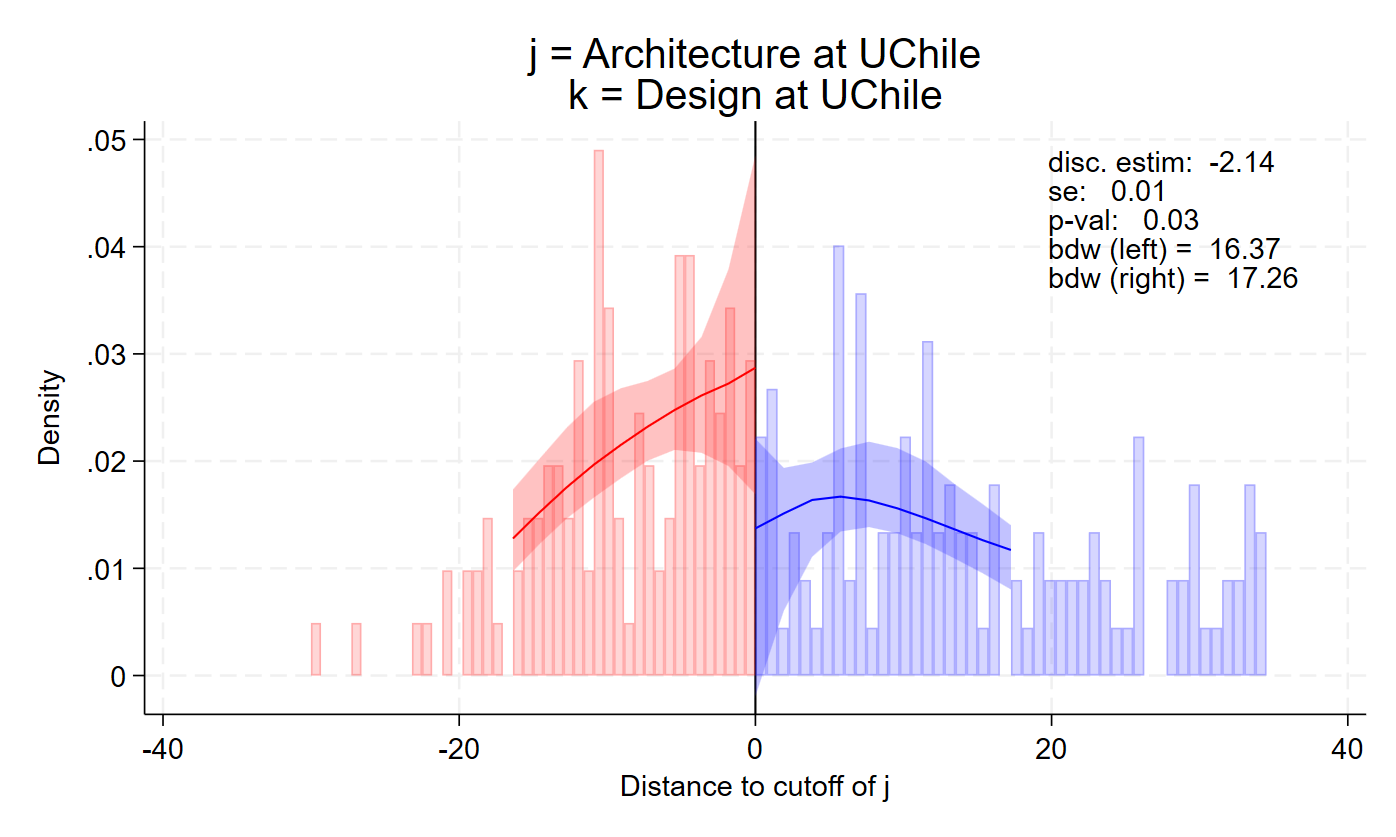}%
\includegraphics[width=0.49\linewidth]{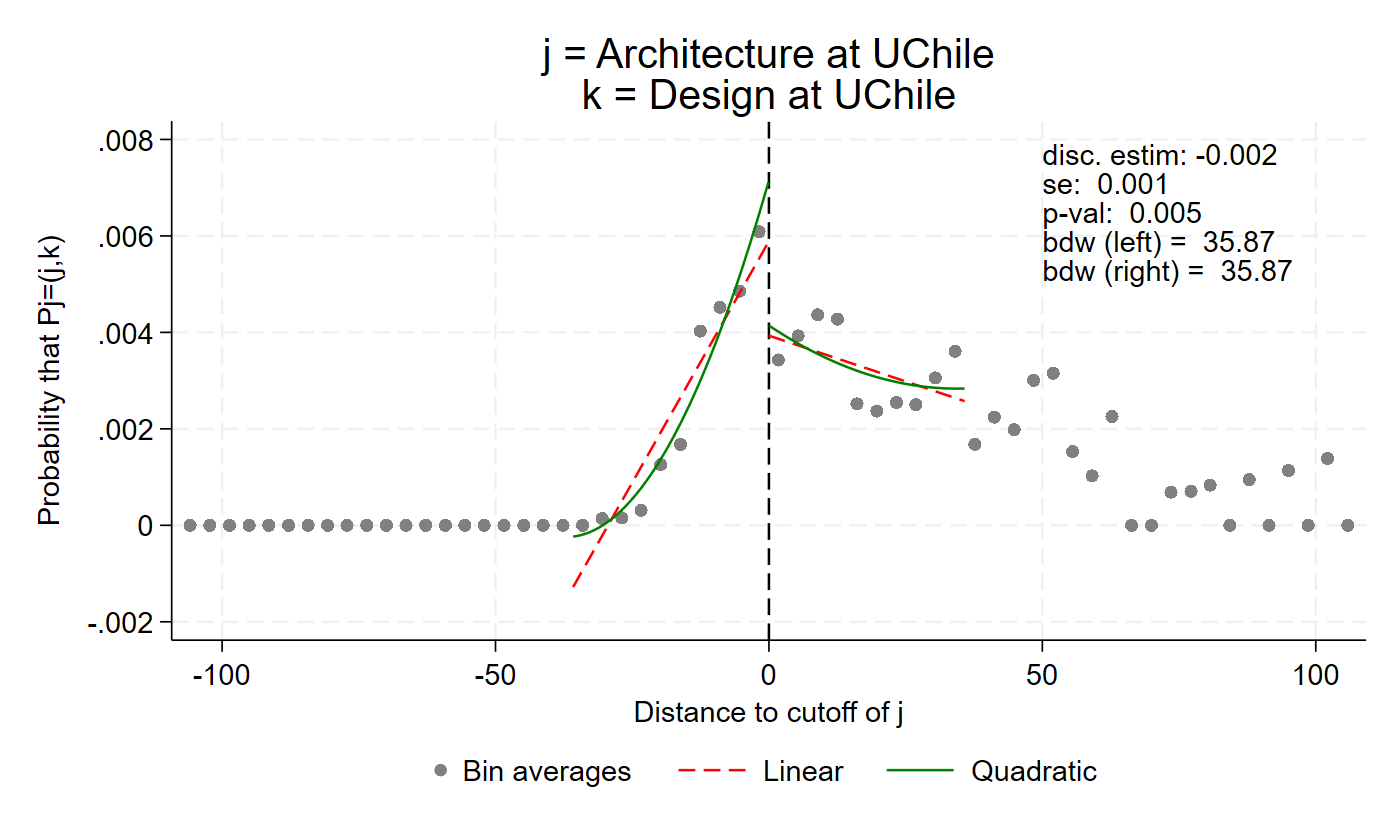}\\

\vspace{.4cm}
\includegraphics[width=0.49\linewidth]{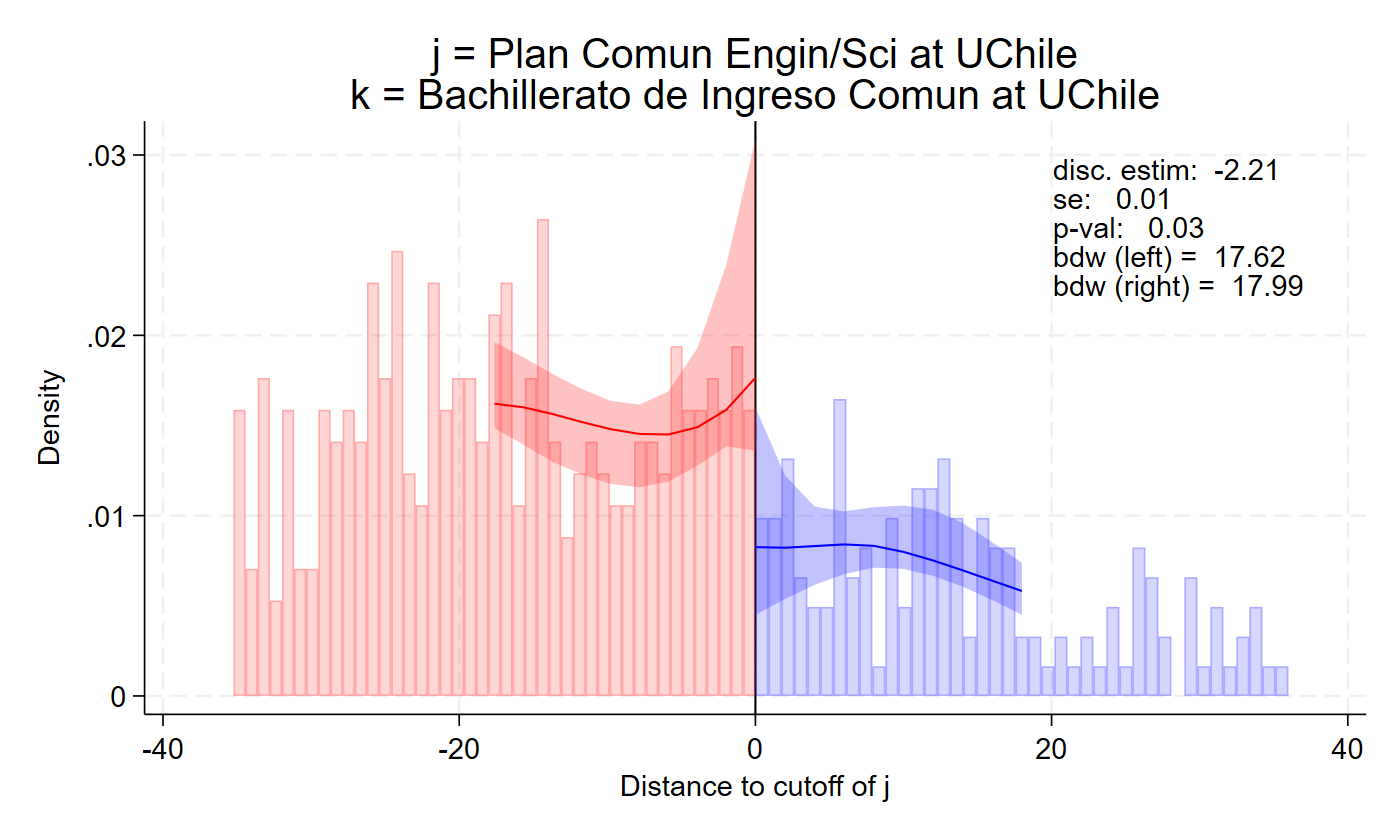}%
\includegraphics[width=0.49\linewidth]{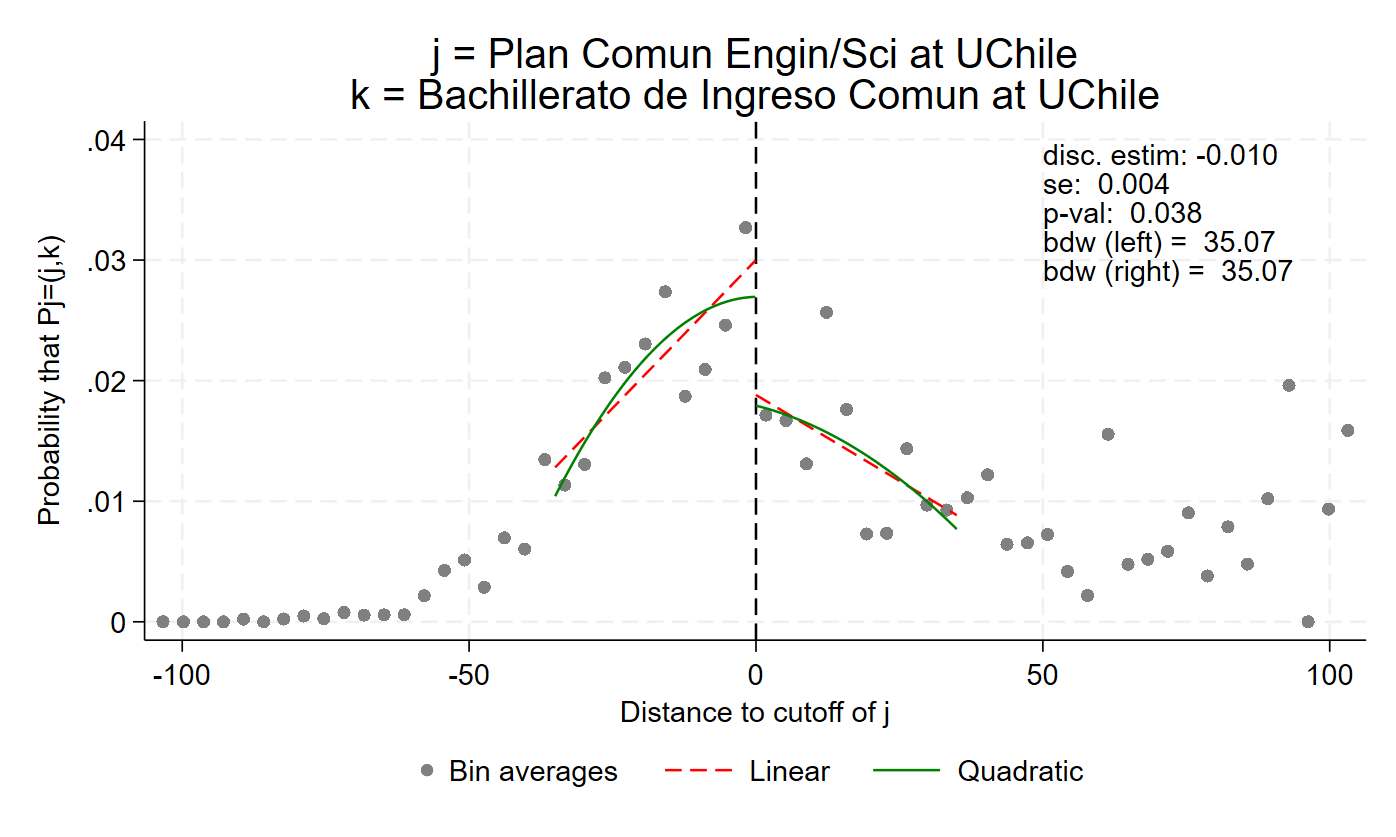}\\

\vspace{.4cm}
\includegraphics[width=0.49\linewidth]{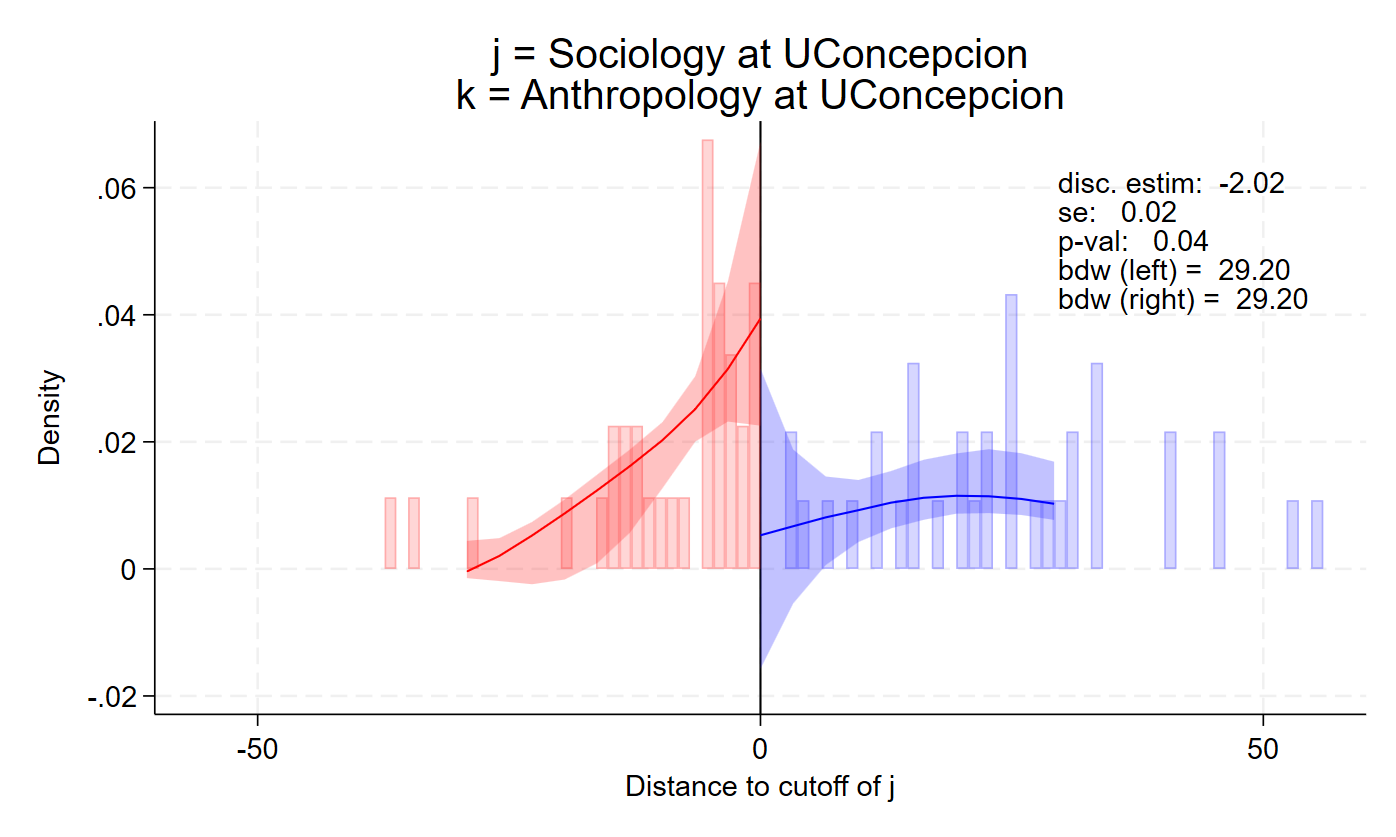}
\includegraphics[width=0.49\linewidth]{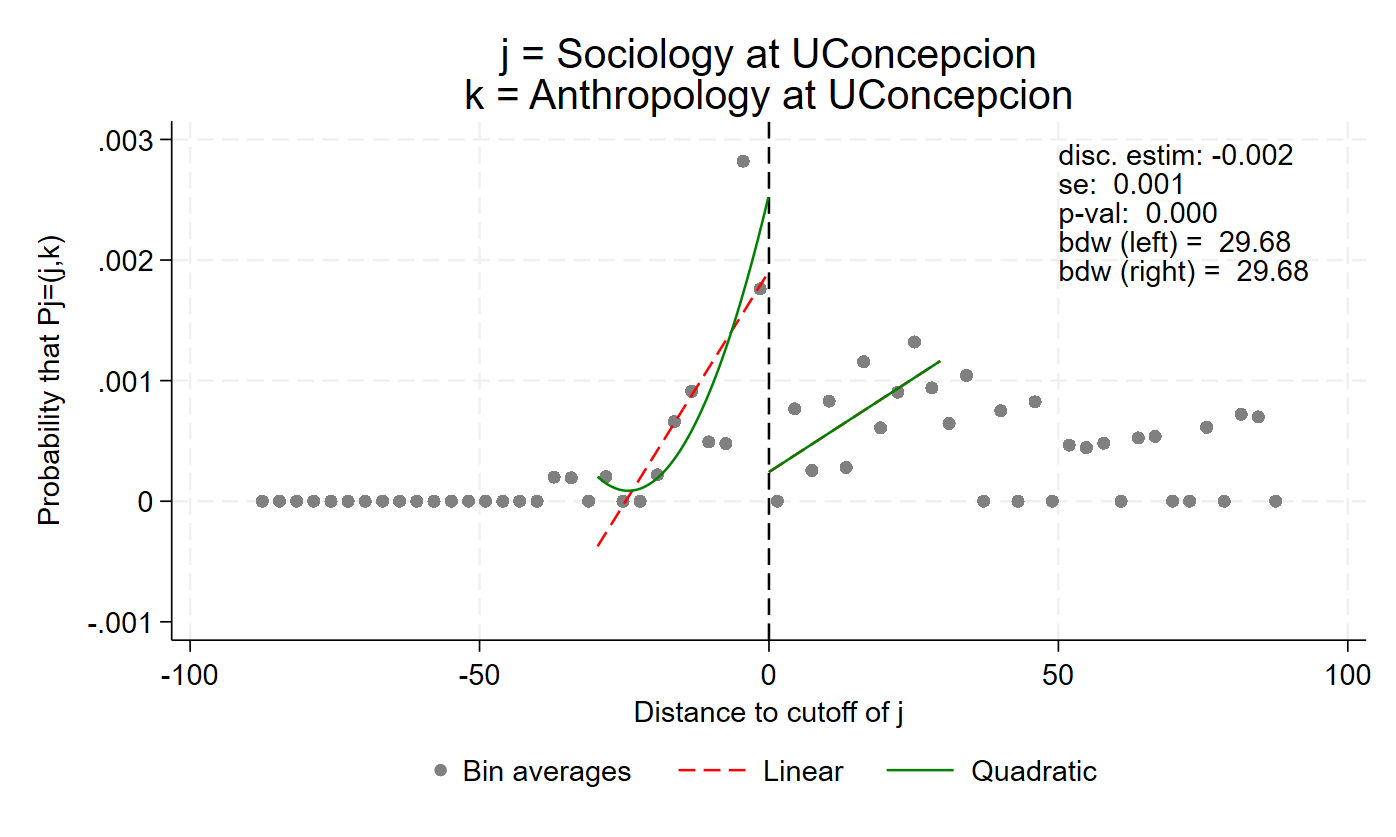}
\begin{tabular}{p{16cm}}
\footnotesize{ This figure provides evidence of strategic behavior of the discontinuous type. Plots on the left show, for three pairs $(j,k)$, the density of the application score $S_j$ among applicants for whom $P_j=(j,k)$. Plots were generated using the \texttt{rddensity} package; the discontinuity estimates provided were obtained using the default MSE-based bandwidth selection method. The graphs on the right show the probability that $P_j=(j,k)$ among all potential applicants as a function of $S_j$. The discontinuity estimates provided were obtained using the \texttt{rdrobust} package and its default MSE-based bandwidth selection method. }
\end{tabular} 
\end{center}
\end{figure}

It may be difficult to disentangle the role of preferences from the role of expectations about admission probabilities when both may enter students' choice of which programs to include in their ROLs (\cite{manski_ecta2004}; \cite{agarwal2018}). The pattern observed in Figure \ref{fig:appliproba_distcutoffmean} could, alternatively, be consistent with students not behaving strategically but preferring programs that are a good fit in terms of quality, 
that is, programs in which their skill level would be close to the marginal skill level. If this were the case, application behavior would not change discontinuously as students’ scores cross a program’s admission cutoff. Although such discontinuities are not required for strategic behavior to arise, examining specific program pairs—rather than pooling all programs as in Figure \ref{fig:appliproba_distcutoffmean}—reveals clear discontinuities (Figure \ref{fig:discontinuities}).
The left panels plot, for three program pairs $(j,k)$, the density of the application score $S_j$ among applicants with $P_j=(j,k)$ and show a significant jump at $S_j=c_j$. The right panels display the probability that $P_j=(j,k)$ among all potential applicants as a function of $S_j$, again exhibiting a sharp discontinuity at the cutoff.
Importantly, the programs $j$ in these pairs have highly predictable cutoffs: the $R^2$ from three-lag autoregressive models forecasting the cutoff at time $t$ exceeds 0.98 in all three cases. This discontinuous type of strategic behavior may lead to issues with internal validity of an RD identification strategy that controls for $P_j$ (see Section \ref{sec:naive} for discussion).

\vspace{-.3cm}

\subsection{Results}\label{sec:empir:results}

\indent 

We are interested in identifying the effects of assignment to a given postsecondary program on college graduation.
College returns are typically thought of as tied to college graduation, motivating our focus on graduation-related outcomes (see \cite{kirkeboen2016} and \cite{altonjiarcidiaconomaurel2016} for further references). In addition, the extent to which students eventually graduate from the program to which they are assigned to (or from programs their assigned program is a pathway to) can be viewed as measure of performance of the assignment mechanism. However, for the average college program, initial enrollment and eventual graduation are far from being perfectly correlated (OECD 2019, \cite{larroucaurios2021}). 
We also consider graduation from a top university as another outcome of interest.
The choice of this second outcome is in line with the literature on returns to education and college choice, which highlights the role of institution quality in driving returns (see \cite{kirkeboen2016} and \cite{altonjiarcidiaconomaurel2016} for more references). In our setting, the set of ``top'' universities consists of Pontificia Universidad Cat\'olica de Chile in Santiago (PUC Santiago; hereafter PUC) and Universidad de Chile (UChile).

We present three sets of results. First, we report estimates of average structural functions, followed by estimates of treatment effects, and then discuss the economic implications our results.   In these two exercises, to keep things simple, we focus on a single program $j$ and the popular next-best programs $k$ associated with it (or a single program $k$ and the popular local first-best $j$ associated with it).
Finally, we further illustrate the importance of our approach by showing that ignoring strategic behavior in identifying treatment effects can lead to misleading conclusions. In this exercise, we estimate bounds on the treatment effects of interest for a large number of pairs $(j,k)$ and present statistics across these pairs. 
The bounds we present throughout this section represent estimates---not inference bounds. Inference is beyond the scope of this paper.\footnote{Beyond the fact that the admission cutoffs are estimated, the fact that the $\delta$s used in the construction of bounds are partially identified must be taken into account when confidence bands are derived---a task that we leave for future research.} As noted earlier, the identification proposed in this paper resembles a sharp RD design and is therefore well suited for studying the effects of assignment to a program on graduation-related outcomes. In contrast, estimating the effects of graduation from a program on earnings typically involves a fuzzy design, which we study in separate ongoing work.\footnote{At the time of writing of the present paper, we do not have access to earnings data and therefore cannot provide bounds here on the intent-to-treat (ITT) effects on earnings.}

\subsubsection{Average Structural Functions and the Importance of Preferences for Graduation Outcomes}

\paragraph{Setup.} We focus on several program pairs 
$(j,k)$ that share the same 
$k$: the Bachillerato de Ingreso Común at UChile. This program is chosen for two reasons: (1) its large applicant pool allows for precise estimation; and (2) its design and purpose make it intrinsically interesting. As a popular entry point at a selective university, the Bachillerato lets students explore multiple disciplines before choosing a major; it is often seen as an alternative route into competitive programs for those without high enough scores for direct admission. This makes it particularly compelling to examine whether assignment to this program affects students’ likelihood of graduating from their original local first-best 
$j$. Accordingly, we examine two graduation outcomes: (1) graduating from one’s local first-best program $j$ when assigned to the next-best alternative, the Bachillerato de Ingreso Común at UChile; and (2) graduating from a top university. Specifically, we show bounds for the average structural function $\mme\left[ Y(k) \left| Q_j=(j,k), S_j=c_j \right. \right]$, as in Eq. \eqref{eq:bounds_ASF_k}. We restrict attention to five of the most common local first-best programs 
$j$ ---Medicine at UChile, Medicine at Universidad de Santiago de Chile, Odontology at UChile, Kinesiology at UChile, Math at PUC.

As a simple procedure to construct a sample local to each of the admission cutoffs $c_j$ of interest, we use a 30-point bandwidth on either side of $c_j$.\footnote{The resulting estimation sample is therefore invariant across outcomes once the pair $(j,k)$ is fixed.} Given our choice of bandwidth, the next step of the exercise is to recover local preferences $Q_j$ for each student at each cutoff $c_j$. 
In the absence of strategic behavior, these preferences can be directly inferred from application lists ---direct observation of $P$ and placement scores allows to compute cutoffs, budget sets, and finally $P_j$, which equals $Q_j$ under truth-telling. 
However, under strategic behavior, $Q_j$ is not fully revealed by the observed $P_j$. Section \ref{sec:id_r:qj_id} shows how to construct the set of possible $Q_j$s, that is, the set $\bQ_j$, under various assumptions. The weak partial order assumption (WPO) implies that any submitted list is a subset of acceptable programs ordered just as in the student's true preferences. 
The strong partial order assumption (SPO) adds that students who submit shorter-than-maximum lists reveal their full set of acceptable programs.
Under SPO, we fully observe $Q_j$ for students who rank strictly fewer than the maximum number of programs (eight, in the case of our empirical application). For students ranking eight choices, $Q_j$ is observed only if the set $\mathbf{Q}_j$ comes out as a singleton. 
The UMAS assumption (Assumption \ref{aspt:umas}, hereafter UMAS) can be used in combination with WPO and SPO to refine $\mathbf{Q}_j$ for the non-singleton cases (Corollary \ref{result:umas_refine}). 
In a context with a large number of alternatives and a small cap $K$, like in Chile's college application system, the $\bQ_j$ constructed under WPO will likely be large. As discussed in Section \ref{sec:id_r:qj_id}, additional context-specific assumptions can be combined with UMAS to reduce the size of $\bQ_j$. In the applied literature, fields of study are widely recognized as key drivers of student preferences—and of the heterogeneity in preferences across students—for post-secondary programs \citep{altonjiarcidiaconomaurel2016}. Building on this insight, we propose the following {\it fields} assumption: 
\begin{assumption}\label{aspt:fields}
Suppose the set of all programs is partitioned into fields of study. Let $\mathcal{F}$ be a set of programs in an arbitrary field of study in this partition. Consider any cutoff $c_j$. Let $\omega$ be a student with score $S_j$ near $c_j$. If no program from field $\mathcal{F}$ appears in the reported list $P(\omega)$, then no program from $\mathcal{F}$ can be part of the student's true local preference at $c_j$, i.e., 
\vspace{-.25cm}
\begin{align*}
    & \forall \mathcal{F}, P(\omega) \cap \mathcal{F} = \emptyset \Rightarrow Q_j(\omega) =(a,b) : a \notin \mathcal{F} \text{ and } b \notin \mathcal{F}.
\end{align*}
\end{assumption}
\noindent  In practice, we partition programs in five fields of study: Health/Medicine; STEM; Economics/Business; Law; and Other. So for instance, if a student's reported preferences include programs in Medicine, STEM, and Economics only, Assumption \ref{aspt:fields} excludes programs in Law and Other from their $\bQ_j$.
For any program $\ell \in \mJ$, let $\mathcal{F}(\ell)$ denote the set of all programs in the field of study of program $\ell$. Researchers may apply Assumption \ref{aspt:fields} to refine the set $\bQ_j(\omega)$ of student $\omega$ as follows:
\vspace{-.25cm}
\[
\bQ_j^*(\omega) = \bQ_j(\omega) 
\setminus 
\left\{(\ell,\ell') \in \m{J} \times \m{J}^0 :  
P(\omega) \cap \mathcal{F}(\ell)=\emptyset ~~\text{or}~~ P(\omega) \cap \mathcal{F}(\ell')=\emptyset \right\},
\]
where we adopt the convention that $\mathcal{F}(0) = \m{J}$.

\begin{figure}[t]
\caption{Graduation Rates if Assigned to Bachillerato de Ingreso Comun at UChile}\label{fig:k1178_asf}
\centering
\includegraphics[width=.49\linewidth]{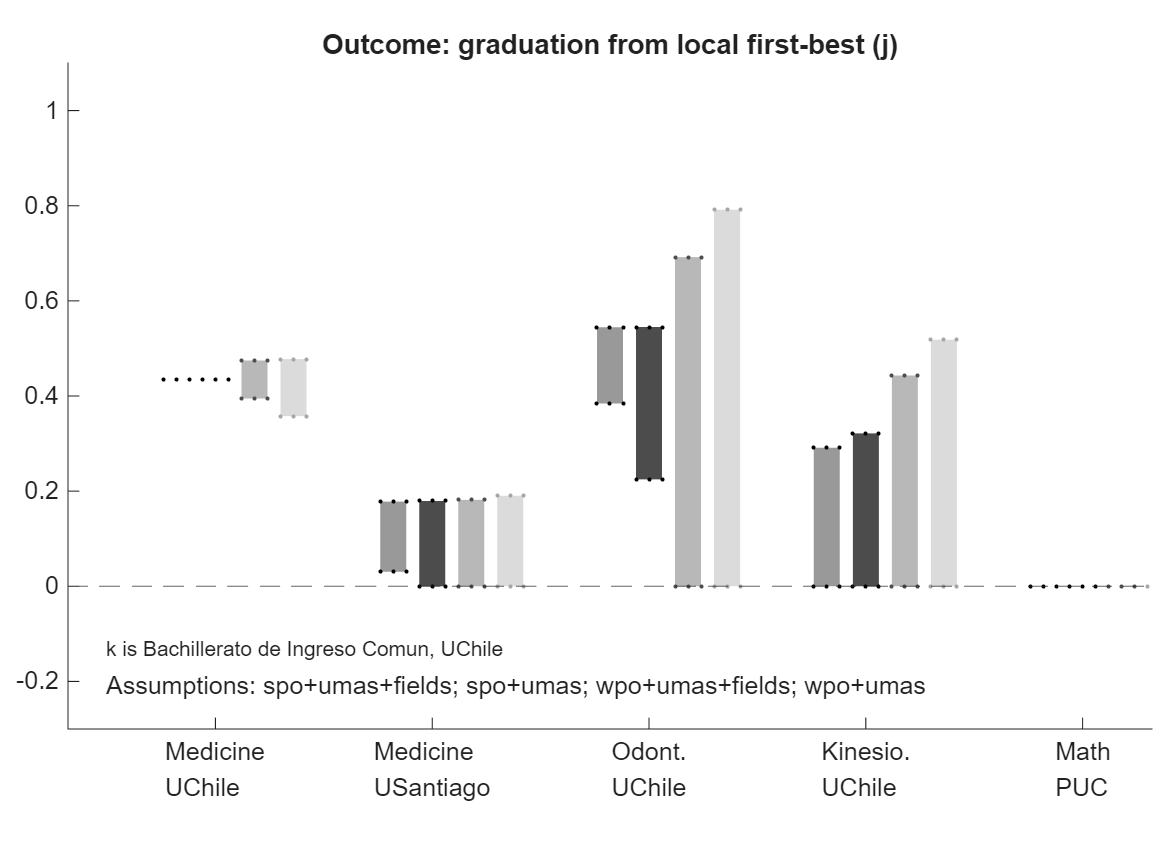}%
\includegraphics[width=.49\linewidth]{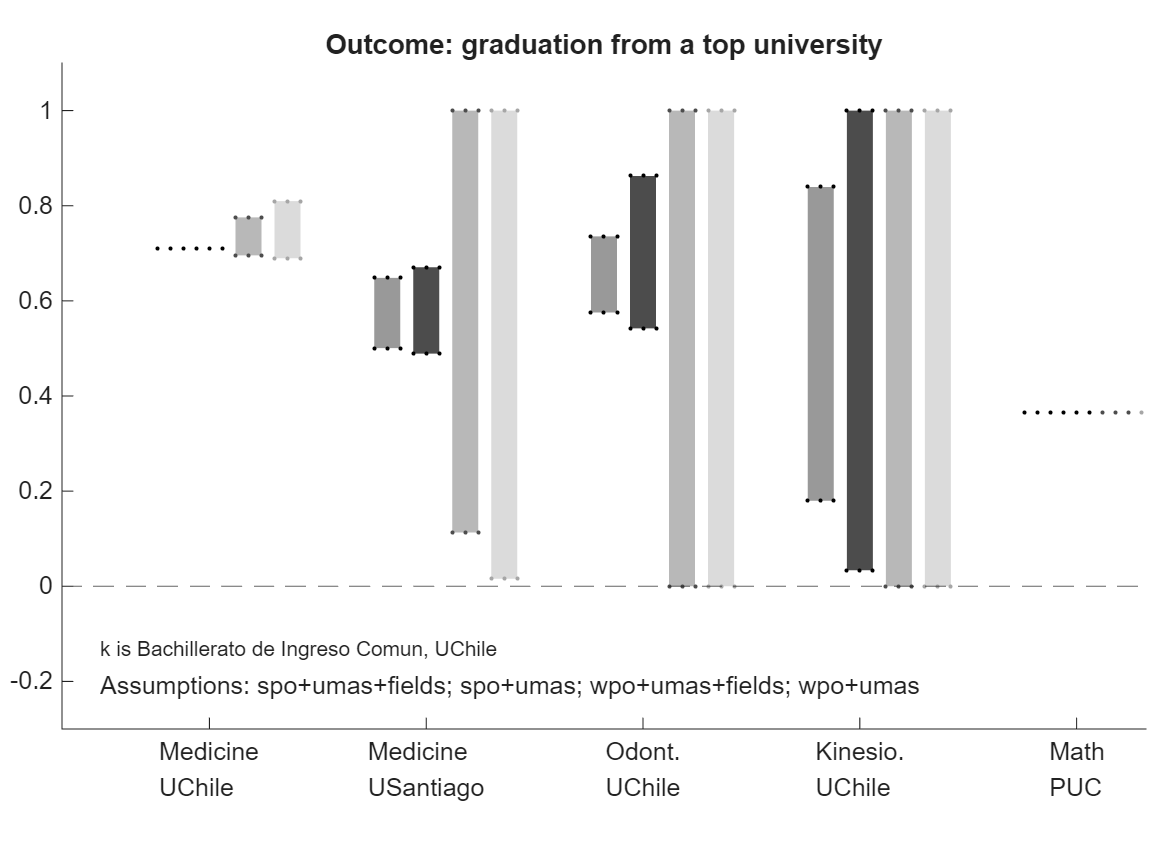}
\begin{tabular}{p{16.3cm}}
\footnotesize{The figure shows estimated bounds for $\mme[Y(k)|Q_j=(j,k),S_j=c_j]$ (y-axis) with $k=$ Bachillerato de Ingreso Comun at UChile for several local first-best programs $j$ (x-axis), two binary graduation outcomes (graduation from the local first-best choice $j$ on the left panel and graduation from a top university, i.e., PUC or UChile, on the right panel), and four different behavioral assumptions used to construct $\bQ_j$ (four bars for each $j$ in each panel). 
The four behavioral assumptions are, from left to right, SPO+UMAS+{\it fields}, SPO+UMAS, WPO+UMAS+{\it fields}, and WPO+UMAS.
The average parameter denotes the probability of graduating (from a top university or from program $j$) among those individuals with $S_j=c_j$ and $Q_j=(j,k)$ if they were all assigned to program $k$. 
All sets of bounds are estimated with a 30-point bandwidth on either side of the assignment cutoff $c_j$ of the program of interest.
Note that these are estimated bounds and not inference bounds.
}
\end{tabular} 
\end{figure}

Figure \ref{fig:k1178_asf} reports estimated bounds on the probability of graduating if assigned to Bachillerato de Ingreso Común at UChile across local first-best programs $j$, for students whose local next best is Bachillerato de Ingreso Comun at UChile. 
The left panel shows the probability of graduation from the local first-best program; the right panel shows the probability of graduation from a top university. 

\paragraph{On the importance of preferences for graduation outcomes.} For both outcomes, we find evidence of heterogeneous effects across different local first-best programs ---in a number of cases, bounds do not overlap across $j$. For instance, in the left panel, students whose local first-best is Medicine at UChile are more likely to graduate from that program after assignment to the Bachillerato than those whose first-best is Medicine at Universidad de Santiago de Chile or Math at PUC. In the right panel, students with Medicine at UChile as their first-best are more likely to graduate from a top university when assigned to the Bachillerato than those with Math at PUC as their first-best.

The observed heterogeneity may arise from two sources. First, differences in preferences—potentially correlated with effort and ability—can influence graduation outcomes. Second, variation in academic preparation or ability, proxied by test scores, could also play a role as we are comparing subpopulations with $S_j=c_j$ for different $j$'s in Figure \ref{fig:k1178_asf}. We show that this second channel is unlikely to explain the differences observed in Figure \ref{fig:k1178_asf}. If admissions across programs $j$ were based on a single priority score, it would suffice to show that the cutoffs $c_j$ are close or that applicant score distributions are similar around them. However, priority scores vary by program, as each $S_j$ is a weighted sum of six primary components with program-specific weights.\footnote{Primary scores: NEM, math, Spanish, science, history, and the max of science and history. Weights can be zero but always sum to 100\%.} Therefore, to assess comparability across cutoffs, we analyze the Euclidean distance between students’ six-dimensional primary score vectors. We show that at any of the cutoffs $c_j$ of interest, students within the bandwidth of the cutoff are, in terms of the Euclidean distance between their vectors of primary scores, on average as close to each other as they are to students around the other cutoffs of interest. Results are shown in Table \ref{tab:Yk_agvdist_1178} and discussed in detail in Appendix \ref{sec:app:empirical}. 

We therefore interpret the results in Figure \ref{fig:k1178_asf} as evidence that students' preferences matter for their graduation outcomes, suggesting that preferences are correlated with effort choices or ability that is not captured by test scores.\footnote{Evidence from \cite{heckmankautz2012} shows that achievement tests do not fully capture soft skills or personality traits that influence labor market outcomes (see also \cite{borghans2008}, \cite{almlund2011}). Structural models support this, showing that factors beyond test scores affect educational choices. For example, \cite{arci2005} find that, even conditional on SAT scores, some individuals have higher college admission chances, financial aid likelihood, labor market earnings without college, and returns to all majors.}

\paragraph{Role of behavioral assumptions.} We next examine how different behavioral assumptions affect the estimated bounds. Table \ref{tab:samplesizes_deltas_k1178} in the appendix provides descriptive statistics and the estimated $\lbar{\delta}_{j,k}^+$ and $\lbar{\delta}_{j,k}^-$ for each pair $(j,k)$ and each set of assumptions. The different assumptions are used to construct $\bQ_j$ for each individual and therefore determine the conditioning set $\{\bQ_j \cap \{(j,k)\} \neq \emptyset\}$ and the values of $\lbar{\delta}_{j,k}^+$ and $\lbar{\delta}_{j,k}^-$. Under WPO, $\bQ_j$ is larger than under SPO, resulting in a larger conditioning set (Table \ref{tab:samplesizes_deltas_k1178}, Columns (1) vs. (3) and (2) vs. (4)). Given a fixed bandwidth around the cutoff $c_j$, this 
means that (i) the denominator in the equations defining $\lbar{\delta}_{j,k}^+$ and $\lbar{\delta}_{j,k}^-$ is larger under WPO than under SPO; and (ii) 
the share of individuals for whom $Q_j$ is found to be a singleton is lower under WPO than it is under SPO, which in turn implies that the numerator in the equations defining $\lbar{\delta}_{j,k}^+$ and $\lbar{\delta}_{j,k}^-$ is smaller under WPO than under SPO. Overall, this leads to $\lbar{\delta}_{j,k}^+$ and $\lbar{\delta}_{j,k}^-$ being smaller under WPO than under SPO. The increase in $\lbar{\delta}_{j,k}^+$ and $\lbar{\delta}_{j,k}^-$ as we impose SPO instead of WPO unambiguously tends to reduce the width of the bounds. The exact extent to which the width of the bounds shrinks as we impose SPO instead of WPO also depends on the joint distribution of the outcome $Y$ and $\bQ_j$, however. Subsequently imposing UMAS and the {\it fields} assumption eliminate certain elements from individuals' $\bQ_j$s and futher increase $\lbar{\delta}_{j,k}^+$ and $\lbar{\delta}_{j,k}^-$ (Columns (1) vs. (2) and (3) vs. (4)).
\footnote{Note that the impact of the {\it fields} assumption depends on the overlap between fields of $j$ and $k$. Under stability, the first coordinate of $P_j$ and the first coordinate of any element of $\bQ_j$ coincide for students just above $c_j$, while the second coordinates of $P_j$ and any element of $\bQ_j$ coincide for students just below $c_j$ 
(Proposition \ref{result:partial_id:nextbest}). 
This means that, absent the {\it fields} assumption, each student in the set $\{\omega: \bQ_j \cap \{(j,k)\} \neq \emptyset\}$ includes a program from either $\mathcal{F}(j)$ or $\mathcal{F}(k)$ in her reported preference. As a consequence, the {\it fields} assumption can only reduce the set $\{\omega: \bQ_j \cap \{(j,k)\} \neq \emptyset\}$ for pairs $(j,k)$ for which $\mathcal{F}(j)\neq \mathcal{F}(k)$. 
}

\subsubsection{Treatment Effects}
We now illustrate our method for estimating treatment effects, as opposed to average structural functions. Specifically, we estimate the effect on graduation outcomes of being assigned to Medicine at PUC Santiago instead of being assigned to a second-best option. We focus on this program for two main reasons. First, its high selectivity and popularity ensure a sufficiently large sample for precise estimation. Second, its competitiveness and highly predictable cutoff make it likely that applicants adjust their submissions strategically around the cutoff—a type of behavior our approach is designed to guard against.  We consider the five most common next-best options to Medicine at PUC: Medicine at UChile, at Universidad de Concepción, and at Universidad de Santiago de Chile; and  Science, and Engineering at PUC. As in earlier analyses, we construct our estimation sample using a 30-point bandwidth around the cutoff $c_j$.

\begin{figure}[t]
\caption{Effects of assignment to Medicine at PUC vs. next-best alternatives} \label{fig:j1258_TE}
\centering
\includegraphics[width=.49\linewidth]{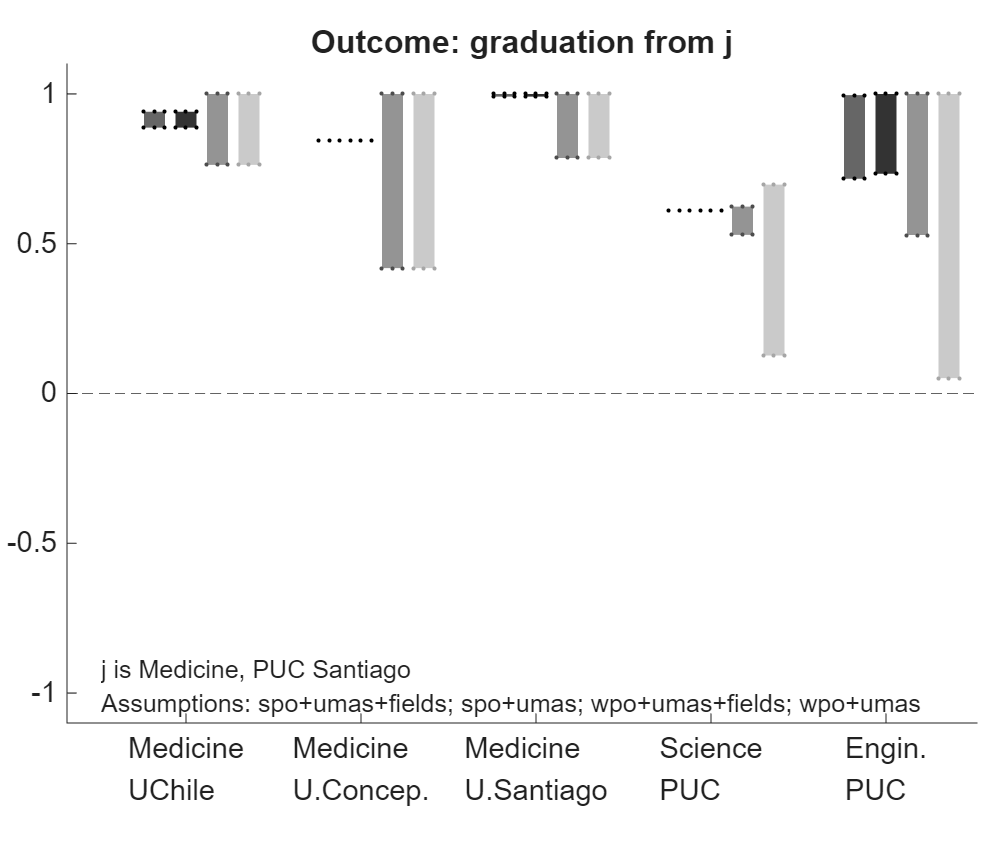}%
\includegraphics[width=.49\linewidth]{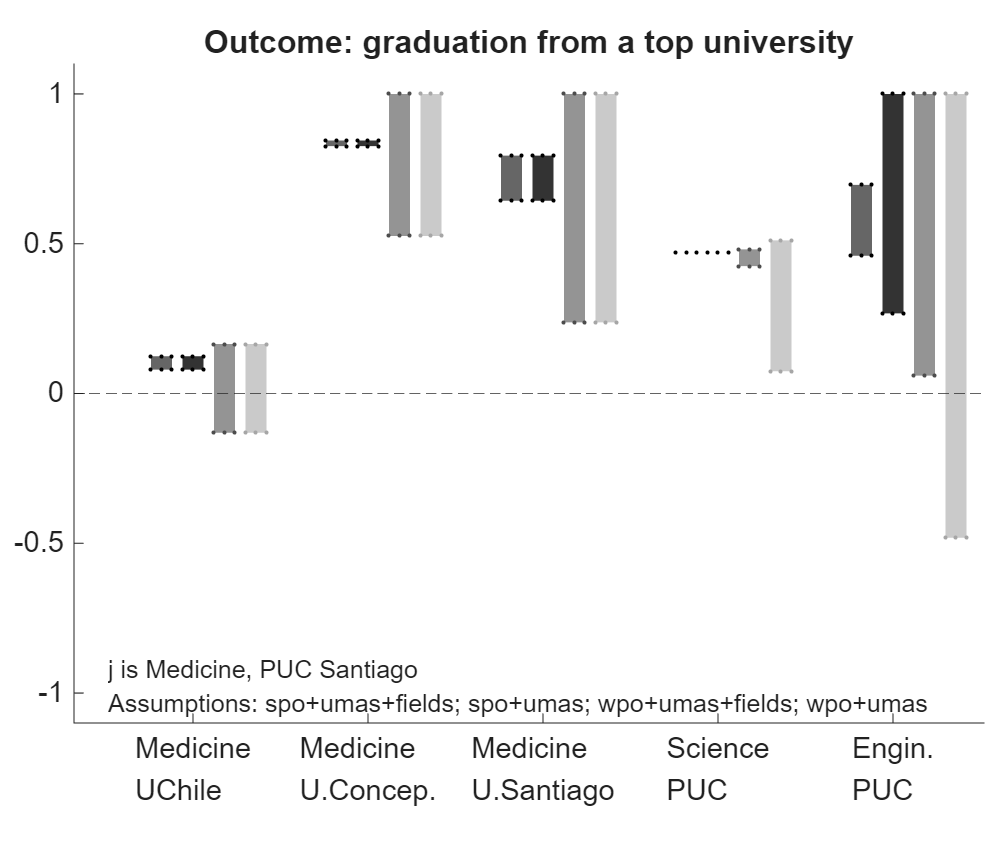}
\begin{tabular}{p{16.3cm}}
\footnotesize{The figure shows estimated bounds for the average effect on the probability of graduation (y-axis) of assigning students to Medicine at PUC as opposed to assigning them to a second-best program. 
Bounds are shown for five different second-best programs (x-axis), two binary outcomes (graduation from Medicine at PUC on the left panel and graduation from a top university, i.e., PUC or UChile, on the right panel), and four behavioral assumptions used to construct $\bQ_j$ (four bars for each second-best program).
The four behavioral assumptions are, from left to right, SPO+UMAS+{\it fields}, SPO+UMAS, WPO+UMAS+{\it fields}, and WPO+UMAS. All bounds are estimated with a 30-point bandwidth on either side of the assignment cutoff of the program of interest (Medicine at PUC). 
Note that these are estimated bounds and not inference bounds.
}
\end{tabular} 
\end{figure}

Figure \ref{fig:j1258_TE} presents the estimated bounds. The left panel shows treatment effects on the probability of graduating from Medicine at PUC Santiago; the right panel shows treatment effects on the probability of graduating from a top university. Starting with the left panel, we find, reassuringly, that assignment to Medicine at PUC increases the probability of graduating from that program. In all five cases, bounds derived under any of the four sets of assumptions exclude zero, identifying a positive treatment effect. Importantly, the magnitude of the effect varies across next-best alternatives, as reflected by non-overlapping bounds. For example, when it comes to graduating from Medicine at PUC, students whose second-best is Science at PUC appear to benefit less from admission to Medicine at PUC than those whose next-best is Medicine at UChile or at Universidad de Santiago de Chile.

The right panel reveals similar patterns. Admission to Medicine at PUC increases the probability of graduating from a top university in many cases, and regardless of whether the next-best option is at a top university itself. For instance, the estimated bounds indicate a positive effect both for students whose next-best program is Science at PUC and for those whose next-best is Medicine at Universidad de Concepción or Universidad de Santiago de Chile. There again, the bounds highlight substantial heterogeneity across next-best alternatives.

As we relax assumptions, the bounds widen and become less informative. This exercise thus helps clarify the identifying power of each assumption, providing researchers with a framework to evaluate the empirical content of different preference assumptions in similar settings.

\subsubsection{On the Importance of Accounting for Strategic Behavior}\label{sec:naive}

Finally, we illustrate our bounding approach more broadly using 302 pairs $(j,k)$ of interest.
We show that ignoring strategic behavior may be misleading when interest lies in a average treatment effect conditional on true preferences as in \eqref{eq:param}.
In what follows, we compare our estimated bounds to point estimates produced by an RD that controls for reported preferences, as the current practice advocates.
Current-practice estimates are consistent for
\vspace{-.25cm}
\[\mu_{j,p} := \mme [Y\mid P_j=(j,k),S_j=c_j^+]-\mme[Y\mid P_j=(j,k),S_j=c_j^-].\]

Under the assumption that every student is a truth-teller at the cutoff $c_j$, 
the parameter
$\mu_{j,p}$ coincides with $\mme [Y(j) - Y(k) \mid Q_j=(j,k),S_j=c_j]$, which is our parameter of interest \eqref{eq:param}.
Since the truth-telling assumption implies our behavioral assumptions, our bounds contain $\mu_{j,p}$ under truth-telling. 
On the contrary, under strategic behavior, $\mu_{j,p}$ does not equal $\mme [Y(j) - Y(k) \mid Q_j=(j,k),S_j=c_j]$ in general, and our bounds may or may not contain $\mu_{j,p}$.
The interpretation of $\mu_{j,p}$ depends on the type of strategic behavior. 
First, as mentioned earlier, strategic behavior may be of the discontinuous type to the point of breaking the continuity of $\mme[Y(d)\mid P_j=(j,k),S_j=s]$ as a function of $s$ at $s=c_j$, for $d=j$ or $d=k$
(Figure \ref{fig:discontinuities} presents evidence consistent with this behavior).
In this case, the RD is not internally valid and $\mu_{j,p}$ does not equal an average treatment effect.
Alternatively, strategic behavior may not be of the discontinuous type but still make $P_j \neq Q_j$.
In this case, $\mu_{j,p}$ equals a weighted average of various average treatment effect parameters, where the weights are functions of the unknown mapping $\omega \mapsto Q_j(\omega)$.

We look at 302 pairs $(j,k)$ which we select as follows. 
First, we consider all pairs $(j,k)$ such that program $j$ is available on the platform at any point between 2004 and 2010 and at least one student lists $k$ right after $j$ in his ROL. 
Second, we count the number of students with score $S_j$ within a 30-point bandwidth of the year-specific $c_j$ who $P_j=(j,k)$.\footnote{
    For a number of $j$'s of interest, there exists another program $\ell$ that uses the same placement score as $j$ and whose cutoff $c_{\ell}$ is within a 30-point distance of $c_j$. 
    In these cases, we use a bandwidth smaller than 30 around the cutoff $c_j$ of interest, so the sample of $c_j$ does not include observations from the sample of $c_{\ell}$ that are subject to a different treatment change.  
    } 
We find 327 pairs $(j,k)$ with a least 50 such students.
These pairs involve 137 distinct programs $j$, each of them associated with a number of $k$'s ranging from one to 18. We exclude the 25 pairs $(j,k)$ associated with three of these programs $j$ because no graduates are recorded in the data for these programs. This yields a final set of 302 pairs involving 134 distinct $j$'s.

Table \ref{tab:stats_results} provides statistics on the estimated bounds that we obtain for our two outcomes of interest, both under the SPO and the SPO+UMAS+fields assumptions. 
First, we see that SPO bounds identify the sign of our average treatment effect parameter for many pairs; the estimated bounds even collapse to point-estimates for a small share of pairs. 
Adding the UMAS and fields assumptions make our estimated bounds narrower, and  they identify the sign of average effects for 70\% of pairs when outcome is graduation from $j$; they collapse to point estimates for  23\% of pairs when outcome is graduation from top university.
Most importantly, we see that current-practice estimates fall outside of bounds in a significant share of cases, 12\% \& 15\% under SPO and 25\% \& 24\% under SPO+UMAS+fields.
This evidence is consistent with lack of truth-telling at the cutoff of these pairs. 
More broadly, this also suggests caution in contexts like the Chilean case, where most but not all students submit $P$ with $|P|<K$ or very few students are assigned to their last-listed option.
It may not be safe to simply assume that everyone is a truth-teller just because the constraint does not bind for most people.
In other words, current-practice estimates may be severely biased for $\mme [Y(j) - Y(k) \mid Q_j=(j,k),S_j=c_j]$ even if the share of constrained students is small.

\begin{table}[t] 
\caption{Bounds at 302 pairs $(j,k)$}\label{tab:stats_results}
\centering
\begin{tabular}{lcc|cc}
\hline\hline
& \multicolumn{2}{c|}{Grad. from $j$} & \multicolumn{2}{c}{Grad. from top uni} 
\\ 
&SPO & SPO+UMAS &SPO & SPO+UMAS 
\\
& & +fields & & +fields 
\\
\cline{2-3}\cline{4-5}
Share of pairs for which:  & & & & 
\\
- bounds identify sign of parameter of interest &0.48&0.70&0.28&0.38
\\
- current-practice estimate is outside of bounds &0.12&0.25&0.15&0.24
\\
- bounds point-identify parameter of interest &0.01&0.03&0.17&0.23
\tabularnewline
[.5em]
\hline\hline
\end{tabular}%
\begin{tabular}{p{16cm}}
\footnotesize{The table summarizes the results of the implementation of our method for 302 pairs $(j,k)$ under the SPO and SPO+UMAS+fields assumptions. 
For each of the two outcomes of interest, the table shows the share of pairs $(j,k)$ for which our estimated bounds (i) identify the sign of the average treatment effect parameter \eqref{eq:param}, (ii) do not contain current-practice estimate $\ha{\mu}_{j,p}$, and (iii) collapse to a point estimate.}
\end{tabular} 
\end{table}
\vspace{-.3cm}

\section{Conclusion}\label{sec:conclu}
\indent

This paper provides a novel approach to partially identify the effects of mechanism assignment on future outcomes that is robust to strategic behavior.
We illustrate our approach using data from Chile, where a DA mechanism assigns about 80,000 students to more than 1,000 university-major programs every year, and for which, in line with previous literature, we find substantial evidence of strategic behavior.  In two high-stakes settings—a selective entry program at the University of Chile and Medicine at PUC Santiago—we find heterogeneous effects of assignment on graduation outcomes across students’ preferred programs and next-best alternatives, consistent with preferences being linked to unobserved traits. 

We illustrate our bounding approach for RD-like parameters at the granular university–program level, where students make choices and matching occurs. In many settings, however, researchers or policymakers may seek more aggregate effects—such as STEM vs. non-STEM—due to limited data near multiple cutoffs or broader policy relevance.
Aggregation of local effects brings additional challenges that we address in separate ongoing work.

\small 

\bibliographystyle{ecta}
\bibliography{biblio}

@article{almlund2011,
	author = {Mathilde Almlund and Angela Duckworth and Jim Heckman and Tim Kautz},
	journal = {Handbook of the Economics of Education},
	pages = {1--188},
	title = {Personality Psychology and Economics},
	volume = {4},
	year = {2011}}

@misc{Arkhangel2025,
  title         = {Evaluating Local Policies in Centralized Markets},
  author        = {Arkhangelsky, Dmitry and Rutgers, Wisse},
  year          = {2025},
  month         = oct,
  note          = {arXiv:2510.20032 [econ.GN]},
  eprint        = {2510.20032},
  archivePrefix = {arXiv},
  primaryClass  = {econ.GN},
  doi           = {10.48550/arXiv.2510.20032},
  url           = {https://arxiv.org/abs/2510.20032}
}

@article{borghans2008,
	author = {Lex Borghans and Angela Duckworth and Jim Heckman and Bas ter Weel},
	journal = {Journal of Human Resources},
	pages = {972--1059},
	title = {The Economics and Psychology of Personality Traits},
	volume = {43},
	year = {2008}}

@article{heckmankautz2012,
	author = {Jim Heckman and Tim Kautz},
	journal = {Labor Economics},
	pages = {451--464},
	title = {Hard Evidence on Soft Skills},
	volume = {19},
	year = {2012}}

@article{arci2005,
	author = {P. Arcidiacono},
	journal = {Econometrica},
	pages = {1477--1524},
	title = {Affirmative Action in Higher Education: How Do Admission and Financial Aid Rules Affect Future Earnings?},
	volume = {73},
	year = {2005}}

@article{abdul_ecta2022,
  title={Breaking Ties: Regression Discontinuity Design Meets Market Design},
  author={Abdulkadiroglu, Atila and Angrist, Joshua D and Narita, Yusuke and Pathak, Parag},
  journal={Econometrica},
  volume={90},
  number={1},
  pages={117--151},
  year={2022},
  publisher={Wiley Online Library}
}

@article{agarwal2018,
  title={Demand analysis using strategic reports: An application to a school choice mechanism},
  author={Agarwal, Nikhil and Somaini, Paulo},
  journal={Econometrica},
  volume={86},
  number={2},
  pages={391--444},
  year={2018},
  publisher={Wiley Online Library}
}

@article{altonjiarcidiaconomaurel2016,
  title={The Analysis of Field Choice in College and Graduate School: Determinants and Wage Effects},
  author={Altonji, Joseph and Arcidiacono, Peter and Maurel, Arnaud},
  journal={R. Hanushek, S. Machin, and L. Worrseman (eds) The Handbook of the Economics of Education},
  volume={V},
  year={2016}
}

@article{andersson_beyond_2026,
	title = {Beyond {Truth}‐{Telling}: {A} {Replication} {Study} on {School} {Choice}},
	journal = {Journal of Applied Econometrics},
	author = {Andersson, Tommy and Kessel, Dany and Lager, Nils and Olme, Elisabet and Reese, Simon},
	month = jan,
	year = {2026},
	pages = {jae.70038},
}

@article{artemovCheHe2017,
author={Artemov, Georgy and Che, Yeon-Koo and He, Yinghua},
title = {Strategic mistakes: Implications for Market Design Research},
journal = {Working paper},
year={2017}
}

@article{azevedo2016,
  title={A Supply and Demand Framework for Two-sided Matching Markets},
  author={Azevedo, Eduardo M and Leshno, Jacob D},
  journal={Journal of Political Economy},
  volume={124},
  number={5},
  pages={1235--1268},
  year={2016},
  publisher={University of Chicago Press Chicago, IL}
}

@article{beresteanu_joe2012,
  title={Partial Identification Using Random Set Theory},
  author={Beresteanu, Arie and Molchanov, Ilya and Molinari, Francesca},
  journal={Journal of Econometrics},
  volume={166},
  number={1},
  pages={17--32},
  year={2012},
  publisher={Elsevier}
}

@article{bertanha_jmp2014,
  title={Regression Discontinuity Design with Many Thresholds},
  author={Bertanha, Marinho},
  journal={Job Market Paper, Stanford University},
  month={november},
  year={2014},
}

@article{bertanha_joe2020,
  title={Regression Discontinuity Design with Many Thresholds},
  author={Bertanha, Marinho},
  journal={Journal of Econometrics},
  volume={218},
  number={1},
  pages={216--241},
  year={2020},
  publisher={Elsevier}
}

@article{bertanha_impossible,
  title={Impossible Inference in Econometrics: Theory and Applications},
  author={Bertanha, Marinho and Moreira, Marcelo J},
  journal={Journal of Econometrics},
  volume={218},
  number={2},
  pages={247--270},
  year={2020},
  publisher={Elsevier}
}

@article{campos_impact_2024,
	title = {The {Impact} of {Public} {School} {Choice}: {Evidence} from {Los} {Angeles}’s {Zones} of {Choice}},
	volume = {139},
	number = {2},
	journal = {The Quarterly Journal of Economics},
	author = {Campos, Christopher and Kearns, Caitlin},
	month = mar,
	year = {2024},
	pages = {1051--1093},
}

@article{chen2026,
  title={Nonparametric Treatment Effect Identification in School Choice},
  author={Chen, Jiafeng},
  journal={Journal of Econometrics},
  volume={253},
  month={January},
  year={2026},
  page={106172},
}

@article{chesher_ecta2017,
  title={Generalized Instrumental Variable Models},
  author={Chesher, Andrew and Rosen, Adam M},
  journal={Econometrica},
  volume={85},
  number={3},
  pages={959--989},
  year={2017},
  publisher={Wiley Online Library}
}

@article{dubins1981,
  title={Machiavelli and the Gale-Shapley Algorithm},
  author={Dubins, Lester E and Freedman, David A},
  journal={American Mathematical Monthly},
  volume={88},
  number={7},
  pages={485--494},
  year={1981},
  publisher={Taylor \& Francis}
}

@article{ergin2002,
  title={Efficient Resource Allocation on the Basis of Priorities},
  author={Ergin, Haluk I},
  journal={Econometrica},
  volume={70},
  number={6},
  pages={2489--2497},
  year={2002},
  publisher={JSTOR}
}

@article{estrada2017,
title = {Benefits to Elite Schools and the Expected Returns to Education: Evidence from Mexico City},
journal = {European Economic Review},
volume = {95},
pages = {168-194},
year = {2017},
author = {Estrada, Ricardo and Gignoux, J\'er\'emie}
}

@article{fack2019,
  title={Beyond Truth-telling: Preference Estimation with Centralized School Choice and College Admissions},
  author={Fack, Gabrielle and Grenet, Julien and He, Yinghua},
  journal={American Economic Review},
  volume={109},
  number={4},
  pages={1486--1529},
  year={2019}
}

@article{haeringer2009,
  title={Constrained School Choice},
  author={Haeringer, Guillaume and Klijn, Flip},
  journal={Journal of Economic Theory},
  volume={144},
  number={5},
  pages={1921--1947},
  year={2009},
  publisher={Elsevier}
}

@article{haeringer2008,
  title={Constrained School Choice},
  author={Haeringer, Guillaume and Klijn, Flip},
  journal={Working paper version},
  year={2008}
}

@article{hastingsneilsonzimmerman2013,
  title={Are Some Degrees Worth More Than Others? Evidence From College Admissions Cutoffs in Chile},
  author={Hastings, Justine and Neilson, Christopher and Zimmerman, Seth},
  journal={NBER Working paper No. 19241},
  year={2013}
}

@article{han_kaido_2025,
  title={Set-Valued Control Functions},
  author={Han, Sukjin and Kaido, Hiroaki},
  journal={arXiv preprint arXiv:2403.00347},
  year={2025}
}

@article{horowitz1995,
  title={Identification and Robustness with Contaminated and Corrupted Data},
  author={Horowitz, Joel L and Manski, Charles F},
  journal={Econometrica: Journal of the Econometric Society},
  pages={281--302},
  year={1995},
  publisher={JSTOR}
}

@article{kapor2024,
  title={Aftermarket Frictions and the Cost of Off-platform Options in Centralized Assignment Mechanisms},
  author={Kapor, Adam and Karnani, Mohit and Neilson, Christopher},
  journal={Journal of Political Economy},
  volume={132},
  number={7},
  pages={2346--2395},
  year={2024},
  publisher={The University of Chicago Press Chicago, IL}
}

@article{kedagni2020,
  title={Discordant Relaxations of Misspecified Models},
  author={K{\'e}dagni, D{\'e}sir{\'e} and Li, Lixiong and Mourifi{\'e}, Isma{\"e}l},
  journal={arXiv preprint arXiv:2012.11679},
  year={2020}
}

@article{KroftMourifieVayalinkal2024,
  author       = {Kory Kroft and Ismael Mourifié and Atom Vayalinkal},
  title        = {Horowitz--Manski--Lee Bounds with Multilayered Sample Selection},
  journal         = {NBER Working Paper No.\ 32952},
  year         = {2024},
  month        = sep,
  revision     = {February 2025},
  doi          = {10.3386/w32952}
}

@article{heckman2005scientific,
  author       = {Heckman, James J.},
  title        = {The Scientific Model of Causality},
  journal      = {Sociological Methodology},
  volume       = {35},
  year         = {2005},
  pages        = {1--97},
}

@article{heckman_vytlacil_2005,
  title        = {Structural Equations, Treatment Effects, and Econometric Policy Evaluation},
  author       = {Heckman, James J. and Vytlacil, Edward J.},
  journal      = {Econometrica},
  volume       = {73},
  number       = {3},
  pages        = {669--738},
  year         = {2005},
}

@article{mogstad_santos_torgovitsky_2018,
  title = {Using Instrumental Variables for Inference about Policy Relevant Treatment Parameters},
  author = {Mogne Mogstad and Andres Santos and Alexander Torgovitsky},
  journal = {Econometrica},
  volume = {86},
  number = {5},
  pages = {1589--1619},
  year = {2018},
  doi = {10.3982/ECTA15463},
}

@unpublished{munro_causal_2025,
	title = {Causal {Inference} under {Interference} through {Designed} {Markets}},
	url = {http://arxiv.org/abs/2504.07217},
	publisher = {arXiv},
	author = {Munro, Evan},
	month = oct,
	year = {2025},
	note = {arXiv:2504.07217 [econ]},
}

@article{Artemov2023,
  title={Stable Matching with Mistaken Agents},
  author={Georgy Artemov and Yeon-Koo Che and YingHua He},
  journal={Journal of Political Economy Microeconomics},
  volume={1},
  number={2},
  pages={270--320},
  year={2023},
  publisher={University of Chicago Press},
  doi={10.1086/722978},
  url={https://www.journals.uchicago.edu/doi/10.1086/722978}
}

@techreport{kirabo_nber2009,
 title = "Ability-grouping and Academic Inequality: Evidence From Rule-based Student Assignments",
 author = "Jackson, C. Kirabo",
 institution = "National Bureau of Economic Research",
 type = "Working Paper",
 series = "Working Paper Series",
 number = "14911",
 year = "2009",
 month = "April",
 doi = {10.3386/w14911},
}

@article{kirkeboen2016,
  title={Field of Study, Earnings, and Self-selection},
  author={Kirkeboen, Lars J and Leuven, Edwin and Mogstad, Magne},
  journal={Quarterly Journal of Economics},
  volume={131},
  number={3},
  pages={1057--1111},
  year={2016},
  publisher={Oxford University Press}
}

@article{kojima2009incentives,
  title={Incentives and Stability in Large Two-sided Matching Markets},
  author={Kojima, Fuhito and Pathak, Parag A},
  journal={American Economic Review},
  volume={99},
  number={3},
  pages={608--627},
  year={2009},
  publisher={American Economic Association}
}

@article{larroucaurios2020,
author={Larroucau, Tomas and Rios, Ignacio},
title = {Do ``Short-List'' Students Report Truthfully? Strategic Behavior in the Chilean College Admissions Problem},
journal = {Working paper},
year={2020}
}

@article{larroucaurios2021,
author={Larroucau, Tomas and Rios, Ignacio},
title = {Dynamic College Admissions},
journal = {Working paper},
year={2021}
}

@article{manski_ecta2004,
  title={Measuring Expectations},
  author={Manski, Charles},
  journal={Econometrica},
  volume={72},
  number={5},
  pages={1329--1376},
  year={2004},
  publisher={Wiley Online Library}
}

@incollection{molinari2020,
title = {Microeconometrics with Partial Identification},
editor = {Steven N. Durlauf and Lars Peter Hansen and James J. Heckman and Rosa L. Matzkin},
series = {Handbook of Econometrics},
publisher = {Elsevier},
volume = {7},
chapter = {5},
pages = {355-486},
year = {2020},
booktitle = {Handbook of Econometrics, Volume 7A},
author = {Francesca Molinari},
}

@techreport{popurquiola2011,
 title = "Going to a Better School: Effects and Behavioral Responses",
 author = "Pop-Eleches, Cristian and Urquiola, Miguel",
 institution = "National Bureau of Economic Research",
 type = "Working Paper",
 series = "Working Paper Series",
 number = "16886",
 year = "2011",
 month = "March",
 doi = {10.3386/w16886},
}

\clearpage
\appendix

\begin{centering}
\LARGE

\textbf{Supplemental Appendix for Online Publication}

\bigskip

\textit{Causal Effects in Matching Mechanisms with Strategically Reported Preferences}

\normalsize 

\bigskip

By Marinho Bertanha, Margaux Luflade, and Ismael Mourifi\'{e}.

\end{centering}

\setcounter{page}{1}
\renewcommand{\thepage}{\mbox{A-\arabic{page}}} 

\renewcommand{\theequation}{A-\arabic{equation}} \setcounter{equation}{0} 
\renewcommand{\thetable}{A-\arabic{table}} \setcounter{table}{0}
\renewcommand{\thefigure}{A-\arabic{figure}} \setcounter{figure}{0}

\section{Additional Results Not in the Main Text}

\subsection{Parameter of Interest}
\label{sec:app:parinteres}

\indent 

This section complements and extends the discussion of our parameter of interest in Section \ref{sec:model} of the main text. 
The motivation behind our causal parameter is counterfactual analysis. 
Counterfactual analysis is a cornerstone of research in education market design, providing essential insights for policy evaluation and recommendation. In empirical applications, several types of counterfactual policies are commonly considered: changes to the assignment mechanism, modifications of priority scores, or adjustments in program capacities.

As detailed in Section V.B of \cite{Artemov2023}, the literature offers two main approaches to counterfactual analysis. The first takes the submitted rank-ordered lists (ROLs), i.e., $P$, as inputs. This approach assumes that the submitted ROLs under the current policy reflect applicants' true preferences and that individuals will submit the same lists under the counterfactual policy. However, \cite{Artemov2023}  show the existence of \textit{robust equilibria} in which a non-negligible fraction of participants do not submit their true preferences, even in a strategy-proof environment.\footnote{Motivated 
by empirical evidence of non-truthful reporting that often has limited payoff consequences, \cite{Artemov2023} develop a novel solution concept—\textit{robust equilibrium}—which relaxes the Bayesian Nash equilibrium by allowing for deviations from truthful reporting when the associated payoff losses are small.} 
They argue that this finding raises important concerns about empirical methods that treat submitted ROLs as truthful representations of preferences—both under the current policy and in counterfactual scenarios.

The second approach, instead, assumes that individuals will report their true preference profile, $Q$ (potentially different from $P$), under the counterfactual policy. This approach requires the researcher to identify the true preference $Q$ in settings where agents behave strategically and misreport. 
\citet{Artemov2023} advocates for this second approach, arguing that a broad class of robust equilibrium concepts yield asymptotically stable matchings. The allocations produced by these equilibria can be well approximated by the stable matching induced under truthful reporting. 
In other words, this implies that for a wide range of reporting behaviors, the ultimate allocation will be stable, allowing us to consider counterfactual assignments as functions of the true preferences.

This paper examines the identification of parameters essential for counterfactual analyses following the second approach advocated in \citet{Artemov2023}.
Let ~$(Y^c,Q^c, S^c)$ denote the counterfactual distribution of outcomes, preferences, and scores in the population.\footnote{For simplicity, we assume that $S_j=S$ for any $j$ in this section.}
Let ~$\varphi^c(q, s)$ denote the program to which a student with true preference ~$Q^c = q$ and grade ~$S^c = s$ is assigned under the counterfactual assignment mechanism.\footnote{This expression is derived under the assumption of a deterministic assignment mechanism, but it extends straightforwardly to cases involving extrinsic tie-breaking through randomization.} For any function ~$g \in \m{G}$ of potential outcomes, we have:
\begin{eqnarray*}
\mathbb E(g(Y^c)) & = & \mathbb E\left( \mathbb E[g(Y^c)\vert Q^c,S^c] \right) \\
& = & \sum_{j\in\mJ^0}\mathbb E(\mathbb E[\mmi\{\varphi^c(Q^c,S^c)=j\}g(Y^c(j))\vert Q^c,S^c])\\
& = & \sum_{j\in\mJ^0}\mathbb E(\mathbb E[\mmi\{\varphi^c(Q,S)=j\}g(Y(j))\vert Q,S])\\
& = & \sum_{j\in\mJ^0}\mathbb E[w^c(j,Q,S)\mathbb E[g(Y(j))\vert Q,S]],\\
\end{eqnarray*}
where the weights ~$w^c(j, q, s) = \mmi\{\varphi^c(q, s) = j\}$ are determined by the counterfactual assignment mechanism  ~$\varphi^c(q, s)$. 
 The third equality holds under the so-called \emph{policy invariance} assumption, i.e.,
\[
(Y^c(0), \ldots, Y^c(J), Q^c, S^c) \sim (Y(0), \ldots, Y(J), Q, S),
\]
which requires that the joint distribution of certain model primitives—such as potential outcomes, scores, and preferences—remains unchanged in the counterfactual world. See, for example, \cite{heckman_vytlacil_2005}, who discuss both the necessity of this assumption and its limitations for certain classes of counterfactuals. 
Notice that, it may be more reasonable to assume that the true preference $Q$ is invariant to changes in the counterfactual world, i.e., $Q^c \sim Q$, rather than assuming that the submitted preferences $P$ remain unchanged. Indeed, some counterfactual scenarios may alter students’ strategic behavior, leading to $P^c \not\sim P$, which could pose challenges for researchers seeking to analyze such counterfactuals using the first approach described in \cite{Artemov2023}.

Analyzing the welfare effects of a policy change involves comparing the counterfactual outcome distribution ~$Y^c$ to the observed (status quo) outcome distribution ~$Y$, through the quantity $\mathbb{E}[g(Y^c)] - \mathbb{E}[g(Y)].$

As can be seen, the key ingredient needed to conduct this type of counterfactual analysis is knowledge of the average structural functions (ASF) ~$s \mapsto \mathbb{E}[g(Y(j)) \mid Q = q, S = s]$ for $j \in \{0,\ldots,J\}$ and $q \in \mathcal{Q}$.
In practice, nonparametrically identifying each of these functions for all ~$s \in \mathcal{S}$ is extremely challenging.
However, an RD design can point or set identify differences of the ASF's at a finite number of points, namely,  
$\mathbb{E}[g(Y(j))-g(Y(k)) \mid Q = q, S = c_j]$
for certain values of $j$, $k$, and $q$.
In data contexts with a large number of cutoffs, \cite{bertanha_joe2020} proposes a consistent and asymptotically normal estimator for counterfactual weighted averages of these differences.
Outside of these contexts, researchers may rely on parametric or semi-parametric assumptions to extrapolate information from finitely many points to the entire domain.

It is important to note that the continuity assumptions utilized by RD can point or set identify ASF's  for finitely many values of $j$ and $q$  but for \textit{infinitely} many values of $s$ in certain subsets of $\m{S}$.
For example, in the SD example of the main text, $\mathbb{E}[g(Y(0)) \mid Q = q, S = s]$ is identified for any $q$ and $s \leq c_1 $.
Researchers may then combine all the identifying information with assumptions on the ASF's to construct bounds on 
$\mathbb{E}[g(Y^c)] - \mathbb{E}[g(Y)].$
Following the ideas proposed by \cite{mogstad_santos_torgovitsky_2018},
one may assume, for example, a Bernstein polynomial basis representation for each ASF:
\[
\mathbb{E}[g(Y(j)) \mid Q = q, S = s] = \sum_{a=0}^{A} \theta_{ja}^q b_{a}(s),
\]
where ~$\{b_{a}\}_{a=0}^{A}$ are known basis functions of $s$, and ~$\theta_j^q := (\theta_{j0}^q, \ldots, \theta_{j A}^q)$ are unknown parameters to be (set) identified.

Finally, due to data constraints inherent to the local nature of our RD estimation, our parameter of interest in the main text does not condition on the full preference profile ~$Q = q$. 
Instead, we follow the intuition of \cite{kirkeboen2016} and condition on what we define in Section \ref{sec:id_ur} as the local preference ~$Q_j$.  
One value of $Q_j$ maps to multiple values of $Q$ for individuals with $S=c_j$. 
It follows that our ASF parameters in the main text
\begin{align*}
& \mathbb{E}[g(Y(j)) \mid Q_j =(j,k), S = c_j] ~\text{ and } ~
\mathbb{E}[g(Y(k)) \mid Q_j =(j,k), S = c_j]
\end{align*}
for $(j,k) \in \m{P}$,
equal weighted averages of $\mathbb{E}[g(Y(j)) \mid Q, S = c_j]$ 
and
$\mathbb{E}[g(Y(k)) \mid Q, S = c_j]$ 
over values of $Q$ (see proof of Lemma \ref{lemma:contqj} in Section \ref{proof:lemma:contqj} of the appendix).
Therefore, they contain identifying information about $\mathbb{E}[g(Y(j)) \mid Q = q, S = s]$
and
$\mathbb{E}[g(Y(k)) \mid Q = q, S = s]$.

As discussed earlier, parameters of this form are essential for implementing a broad class of  counterfactual analyses. 
Their identification is generally more challenging than that of parameters conditioned on $P$, 
such as
$\mathbb{E}\left[ g(Y(j)) \,\middle|\, P_j = (j,k), \; S_j = c_j \right].$
We do not consider these parameters because they are not directly useful to the counterfactual approach advocated by \cite{Artemov2023} for the reasons given above.

\subsection{Partial Orders Dominate Nonpartial Orders}
\label{sec:app:partial_order}

\indent 

First, we state some definitions. 
The strategy for each student $\omega$ is an ordered list of schools $P(\omega) \subseteq \m{J}$ that has at least one and at most $K<J$ schools in it.
The strategy profile of the economy is a mapping $P$ from $\Omega$ to an ordered subset of $\m{J}$.
We refer to this mapping as $P: \Omega \rightrightarrows \m{J}$ in a slight abuse of notation. 
The score profile of individuals in the economy is denoted by the random vector $\bS:\Omega \rightarrow \m{S}$.
A mechanism $\varphi$ takes the whole correspondence $P$ and function $\bS$ as givens and produces school assignments for each individual $\omega\in \Omega$,
such that $\varphi(P,\bS):\Omega \to \m{J}^0$.
The assignment of student $\omega$ for profiles $(P,\bS)$ is $\varphi(P,\bS)[\omega]$.

For this proof, it is convenient to focus on the assignment of an individual $\omega_0$ as a function of her individual ranking submission $p \subseteq \m{J}$,
the ranking submissions of others $P^{-\omega_0}:\Omega \setminus \{\omega_0\} \rightrightarrows \m{J}$,
and everyone’s scores $\bS:\Omega \rightarrow \m{S}$.
We write this assignment as $\varphi((p,P^{-\omega_0}),\bS)[\omega_0]$.

For individual $\omega_0$ with true preferences $Q(\omega_0)$,
we say 
$p'$ \textbf{weakly dominates} $p$ if \\ 
$\varphi((p', P^{-\omega_0}),\bS)[\omega_0] ~ \bar Q(\omega_0) ~ \varphi((p, P^{-\omega_0}),\bS)[\omega_0] $
for every $P^{-\omega_0}$.
In addition, we say $\varphi$ is strategy proof with unrestricted lists if 
submitting the true list 
of acceptable schools weakly dominates submitting anything else for every individual $\omega$.

\begin{lemma}\label{result:partial_order}
Assume $\varphi$ is strategy proof with unrestricted lists.
Consider student $\omega$ with true preference $Q(\omega)$.
Fix an arbitrary ranking of schools $p \subseteq \m{J}$.
For this student, $p$ is weakly dominated by any weak partial order $p'$ of $Q(\omega)$ that contains all the acceptable schools in $p$. In turn, $p'$ is weakly dominated by any strong partial order $p''$ of $Q(\omega)$
that contains all the acceptable schools in $p'$.

Moreover, suppose the number of acceptable schools in $Q(\omega)$ is less than or equal to $K$.
Then, the dominant strategy is to submit the unique strong partial order of 
$Q(\omega)$ that equals the true list of acceptable schools. 
\end{lemma}

\noindent \textbf{Proof of Lemma \ref{result:partial_order}: }

This proof is from \cite{haeringer2008}, Lemma 4.2.
We expand it here in terms of our framework and definitions of partial order.

From the main text, everyone has at least one acceptable school. 
Take student $\omega$, and consider an arbitrary list of schools $p$ that has at least one acceptable school for that student. 
If $p$ does not have any acceptable schools, then it is clearly dominated by any weak partial order.

First, remove the unacceptable schools from $p$ (if any), 
keep the relative ordering of the acceptable schools, and call the resulting list $\bar{p}$.
It follows that $\bar{p}$ weakly dominates $p$ for student $\omega$.

Second, let $p'$ be a weak partial order of $Q(\omega)$ that contains the schools listed in $\bar{p}$. 
Construct a new ``true'' preference ranking $q \subseteq \m{J}^0$ as follows: 
(a) take the acceptable schools of $p'$, and place them first in $q$ in the same order as they appear in $p'$;
(b) add a $0$ to $q$ after the last school in part (a);
(c) fill the remaining positions below $0$ in $q$ with the schools not listed in $p'$, in any order.
Note that $p'$ equals the true list of acceptable schools from $q$, $p'$ is a weak partial order of $q$, and $p'$ is the unique strong partial order of $q$.

Third, suppose for a moment that the true preference of individual $\omega$ were $q$ instead of $Q(\omega)$.
In that case, strategyproofness of $\varphi$ implies that 
\[
\varphi((p', P^{-\omega}),\bS)[\omega] ~ \bar q ~ \varphi((\bar{p}, P^{-\omega}),\bS)[\omega] \text{ for every } P^{-\omega}.
\]
Given that $p'$ is a weak partial order of both $Q(\omega)$ and $q$,
for any two options $d,d'$ in $p' \cup \{0\}$, we have $ d' ~\bar q ~ d$ implies $d' \bar Q(\omega) d $.
Therefore, 
\[
\varphi((p', P^{-\omega}),\bS)[\omega] ~ \bar Q(\omega) ~ \varphi((\bar{p}, P^{-\omega}),\bS)[\omega] \text{ for every } P^{-\omega}.
\]
It then follows that $p'$ weakly dominates $\bar{p}$, which weakly dominates $p$,
so
$p'$ weakly dominates $p$.

It follows that any strong partial order $p''$ of $Q(\omega)$ that has all the acceptable schools from $p'$ weakly dominates $p'$.
To see this, repeat the argument above by replacing $p$ with $p'$ and replacing $p'$ with $p''$.

The second claim of the lemma follows from the strategyproofness of $\varphi$ since the cap constraint is not binding for an individual whose number of acceptable schools is less than or equal to $K$ and submission of the true list
of acceptable schools is feasible. 
\\*
$\square$

\subsection{Sharp Bounds on Treatment Effects}
\label{sec:app:sharp_bounds}

\indent 

In this section, we utilize Artstein's inequality (Theorem A.1 from \cite{molinari2020}) to characterize sharp
bounds on the joint distribution of potential outcomes and true local preferences.
That set of distributions produces sharp bounds on the averages of the treatment effects
$Y(j)-Y(k)$ conditional on $Q_j=(j,k)$ and $S_j=c_j$ for any comparable pair $(j,k) \in \m{P}$.
Recall that we denote the space of possible outcomes $Y$ as $\m{Y}$.
Let $2^{\m{J}^0 \times \m{J}^0}$ denote the power set of ${\m{J}^0 \times \m{J}^0}$
and
$2^{\m{Y}}$ denote the power set of ${\m{Y}}$.

\begin{theorem}\label{result:sharp_outcome_Qj}
Suppose Assumptions \ref{aspt:continuity}, \ref{aspt:sharp_bQj}, and \ref{aspt:dist_bQj} hold.
Assume $\m{Y}$ is compact.
Consider a pair $(j,k) \in \m{P}$.
Then, the inequalities below characterize the sharp set of all probability values
of
$\mmp \left[ Y(d) \in A, Q_j=(b,b') | S_j=c_j \right]$ for $A \subseteq \m{Y}$, 
$(b,b') \in \m{J}^0 \times \m{J}^0$,
and 
$d \in \{b, b'\}$:
\begin{align*}
& \mmp\left[ Y \in A, \bQ_j \subseteq B | S_j=c_j^+ \right] 
\leq
\sum_{(b,b') \in B} \mmp \left[ Y(b) \in A, Q_j=(b,b') | S_j=c_j \right]
\;\; \;\forall A \in 2^{\m{Y}}, \;\ B \in \bLambda^{\cup+}_{j},
\\
\\
& \mmp\left[ Y \in A, \bQ_j \subseteq B | S_j=c_j^- \right] 
\leq
\sum_{(b,b') \in B} \mmp \left[ Y(b') \in A, Q_j=(b,b') | S_j=c_j \right]
\;\; \; \forall A \in 2^{\m{Y}}, \;\ B \in \bLambda^{\cup-}_{j}.
\end{align*}
\end{theorem}

\noindent \textbf{Proof of Theorem \ref{result:sharp_outcome_Qj}: }
The random set $(\{Y\} \times \bQ_j)$ 
is a measurable map from $\Omega$ to $\m{Y} \times \m{J}^0 \times \m{J}^0$, so it is compact valued. 
Following Definition A.1 from \cite{molinari2020}, we say that $(\{Y\} \times \bQ_j)$ is a random closed set because, for every compact set $\mathcal K \in \mmr^3$, 
the set
$\{\omega \in \Omega: (\{ Y(\omega) \} \times \bQ_j (\omega)) \cap \mathcal K\neq \emptyset\}$ is a measurable event. 
By Assumption \ref{aspt:sharp_bQj}(i), 
the random vector $(Y,Q_{j})$ and the random set $(\{Y\} \times \bQ_j)$ are measurable maps on the same probability space,
and $\mmp\left[ (Y,Q_{j}) \in (\{Y\} \times \bQ_j) |S_j =s \right]=1$ for any $s \in \m{S}_j$.

Artstein's inequality (Theorem A.1 from \cite{molinari2020}) characterizes the sharp set of all possible probability distributions
for $(Y,Q_{j})$ that are consistent with our observation of $(\{Y\} \times \bQ_j)$ and 
the fact that
$\mmp [ (Y,Q_{j}) \in (\{Y\} \times \bQ_j) |S_j =s ]=1$.
For any $s \in \m{S}_j$, the inequality says that
\begin{eqnarray}\label{eq:Artstein}
\mmp\left[ Y \in A, \bQ_j \subseteq B | S_j=s \right] 
\leq \mmp \left[ Y \in A, Q_j \in B|S_j=s \right] 
\;\; \;\forall A \in 2^{\m{Y}}, \;\ B \in 2^{\m{J}^0 \times \m{J}^0}.
\end{eqnarray}

Next, we use the cutoff characterization and rewrite the right-hand side of \eqref{eq:Artstein} as a sum.
For $s \geq c_j$, 
\begin{align}
& \mmp \left[ Y \in A, Q_j \in B | S_j=s \right]
= \sum_{(b,b') \in B} \mmp \left[ Y \in A, Q_j=(b,b') | S_j=s \right]
\notag
\\
& = \sum_{(b,b') \in B} \mmp \left[ Y(b) \in A, Q_j=(b,b') | S_j=s \right].
\label{eq:Artstein_rhs}
\end{align}
Substitute \eqref{eq:Artstein_rhs} into \eqref{eq:Artstein} and take the limit as $s \downarrow c_j$ on both sides,
\begin{align}
& \mmp\left[ Y \in A, \bQ_j \subseteq B | S_j=c_j^+ \right] 
\leq
\sum_{(b,b') \in B} \mmp \left[ Y(b) \in A, Q_j=(b,b') | S_j=c_j \right]
\;\; \;\forall A \in 2^{\m{Y}}, \;\ B \in 2^{\m{J}^0 \times \m{J}^0},
\label{eq:Artstein_rhs2}
\end{align}
where the limits of the left- and right-hand sides of the inequality are well defined by Assumptions \ref{aspt:dist_bQj} and 
\ref{aspt:continuity}, respectively.

Similarly, for $s < c_j$,
\begin{align}
& \mmp \left[ Y \in A, Q_j \in B | S_j=s \right]
= \sum_{(b,b') \in B} \mmp \left[ Y \in A, Q_j=(b,b') | S_j=s \right]
\notag
\\
& = \sum_{(b,b') \in B} \mmp \left[ Y(b') \in A, Q_j=(b,b') | S_j=s \right].
\label{eq:Artstein_lhs}
\end{align}
Use \eqref{eq:Artstein_lhs} into \eqref{eq:Artstein}, and take the limit as $s \uparrow c_j$ on both sides,
\begin{align}
& \mmp\left[ Y \in A, \bQ_j \subseteq B | S_j=c_j^- \right] 
\leq
\sum_{(b,b') \in B} \mmp \left[ Y(b') \in A, Q_j=(b,b') | S_j=c_j \right]
\;\; \; \forall A \in 2^{\m{Y}}, \;\ B \in 2^{\m{J}^0 \times \m{J}^0}.
\label{eq:Artstein_lhs2}
\end{align}

Lemma 1 of \cite{chesher_ecta2017} applied to our case shows that the inequalities \eqref{eq:Artstein_rhs2} and \eqref{eq:Artstein_lhs2} are equivalent to:
\begin{align*}
& \mmp\left[ Y \in A, \bQ_j \subseteq B | S_j=c_j^+ \right] 
\leq
\sum_{(b,b') \in B} \mmp \left[ Y(b) \in A, Q_j=(b,b') | S_j=c_j \right]
\;\; \;\forall A \in 2^{\m{Y}}, \;\ B \in \bLambda^{\cup+}_{j},
\\
\\
& \mmp\left[ Y \in A, \bQ_j \subseteq B | S_j=c_j^- \right] 
\leq
\sum_{(b,b') \in B} \mmp \left[ Y(b') \in A, Q_j=(b,b') | S_j=c_j \right]
\;\; \; \forall A \in 2^{\m{Y}}, \;\ B \in \bLambda^{\cup-}_{j}.
\end{align*}
\\*
$\square$

\bigskip

Theorem \ref{result:sharp_outcome_Qj} characterizes the sharp set of all possible probability
values of \\ $\mmp \left[ Y(d) \in A, Q_j=(b,b') | S_j=c_j \right]$ for $A \subseteq \m{Y}$, 
$(b,b') \in \m{J}^0 \times \m{J}^0$,
and 
$d \in \{b, b'\}$.
For a fixed $g\in\m{G}$ of Assumption \ref{aspt:continuity},
that set of distributions allows us to define the sharp set of all possible means of potential outcomes,
\[
\mme \left[ g(Y(d)) | Q_j=(j,k), S_j=c_j \right],
\]
for $(j,k) \in \m{P}$ such that $\lbar{p}_{j,k}>0$ and 
$d \in \{j, k\}$.
The set of possible means in turn allows us to define the sharp set of all average treatment effects of the form
\[
\mme \left[ g(Y(j)) - g(Y(k)) | Q_j=(j,k) , S_j=c_j \right].
\]

In case of continuous $Y$, it is impossible to directly evaluate all inequalities of Theorem \ref{result:sharp_outcome_Qj}
because there are uncountably many sets $A \in 2^{\m{Y}}$.
This is one of the drawbacks of the Artstein inequality approach, which has been extensively discussed by 
\cite{beresteanu_joe2012}.
In case $Y$ takes finitely many values, e.g., when $Y$ is binary, the number of such inequalities is feasible to evaluate
because $2^{\m{Y}} \times \bLambda^{\cup+}_{j}$ (or $2^{\m{Y}} \times \bLambda^{\cup-}_{j}$) has finitely many elements.
In fact, the number of inequalities is slightly higher than the number of inequalities in Proposition \ref{result:partial_id:nextbest-dist} of the main text, 
which we utilize to compute lower bounds on $\delta_{j,k}^+$ and $\delta_{j,k}^-$.

\subsection{Falsification Test}
\label{sec:app:falsification}

\indent

It is useful to characterize a falsification test based on the assumptions that we employ to construct $\bQ_j$.
Such a test can rely on the fact that the right-hand sides of the inequalities in Proposition \ref{result:partial_id:nextbest-dist}
must provide a lower bound for a probability mass function if the model assumptions are correct.
For any partition $\m{A}$ of $\m{J}^0 \times \m{J}^0$, we must have 
$\sum_{ A \in \m{A}} \mmp\left[Q_j \in A | S_j=c_j \right]=1$
for any given distribution in the sharp set of Proposition \ref{result:partial_id:nextbest-dist}.
Thus, the same sum applied to the right-hand sides of the inequalities above must be less than or equal to one.

\begin{corollary}[Model's Falsification Test]\label{result:fals-test}
Assume the setup of Proposition \ref{result:partial_id:nextbest-dist},
which presupposes 
that
the
model assumptions utilized to construct $\bQ_j$ are true.
Then, for any partition $\m{A}$ of $\m{J}^0 \times \m{J}^0$, we have that
\vspace{-.25cm}
\begin{align*}
\sum_{ A \in \m{A}} \Bigg\{ \;\;\;  &  \mmi\left\{A \in \bLambda^{\cup+}_{j} \cap \bLambda^{\cup-}_{j} \right\}
						\max \left\{~
							\mmp\left[ \bQ_j \subseteq A | S_j= c_j^+ \right]
							~;~
							\mmp\left[ \bQ_j \subseteq A | S_j= c_j^- \right]
						~\right\}		
\\
 					& +\mmi\left\{A \in  \bLambda^{\cup+}_{j} \setminus \bLambda^{\cup-}_{j} \right\}
						\mmp\left[ \bQ_j \subseteq A | S_j= c_j^+ \right]
\\
 					& +\mmi\left\{A \in  \bLambda^{\cup-}_{j} \setminus \bLambda^{\cup+}_{j} \right\}
						\mmp\left[ \bQ_j \subseteq A | S_j= c_j^- \right]
					\;\;\; \Bigg\} \leq 1.
\end{align*}
\end{corollary}

\subsection{Agents that Maximize Expected Utility}
\label{sec:app:uncertainty}

\indent 

This section micro-founds the strategic reporting behavior of agents by describing in detail the problem they have to solve. 
We build on \cite{agarwal2018} and \cite{fack2019}, who propose expected-utility maximizing agents that calculate probabilities of admission for each possible choice of rank-ordered list (ROL or $P$).
In their settings, each agent knows her type, the mechanism, the distribution of other agents' types, but does not know the individual types of other agents and the $P$'s they decide to submit. 
Thus, the uncertainty about an agent's match  comes from this lack of information about other agents.

Uncertainty about other people's types and actions ultimately translates to uncertainty about 
the admission cutoffs \textit{ex-post} the match.
From an individual agent's perspective, uncertainty about cutoffs is all that matters for the uncertainty of her final match. 
Therefore, we take a more ``reduced-form'' approach to the source of uncertainty in the agent's problem and simply assume each agent has beliefs on \textit{ex-post} cutoffs.
This assumption not only simplifies the problem of the agent but is also more in line with the real-world decision process of students in Chile, who form expectations about cutoffs based on historical data and submit schools accordingly.

We now describe the problem of the agents (see Section \ref{sec:app:control:P} for a numerical illustration). 
Before the matching mechanism is run, 
the agent knows her placement scores $\boup{s}=(s_1,\ldots, s_J)$ and her true preferences $Q$ but does not know what the admission cutoffs will be after the matching is run.
The agent sees the admission cutoffs as random variables $(C_1,\ldots,C_J)$.
The strict preference relation $Q$ is represented by a vector of distinct utility values
$(U_0 ,U_1 ,...,U_J)$
so that $a Q b ~ \Leftrightarrow ~ U_a > U_b$ for any $a,b \in \m{J}^0$. 
We normalize $U_0=0$ for simplicity.

The agent has to decide on a ranking of acceptable schools $P$ to submit.
The number of schools ranked in $P$ is $|P|$, and 
a feasible ranking has $1\leq |P| \leq K$.
The set of all feasible rankings is defined as $\Delta P$.
We let $P^u$ denote the $u$-th school listed in $P$, for $u=1, \ldots, |P|$.
We define $L^{P}_u$ as the agent's expected probability of being assigned to school $u$ when submitting ranking $P$, 
$u=1,\ldots, |P|$.
$L^{P}_0$ denotes the expected probability of remaining unassigned, that is, of being matched to the outside option. 
Naturally, $L^{P}_u \geq 0$ for every $u$, and $\sum_{u=0}^{|P|} L^{P}_u = 1$. 
Cutoff characterization wrt $P$ (Assumption \ref{aspt:cutoff2}) implies:
\begin{align*}
& L^{P}_0 = \mmp\left[ \cap_{v=1}^{|P|} \left\{ s_{P^v} < C_{P^v} \right\} \right], 
\\
& L^{P}_u = \mmp\left[ 
\cap_{v=1}^{u-1} \{ s_{P^v} < C_{P^v} \} ~ 
\cap ~ \{ s_{P^u} \geq C_{P^u} \}
\right], 
~~~ u=1, \ldots, |P|,
\end{align*}
where we adopt the convention that $\cap_{v=1}^{u-1} \{ s_{P^v} < C_{P^v} \} ~ \cap A = A$ for any measurable set $A$ if $u=1$;
in other words, if the intersection $\cap_{v \in V}$ is to be computed over an empty set of indices, $V=\emptyset$, then
$ \mmp\left[ 
\cap_{v\in V} \{ s_{P^v} < C_{P^v} \} ~ 
\cap ~ A
\right]
=
\mmp\left[ A \right],
$
for any measurable set $A$.

The agent's optimal ranking to be submitted is the solution to the following problem,
\[ 
\max\limits_{P \in \Delta P} \sum^{ |P| }_{u=1}
U_{P^u} L^{P}_{u}. 
\]

\subsection{Issues When Controlling for $P$}
\label{sec:app:control:P}

\indent 

This section complements the discussion after Part III of the SD Example (Section \ref{sec:id_r:qj_id}) in the main text, where we point out the issue of controlling for $P_4=(4,2)$ in RD identification.
We present an example of a continuum economy with agents that face cutoff uncertainty, maximize expected utility, and exhibit the type of discontinuous behavior that invalidates identification. 

We build on the continuum economy of Section \ref{sec:model} and the agent optimization problem delineated in Section \ref{sec:app:uncertainty}.
Part III of the SD Example (Section \ref{sec:id_r:qj_id}) considers an economy with $J=4$ schools and quota $K=3$ on submissions.
Assume $S_1 \sim U[0,1]$ and school capacities are 
$q_1=0.24$,
$q_2=0.22$,
$q_3=0.18$,
and
$q_4=0.16$.
The outside option has unlimited capacity, i.e., $q_0=1$.
The distribution of true preferences $Q$ is independent from $S_1$ and uniform over all permutations of $\{1,2,3,4\}$ with $0$ in the last position  (i.e., all schools are acceptable).

We first assume agents submit their true preferences and compute the four cutoffs by simulation.
We draw $n=100,000$ agents from the population distribution of $(Q,S_1)$ and run the SD algorithm.
We obtain cutoffs $c_1=0.1988$, $c_2=0.2203$, $c_3=0.2958$, and $c_4=0.3623$.
We call these ``truth-telling cutoffs.'' 

True preferences can no longer be submitted in the presence of the quota $K=3$.
Instead of listing all four acceptable schools, agents must think strategically and decide to drop at least one school from their lists. 
Agents have expectations about \textit{ex-post} cutoffs and submit a partial order $P$ that maximizes their expected utilities (Section \ref{sec:app:uncertainty}).
They submit a strong partial order of their true preferences, which implies that $|P|=3$. 
Expectations about \textit{ex-post} cutoffs are modeled as random variables $C_j$, $j=1,\ldots, 4$.
We specify these distributions to be independent across $j$, each with a support containing the corresponding truth-telling cutoff $c_j$.
All supports include a range of values so that no agent is certain about the \textit{ex-post} cutoff. 
More specifically, there are two types of expectations and cardinal utility functions in this economy: those of agents with $Q = (4,3,2,1,0)$ and those of agents with 
$Q \neq (4,3,2,1,0)$.

Agents with $Q = (4,3,2,1,0)$ have expectations that are mixed continuous-discrete distributions. 
The discrete part of the distribution has one mass point at $c_j$ and the continuous part of the distribution is uniform over a closed interval.
In particular, the mass points are $\mmp[C_1=c_1]=0.6$, $\mmp[C_2=c_2]=0.4$,  $\mmp[C_4=c_4]=0.2$, while $C_3$ has no mass point;
the continuous parts are uniform distributions $U[0;0.9]$ for $C_1$ and $C_2$, $U[0;1]$ for $C_3$, and $U[c_4;0.85]$ for $C_4$.
Their cardinal utilities are $4>3>2.99>2.98>0$, where $4$ is the cardinal utility of the first-ranked option in $Q$ and $0$ is the cardinal utility of the last-ranked option in $Q$, that is, the outside option. 
All remaining agents, that is, those with $Q \neq (4,3,2,1,0)$, have expectations that are continuous distributions 
$C_j \sim U[c_j-0.15,c_j+0.15]$, for $j=1,2,3,4$.
Their cardinal utilities are $4>3>2>1>0$, respectively, for the first-, second-, third-, fourth-, and fifth-ranked options in whatever $Q$ the agent has.

We compute optimal $P$s and \textit{ex-post} cutoffs by simulation in the economy with strategic agents. 
First, we draw $n=100,000$ agents from the population distribution of $(Q,S_1)$.
Second, we compute the optimal preference submission $P$ of each agent drawn.
Each individual chooses a strong partial order $P$ of her true preference ranking  that maximizes her expected utility.
The expected utility is the sum of the cardinal utilities weighted by assignment probabilities to each option listed in $P$.
For an agent $i$ with score $s_i$ and cutoff expectations $C^i_j$, $j=1,2,3,4$, the probability of admission to the first-ranked option in a submission $P$ is
$\mmp[s_i \geq C_{P^1} ]$; 
the probability of admission into the second-ranked option is $ \mmp[s_i < C_{P^1} ]  \mmp[s_i \geq C_{P^2} ]$, etc, where $P^u$ denotes the $u$-th ranked option listed in $P$.
Third and finally, we run the SD algorithm using $n$ observations of $(P,Q,S_1)$.
We obtain the same stable matching and cutoffs as in the truth-telling economy above.

Overall, utilities and expectations change with $Q$ but do not vary with $S_1$ once we condition on $Q$.
By design, expectations and utilities of agents with $Q = (4,3,2,1,0)$ are different from the rest in order to have them  
behave in the discontinuous manner indicated in Part III of the SD Example.
They optimally choose $P=(3,2,1)$ if $S_1<c_4$ but $P=(4,2,1)$ when $S_1 \geq c_4$.
Thus, if we control for $P_4=(4,2)$, the fraction of individuals with $Q_4=(4,3)$ jumps from zero to a positive number at the cutoff $c_4$, which invalidates the RD identification strategy.
In fact, we find in the simulated economy
\begin{align*}
&\mmp[Q_4=(4,3)|S_1 \in [c_4, c_4+0.01], P_4=(4,2)] = 0.3478
\\
&\mmp[Q_4=(4,3)|S_1 \in [c_4-0.01, c_4), P_4=(4,2)] = 0.
\end{align*}

There are two key ingredients that cause the discontinuous behavior of agents with $Q = (4,3,2,1,0)$.
First, the mass points in their expectations indicate they possess a good amount of certainty about the \textit{ex-post} values of the cutoffs of schools $1$, $2$, and $4$.
Second, their utility indicates that they are close to being indifferent between schools $1$, $2$, and $3$.
As the score increases and crosses the value $c_4$, the expected probability of having school $4$ in the budget set jumps from zero to 20\%.
The agent then decides to list school $4$ and rank it first, as it has the highest utility. 
There are three acceptable schools left for this agent, $3$, $2$, and $1$, but only two spots left in her submission list given the quota   $K=3$.
The agent decides to skip school $3$ and list schools $2$ and $1$.
The reason is twofold.
First, the utility of $3$ is just a little higher than the utility of $2$ or $1$.
Second, the probability of being assigned to either $2$ or $1$ when the agent submits $P=(4,2,1)$ is sufficiently higher than the probability of being assigned to either $3$ or $2$.
This is the case because an agent with score $S_1=c_4$ is more certain to be above the cutoffs of schools $1$ or $2$ than to clear the cutoff of school $3$. 
Thus, the loss in expected utility of skipping school $3$ is more than compensated for by the gain of listing schools $2$ and $1$. 
Figure \ref{fig:app:discont} below illustrates these two points graphically.

\begin{figure}[h] 
\caption{Expected Utilities and Probabilities of Assignment} \label{fig:app:discont}   
\begin{center} 

    \begin{subfigure}[b]{0.45\textwidth} 
        \centering
        \includegraphics[width=\textwidth]{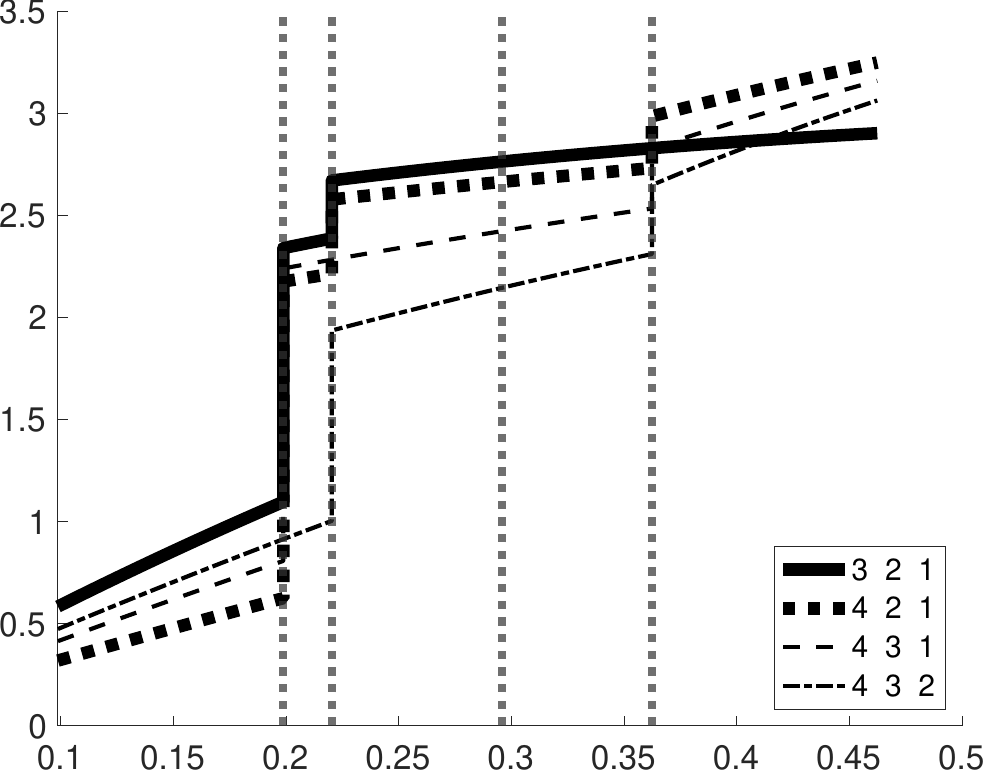} 
        \caption{}
        \label{fig:app:discont:exputil} 
    \end{subfigure}
    \hfill 
    \begin{subfigure}[b]{0.45\textwidth} 
        \centering
        \includegraphics[width=\textwidth]{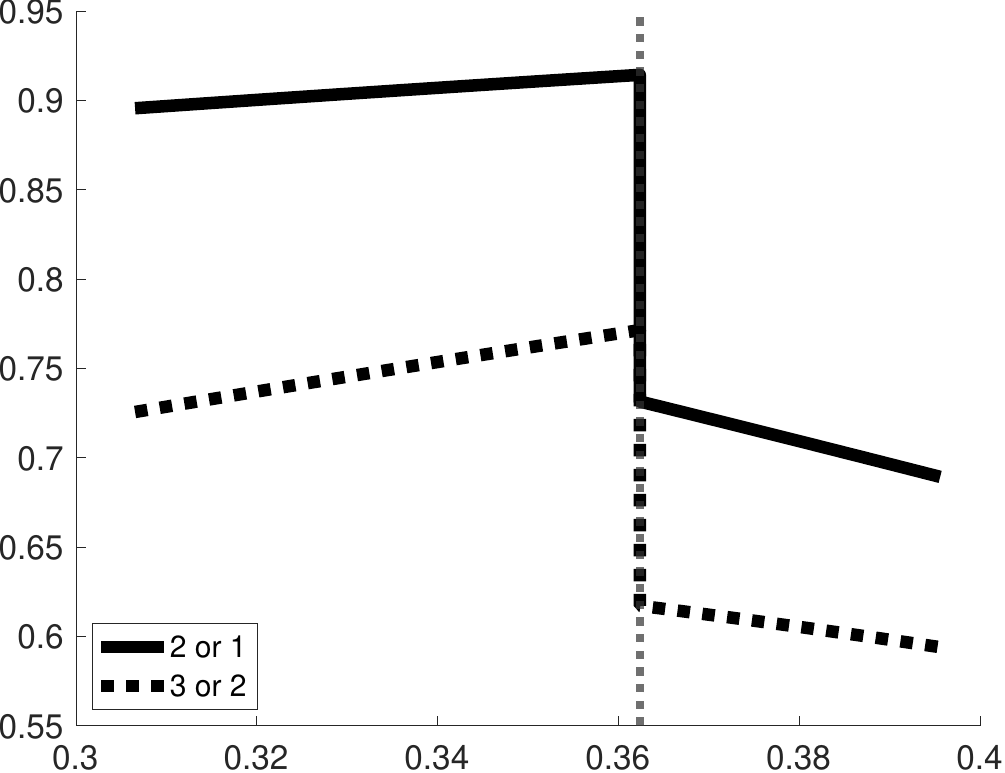} 
        \caption{}
        \label{fig:app:discont:probadm} 
    \end{subfigure}
    
\end{center}

    \footnotesize{Panel (a) depicts the expected utilities (y axis) from submitting four different strong partial orders as functions of the score $S_1$ (x axis) for an agent with $Q=(4,3,2,1,0)$.
    The dashed vertical lines represent the four cutoffs.
    The optimal submission is $P=(4,2,1)$ if $S_1 \geq c_4$ and $P=(3,2,1)$ if $S_1 < c_4$.
    For that same agent, panel (b) compares the probability of assignment (y axis) into schools $2$ or $1$  if  the agent submits $P=(4,2,1)$ 
    with the probability of assignment into schools $3$ or $2$ if the agent submits $P=(4,3,2)$. 
    The probabilities are plotted as functions of the score $S_1$ (x axis) and the dashed vertical line represents cutoff $c_4$.
    These probabilities jump down at $c_4$ because the probability of being assigned to the first-best option $4$ jumps from zero to 20\% at $c_4$ regardless if the agent submits $P=(4,2,1)$ or $P=(4,3,2)$. 
    Panel (b) indicates that the agent expects a higher chance of being assigned to a school by submitting $P=(4,2,1)$ instead of $P=(4,3,2)$, thus the higher expected utility of $P=(4,2,1)$.
    }
    
\end{figure}

\indent

\renewcommand{\theequation}{B-\arabic{equation}} \setcounter{equation}{0} 
\renewcommand{\thetable}{B-\arabic{table}} \setcounter{table}{0}
\renewcommand{\thefigure}{B-\arabic{figure}} \setcounter{figure}{0}

\section{Proofs of Results in the Main Text}

\subsection{Proof of Lemma \ref{lemma:contqj}}
\label{proof:lemma:contqj}

\indent 

Consider a school $j$ with cutoff $c_j$ in the interior of the support $\m{S}_j$.
Once you fix an event $A \in \boup{A}_{-j}$, you fix the availability of those schools with non-$j$ scores $\bS_{-j}$.
The right- and left-counterfactual budget sets $B^-_j(\bS)$ and 
$B^+_j(\bS)$
become
fixed (i.e., nonrandom), regardless of the value of $S_j$.
Once we fix $S_j$ on top of that, the actual budget set $B(\bS )$ is also fixed.

Fix an event $A \in \boup{A}_{-j}$ and $S_j$ in a small neighborhood of $c_j$.
Pick schools $k,l \in \m{J}^0$ such that $\mmp[Q_j=(k,l)|S_j=c_j]>0$.
Define $\mmq(A)$ to be the set of all preference relations $Q \in \m{Q}$ 
such that $Q(B^+_j(\bS))=k$
and
$Q(B^-_j(\bS))=l$.
Note that $\mmq(A)$ is a fixed set of preferences (i.e., nonrandom).

Enumerate the mutually exclusive events in $\boup{A}_{-j}$ as
$A_1, \ldots, A_M$, where $M = 2^{J-1}$.
We have that $Q_j=(k,l)$ is equivalent to 
$Q \in \mmq(A_1)$ if $A_1$, $\ldots$, $Q \in \mmq( A_{M})$ if $A_{M}$. 
Therefore, for $s$ in a small neighborhood of $c_j$,
\begin{align*}
&\mme[ g (Y(d))  \mathbb{I}\{ Q_j=(k,l) \}  |   S_j = s ]
\\
&=\sum_{l=1}^M \mme[ g (Y(d))  \mathbb{I}\{   Q_j=(k,l), \boup{S}_{-j} \in A_l  \}  |   S_j = s]
\\
&=\sum_{l=1}^M \mme[ g (Y(d))  \mathbb{I}\{   Q \in \mmq(A_l),  \boup{S}_{-j} \in A_l \}  |   S_j = s  ]
\\
&=\sum_{l=1}^M \sum_{Q_0 \in \mmq(A_l)} \mme[ g (Y(d))  \mathbb{I}\{   Q = Q_0,  \boup{S}_{-j} \in A_l \}  |  S_j = s  ]. 
\end{align*}

It follows that $\mme[ g (Y(d))  \mathbb{I}\{   Q_j=(k,l) \}  |   S_j = s ]$ is a continuous function of  $s$ at $s=c_j$
because 
$\mme[ g (Y(d))  \mathbb{I}\{   Q = Q_0,  \boup{S}_{-j} \in A_l \}  |  S_j = s  ]$
 is continuous at $s=c_j$ for every $l$ and $Q_0$ by assumption.

Likewise,
\begin{align*}
&\mmp[Q_j=(k,l) | S_j=s]  
\\
&=\sum_{l=1}^M \mmp[     Q_j=(k,l), \boup{S}_{-j} \in A_l    |   S_j = s]
\\
&=\sum_{l=1}^M \mmp[    Q \in \mmq(A_l),  \boup{S}_{-j} \in A_l  |   S_j = s  ]
\\
&=\sum_{l=1}^M \sum_{Q_0 \in \mmq(A_l)} \mmp[     Q = Q_0,  \boup{S}_{-j} \in A_l  |  S_j = s  ], 
\end{align*}
which is continuous at $s=c_j$ for every $l$ and $Q_0$ by assumption. 

Therefore,
\begin{align*}
& \mme[ g (Y(d))    | Q_j=(k,l),  S_j = s ]
\\
& = \frac{\mme[ g (Y(d))  \mathbb{I}\{   Q_j=(k,l) \}  |   S_j = s ]}{ \mmp[  Q_j=(k,l) | S_j=s ]}
\end{align*}
is continuous at $s=c_j$
\\*
$\square$

\subsection{Proof of Proposition \ref{result:truth:identif}}

Take $(j,k) \in \mathcal P$.
Start with school $j$, $s \geq c_j$,
\begin{align*}
\lim_{s \downarrow c_j} \mathbb E [ g(Y )  \vert P_j=(j,k),S_j=s ] 
= 
& \lim_{s \downarrow c_j}  \mathbb E [ g( Y )  \vert Q_j=(j,k),S_j=s]
\\
= 
& \lim_{s \downarrow c_j}  \mathbb E [ g( Y(j) )  \vert Q_j=(j,k),S_j=s]
\\
 & = \mathbb E [ g( Y(j) )  \vert Q_j=(j,k),S_j=c_j].
\end{align*}

For school $k$, $s<c_j$,
 \begin{align*}
 \lim_{s \uparrow c_j} \mathbb E [ g(Y )  \vert P_j=(j,k),S_j=s ] 
 = & \lim_{s \uparrow c_j} \mathbb E [ g( Y )  \vert Q_j=(j,k),S_j=s]
 \\
 = & \lim_{s \uparrow c_j} \mathbb E [ g( Y(k) )  \vert Q_j=(j,k),S_j=s]
 \\ 
= & \mathbb E [ g( Y(k) )  \vert Q_j=(j,k),S_j=c_j].
\hspace{1cm} \square
\end{align*}

\subsection{Proof of Proposition \ref{result:partial_id:nextbest}}
\indent

The true preference list $Q$ is unobserved.
The submitted preference list $P$ is observed, and $P$ is a weak partial order of $Q$.
Note that all the schools listed in $P$ appear in $Q$ ranked before $0$ (i.e., as acceptable schools).
There may be other elements in $Q$ not listed in $P$.
These remaining schools might appear anywhere in $Q$ as long as the relative ordering of schools in $P$ is preserved in $Q$.
Our focus is on students with $P_j=(a,b)$, so we know that $a \bar{Q}b$, where $a Q b$ if $a \neq b$.

We consider all possibilities of $Q$ that are consistent with the observed $P$ and assumptions and that affect $Q_{j}$.
In such cases, the acceptable schools in $Q$ include all the schools in $P$ and possibly more;
in fact, since $|P| \leq K < J$, the $J-|P|>0$ unlisted schools in $P$ may appear as acceptable in $Q$. 
The additional acceptable schools in $Q$ may be schools that are feasible or infeasible within the budget set of the individual.

The proposition defines two sets of unlisted feasible schools for an individual with scores $\bS$ and submitted preferences $P$:
$N_j^+ = B_j^+( \bS) \setminus \left\{ P \cup \{0 \} \right\}$
and
$N_j^- = B_j^-( \bS ) \setminus \left\{ P \cup \{0 \} \right\}$.
The only difference between $B^+_j$ and $B^-_j$ is the set of schools whose priority scores equal $S_j$ and cutoffs equal $c_j$.

The rest of the proof builds on the following reasoning. 
Ignore Assumption \ref{aspt:cutoff2} for a moment. 
The first coordinate of $Q_j$ depends on $P_j$ and the set $N_j^+$.
In fact, the first coordinate of $P_j$ is the best option in $B_j^+( \bS )$ according to $P$: that is option $a$.
The best option in $B_j^+( \bS )$ according to $Q$ may also be $a$ as long as there are no unlisted options in $P$ that are available in $B_j^+( \bS )$ and rank higher than $a$ in $Q$.
A similar argument applies to the second coordinate of $Q$: it is a function of $P_j$ and the set $N_j^-$.
Now, Assumption \ref{aspt:cutoff2} further restricts $Q_j$ because it implies that
$P( B( \bS) ) = Q( B( \bS) )$ and $P( B( \bS) )$ is observed.

Regarding the outside option $0$, the only way that $Q_j$ will have a zero is if $P_j$ has a zero.
In fact, if $a \neq 0$ and $b\neq 0$, we have that $a \bar{P} b P 0$, which implies $a \bar{Q} b Q 0$, so none of the coordinates of $Q_j$ will be zero.
This is why the sets of unlisted feasible options, $N_j^+$ and $N_j^-$, do not contain zero.

\noindent \textbf{Case 1:} $S_j \geq c_j$.

By Assumption \ref{aspt:cutoff2}, we have that the mechanism assignment $\mu$ equals the best option according to $P$ in the set 
$B(\bS) = B^+_j( \bS)$.
An individual with $P_{j}=(a,b)$ has $a= P(B^+_j( \bS) ) = P( B( \bS) )$; therefore, $\mu=a$.
The cutoff characterization dictates that $a=Q(B( \bS)) = Q(B^+_j( \bS) ) $, so the first coordinate of $Q_j$ equals $a$.
It remains to us to determine the second coordinate of $Q_j$, which depends on $P_j$ and the set $N_j^-$.

\noindent \textbf{Case 1.1:} $N_j^- = \emptyset$.

None of the unlisted schools in $P$ are feasible in the counterfactual below the cutoff.
These unlisted schools may rank higher than $b$ in $Q$, but none of them will ever be the best feasible option 
in the counterfactual below the cutoff.
Thus, the 2nd coordinate of $Q_j$ equals $b$, and $\bQ_j=\{ P_j \}$.

\noindent \textbf{Case 1.2:} $N_j^- \neq \emptyset$.

For any option $d \in N_j^-$, 
we have that $d \neq a$, $d \neq b$, $d \in N_j^+$, and $aQd$. 

\noindent \textbf{Case 1.2.1:} If $a \neq b$, we have $a Q b$.
We can always find a $Q$ such that $d Q b$ and the second coordinate of $Q_j$ equals $d$;
and we can always find another $Q$ such that $b Q d$ and the second coordinate of $Q_j$ equals $b$.
Therefore, $\bQ_j=\{ P_j \} \cup \{ (\{a\}  \times N_j^-) \}$.

\noindent \textbf{Case 1.2.2:} If $a = b$, then we have $bQd$ because $aQd$. 
Thus, $\bQ_j=\{ P_j \}$.

\bigskip

\noindent \textbf{Case 2:} $S_j < c_j$. 

By Assumption \ref{aspt:cutoff2} we have that the mechanism assignment $\mu$ equals the best option according to $P$ in the set 
$B(\bS) = B^-_j( \bS)$.
An individual with $P_{j}=(a,b)$ has $b= P(B^-_j( \bS) ) = P( B( \bS) )$; therefore, $\mu=b$.
The cutoff characterization dictates that $b=Q(B( \bS)) = Q(B^-_j( \bS) ) $, so the second coordinate of $Q_j$ equals $b$.
It remains to us to determine the first coordinate of $Q_j$, which depends on $P_j$ and the set $N_j^+$.

\noindent \textbf{Case 2.1:} $N_j^+ = \emptyset$.

None of the unlisted options in $P$ are feasible in the counterfactual above the cutoff.
These unlisted options may rank higher than $a$ in $Q$, but none of them will ever be the best feasible option 
in the counterfactual above the cutoff.
Thus, the first coordinate of $Q_j$ equals $a$, and $\bQ_j=\{ P_j \}$.

\noindent \textbf{Case 2.2:} $N_j^+ \neq \emptyset$

For any option $d \in N_j^+ \cap N_j^-$, 
we have that $d\neq a$, $d \neq b$, and $d$ is in $B_j^-(\bS)$, but $bQd$ since the 2nd coordinate of $Q_j$ equals $b$. 
We have that $a \bar{Q} b$, and it follows that $a Q d$.
In this case, we can never find a $Q$ such that the best choice in $B_j^+(\bS)$ is $d$.

Consider there is an option $d \in N_j^+ \setminus N_j^-$.
We have that $d \neq a$, $d \neq b$, $d$ is in $B_j^+(\bS)$ but not in $B_j^-(\bS)$.

It is possible to find $Q$ such that $d Q a$, in which case the best choice in $B_j^+(\bS)$ is $d$.
It is also possible to find another $Q$ such that $a Q d$, in which case the best choice in $B_j^+(\bS)$ is $a$.
Therefore, $\bQ_j=\{ P_j \} \cup ( (N_j^+ \setminus N_j^-) \times \{ b \})$.

\bigskip

Moreover, assume $P$ is a strong partial order of $Q$.
This implies that $|P| = \min\{K, | \{d \in Q: d Q 0 \} | \}$ in addition to $P$ being a subset of $ \{d\in Q: d Q 0 \} $.
If $|P|<K$, then $|P| = | \{d\in Q: d Q 0 \} | $, and $P$ equals the true list of acceptable schools in $Q$.
Therefore, $Q_j=P_j$, and $\bQ_j=\{ P_j \}$.
On the other hand, if $|P|=K$, $P$ may or may not be the true list of acceptable schools in $Q$, 
and $\bQ_j$ continues to be as defined in the case of weak partial order.

\bigskip

Finally, the proof above is constructive as it considers all possibilities of $Q$ given $P$ and $\bS$ that are consistent with the assumptions.
Thus, it leads to the sharp set of possible $Q_j$s under the full support assumption.
\\*
$\square$

\subsection{Proof of Proposition \ref{result:umas} }
\label{proof:result:umas}

\indent 

The proof is based on the expected-utility maximization problem of Section \ref{sec:app:uncertainty} and we refer the reader to that section. 
From now on, let $P$ be the optimal choice of an arbitrary agent with placement scores $\boup{s}=(s_1,\ldots, s_J)$, true preferences $Q$, utilities $(U_0 ,U_1 ,...,U_J)$, where $U_0=0$,
and cutoff beliefs denoted by the random vector $(C_1,\ldots,C_J)$.

Suppose there is a pair of schools $(d,e) \in UMAS \cap \left(P \times P^c \right)$.
Let $n$ be the position in $P$ where $d$ appears, i.e., $P^n = d$.
Construct $\ti{P}$ by taking $P$ and replacing option $d$ with option $e$.
This implies that $\ti{P}^u = P^u$ for every $u\neq n$ and 
$\ti{P}^n = e$.

Suppose $e Q d \Leftrightarrow U_e > U_d$ by contradiction.
In what follows, we compare the expected utility of submitting $P$ to the expected utility of submitting $\ti{P}$
for this agent and show a contradiction.
To do this, we first establish some implications of Assumption \ref{aspt:umas} for the probabilities of admission:
\begin{gather}
L^{P}_{u} = L^{\ti{P}}_{u} \text{ for any } u \text{ such that } 1 \leq u < n \text{ if }n> 1,
\label{eq:proof:result:umas:prob1}
\\
L^{P}_{n} \leq L^{\ti{P}}_{n}, 
\label{eq:proof:result:umas:prob2}
\\
L^{P}_{u}
\geq 
L^{\widetilde P}_{u}
\text{ for any } u \text{ such that } n < u \leq |P| \text{ if }n< |P|,
\label{eq:proof:result:umas:prob3}
\\
L^{P}_{0} 
\geq 
L^{\widetilde P}_{0},
\label{eq:proof:result:umas:prob4}
\end{gather}
where at least one of the inequalities \eqref{eq:proof:result:umas:prob2}--\eqref{eq:proof:result:umas:prob4} 
is strict if $n< |P|$; or, if $n = |P|$, at least one of the inequalities \eqref{eq:proof:result:umas:prob2} and \eqref{eq:proof:result:umas:prob4} is strict.
Below, we prove \eqref{eq:proof:result:umas:prob1}--\eqref{eq:proof:result:umas:prob4}.

Equation \ref{eq:proof:result:umas:prob1} comes from the fact that $\ti{P}^u = P^u$ for every $1 \leq u < n$ if $n>1$.
The admission probability depends on $u$ being the lowest ranked school for which the agent qualifies.
Thus,
\[
\mmp\left[
\cap_{v=1}^{u-1} \{ s_{P^v} < C_{P^v} \} ~ 
\cap ~ \{ s_{P^u} \geq C_{P^u} \}
\right]
=
\mmp\left[
\cap_{v=1}^{u-1} \{ s_{\ti P^v} < C_{\ti{P}^v} \} ~ 
\cap ~ \{ s_{\ti{P}^u} \geq C_{\ti{P}^u} \}
\right],
\]
for every $1 \leq u < n$ if $n>1$.

For Equation \ref{eq:proof:result:umas:prob2}, we have that 
$\mmp[ s_{d} \geq C_{d} ] \leq \mmp[ s_{e} \geq C_{e} ]$
is implied by the first condition of Assumption \ref{aspt:umas}.
This further implies that
\[
\mmp\left[ 
\cap_{v=1}^{n-1} \{ s_{P^v} < C_{P^v} \} ~ 
\cap ~ \{ s_{d} \geq C_{d} \}
\right]
\leq
\mmp\left[ 
\cap_{v=1}^{n-1} \{ s_{P^v} < C_{P^v} \} ~ 
\cap ~ \{ s_{e} \geq C_{e} \}
\right], 
\]
which is equivalent to $L^{P}_{n} \leq L^{\ti{P}}_{n}$.

For Equation \ref{eq:proof:result:umas:prob3}, note that 
the first condition of Assumption \ref{aspt:umas}
implies 
$\mmp[ s_{d} < C_{d} ] \geq \mmp[ s_{e} < C_{e} ]$.
This further implies that 
\begin{align*}
L^{P}_{u} = &\mmp\left[ \cap_{v=1}^{n-1} \{ s_{P^v} < C_{P^v} \} 
~ \cap ~
\{ s_{d} < C_{d} \} 
~ \cap_{v=n+1}^{u-1} \{ s_{P^v} < C_{P^v} \}
~ \cap ~ 
\{ s_{P^u} \geq C_{P^u} \} 
\right]
\\
&\geq
\\
&\mmp\left[ \cap_{v=1}^{n-1} \{ s_{P^v} < C_{P^v} \} 
~ \cap ~
\{ s_{e} < C_{e} \} 
~ \cap_{v=n+1}^{u-1} \{ s_{P^v} < C_{P^v} \}
~ \cap ~ 
\{ s_{P^u} \geq C_{P^u} \} 
\right]
=
L^{\ti{P}}_{u}
\end{align*}
for $u$ such that $n<u\leq |P|$ if $n<|P|$.

For Equation \ref{eq:proof:result:umas:prob4},
again we have that $\mmp[ s_{d} < C_{d} ] \geq \mmp[ s_{e} < C_{e} ]$
implies 
\begin{align*}
L_0^P = &\mmp\left[ \cap_{v=1}^{n-1} \{ s_{P^v} < C_{P^v} \} 
~ \cap ~
\{ s_{d} < C_{d} \} 
~ \cap_{v=n+1}^{|P|} \{ s_{P^v} < C_{P^v} \}
\right]
\\
&\geq
\\
&\mmp\left[ \cap_{v=1}^{n-1} \{ s_{P^v} < C_{P^v} \} 
~ \cap ~
\{ s_{e} < C_{e} \} 
~ \cap_{v=n+1}^{|P|} \{ s_{P^v} < C_{P^v} \}
\right]
= L_0^{\ti P}.
\end{align*}

Finally, the second condition in Assumption \ref{aspt:umas} (relevance condition) implies that 
at least one of the inequalities \eqref{eq:proof:result:umas:prob2}--\eqref{eq:proof:result:umas:prob4} 
is strict if $n< |P|$ or, if $n = |P|$, at least one of the inequalities \eqref{eq:proof:result:umas:prob2} and \eqref{eq:proof:result:umas:prob4} is strict.
Having established these facts, we now move on to compare the expected utility of submitting $P$ to the expected utility of submitting $\ti{P}$.

Define $\epsilon = U_e - U_d = U_{\ti P^n} - U_{P^n} >0$.
Note that
$L_0^P = 1 - \sum_{u=1}^{|P|} L^P_u$ and
$L_0^P - L_0^{\ti P}$ $= \sum_{u=1}^{|P|} \left( L^{\ti P}_u - L^P_u \right)$ 
$= \left( L^{\ti P}_n - L^P_n \right) + \sum_{u=n+1}^{|P|} \left( L^{\ti P}_u - L^P_u \right)$,
where we adopt the convention that a sum over an empty set of indices equals zero, i.e.,
$\sum_{u=n+1}^{|P|} \left( L^{\ti P}_u - L^P_u \right) =0$ if $n+1>|P|$.
This leads to 
$\left( L^{\ti P}_n - L^P_n \right) 
= \left( L_0^P - L_0^{\ti P} \right) - \sum_{u=n+1}^{|P|} \left( L^{\ti P}_u - L^P_u \right)$.
Next, we combine these definitions with the inequalities in Equations \ref{eq:proof:result:umas:prob2}--\ref{eq:proof:result:umas:prob4}
to evaluate the difference between the expected utility of submitting $P$ and the expected utility of submitting $\ti{P}$:
\begin{align*}
& \sum^{|P|}_{u=1} U_{P^u} L^{P}_{u} - U_{\ti P^u} L^{\ti P}_{u} 
=
U_{P^n} L^{P}_{n} - U_{\ti P^n} L^{\ti P}_{n}
+\sum^{|P|}_{u=n+1} U_{P^u} ( L^{P}_{u}-L^{\ti P}_{u} )
\\
& =U_{P^n} ( L^{P}_{n}-L^{\ti P}_{n}) 
- \epsilon L^{\ti P}_{n}
+ \sum^{|P|}_{u=n+1} U_{P^u} ( L^{P}_{u}-L^{\ti P}_{u} ) 
\\
&=U_{P^n} \left[ - \left( L_0^P - L_0^{\ti P} \right) + \sum_{u=n+1}^{|P|} \left( L^{\ti P}_u - L^P_u \right) \right]
- \epsilon L^{\ti P}_{n}
+ \sum^{|P|}_{u=n+1} U_{P^u} ( L^{P}_{u}-L^{\ti P}_{u} ) 
\\
& = - U_{P^n} \left( L_0^P - L_0^{\ti P}\right) 
- U_{P^n} \sum_{u=n+1}^{|P|} \left( L^{P}_u - L^{\ti P}_u \right)
- \epsilon L^{\ti P}_{n}
+ \sum^{|P|}_{u=n+1} U_{P^u} \left( L^{P}_{u}-L^{\ti P}_{u} \right) 
\\
& = - U_{P^n} \left( L_0^P - L_0^{\ti P}\right) 
- \epsilon L^{\ti P}_{n}
- \sum^{|P|}_{u=n+1} \left( U_{P^n} - U_{P^u} \right) \left( L^{P}_{u}-L^{\ti P}_{u} \right) < 0,
\end{align*}
where we use 
the fact that weak partial order (Assumption \ref{aspt:weakpo}) implies $U_{P^1} > \ldots > U_{P^{|P|}}>0$,
$\epsilon>0$,
and at least one of $\left( L_0^P - L_0^{\ti P}\right)$,
$L^{\ti P}_{n}$,
and 
$\left( L^{P}_{u}-L^{\ti P}_{u} \right)$ for $u>n$
is strictly positive if $n<|P|$.
If $n = |P|$, at least one of $\left( L_0^P - L_0^{\ti P}\right)$ and
$L^{\ti P}_{n}$ is strictly positive. 
The inequality above shows that submitting $\ti P$ increases the expected utility relative to that from submitting $P$, which contradicts $P$ being the optimal choice.
Therefore, $dQe$.
\\*
$\square$

\subsection{Proof of Corollary \ref{result:umas_refine}}
\label{proof:result:umas_refine}

\indent

Let $(d,e)$ be an arbitrary pair of distinct schools such that $e$ is uniformly more accessible than school $d$;
$d \in P$, $e \notin P$.
Call $(a,b)=P_j$.
Proposition \ref{result:umas} implies $d Q e$.
This fact weakly decreases the set of possible $Q$s each individual has and thus potentially affects only the nonsingleton cases in the definition of $\bQ_j$ in Proposition \ref{result:partial_order}.
These correspond to cases 1.2.1 and 2.2 in the proof of Proposition \ref{result:partial_order}.
We re-examine these cases below for the arbitrary pair $(d,e)$.

\noindent \textbf{Case 1.2.1:} $S_j \geq c_j$, $N_j^- \neq \emptyset$, and $a \neq b$.

We have $aQb$.
For any option $f \in N_j^-$, 
we have that $f\neq a$, $f \neq b$, $f \in B_j^-(\bS)$, and $f \in B_j^+(\bS)$.

\noindent \textbf{Case 1.2.1(a):} Suppose $b \bar{P} d$.

$b \bar{P} d$ implies that $b \bar{Q} d$.
Given that $d Q e$, we have $b Q e$.
In case $e \in N_j^-$, it is no longer true that we can construct $Q$ such that $f Q b$ for any $f \in N_j^-$.
We can do so only for $f \neq e$.
Therefore, $Q_j=(a,e)$ does not belong to $\bQ_j$.

\noindent \textbf{Case 1.2.1(b):} Suppose $d P b$.

This implies that $d Q b$. 
The fact that $d Q e$ does not restrict us from having two possibilities: $d Q e Q b$ or $d Q b Q e$. 
It is again possible to construct $Q$ such that $fQb$ for any $f \in N_j^-$, including $f=e$ if $e \in N_j^-$.
Therefore, $Q_j=(a,e)$ does belong to $\bQ_j$ in this case.

\bigskip

\noindent \textbf{Case 2.2:} $S_j < c_j$, $N_j^+ \neq \emptyset$ 

Consider there is an option $f \in N_j^+ \setminus N_j^-$.
We have that $f\neq a$, $f \neq b$, $f$ is in $B_j^+(\bS)$ but not in $B_j^-(\bS)$.

\noindent \textbf{Case 2.2(a):} Suppose $a \bar{P} d$. 

$a \bar{P} d$ implies that $a \bar{Q} d$.
Given that $d Q e$, we have that $a {Q} e$.
If $f=e$, we cannot construct $Q$ such that $f Q a$.
Thus, it is no longer true that we can find $Q$ such that $f Q a$ for every $f \in N_j^+ \setminus N_j^-$; 
this is true only
for $f \neq e$.
Therefore, $Q_j=(e,b)$ does not belong to $\bQ_j$ in this case.

\noindent \textbf{Case 2.2(b):} Suppose $d {P} a$.

$d P a$ implies $d Q a$.
The fact that $d Q e$ does not restrict us from having two possibilities: $d Q e Q a$ or $d Q a Q e$. 
It is again possible to construct $Q$ such that $f Q a$ for every $f \in N_j^+ \setminus N_j^-$, including $f=e$
if $e \in N_j^+ \setminus N_j^-$.
Therefore, $Q_j=(e,b)$ does belong to $\bQ_j$ in this case.

\bigskip

Sharpness follows because we remove all $Q$s that violate the implication of Prop. \ref{result:umas}. $\square$

\subsection{Proof of Proposition \ref{result:partial_id:nextbest-dist}}
\label{proof:result:partial_id:nextbest-dist}

\indent

The random set $\bQ_j$ is a measurable map from $\Omega$ to $\m{J}^0 \times \m{J}^0$, so it is compact valued. 
Following Definition A.1 by \cite{molinari2020}, we say that $\bQ_j$ is a random closed set because for every compact set $\mathcal K \in \mmr^2$, 
the set
$\{\omega \in \Omega:  \bQ_j (\omega) \cap \mathcal K\neq \emptyset\}$ is a measurable event. 
By Assumption \ref{aspt:sharp_bQj}(i), 
the random vector $Q_{j}$ and the random set  $\bQ_j$ are measurable maps on the same probability space,
and $\mmp\left[  Q_{j} \in \bQ_j |S_j =s \right]=1$ for any $s \in \m{S}_j$.

Artstein's inequality (Theorem A.1 by \cite{molinari2020}) characterizes all possible probability mass functions 
$\mmp\left[  Q_{j}=(a,b) |S_j =s \right]$ over $(a,b) \in \m{J}^0 \times \m{J}^0$
for any $Q_j$ such that $\mmp\left[  Q_{j} \in \bQ_j |S_j =s \right]=1$ for any $s \in \m{S}_j$.
Since $\bQ_j$ is sharp (Assumption \ref{aspt:sharp_bQj}(ii)), the Artstein's inequality yields the sharp set of all 
probability mass functions 
$\mmp\left[  Q_{j}=(a,b) |S_j =s \right]$.
For any $s \in \m{S}_j$, the inequality says that
\begin{eqnarray}\label{eq:primal-Arstein}
  \mmp\left[ \bQ_j \subseteq A | S_j=s \right]  
  \leq   \mmp \left[ Q_j \in  A|S_j=s \right] 
  \;\; \;\forall A \in 2^{\m{J}^0 \times \m{J}^0},
\end{eqnarray}
where $2^{\m{J}^0 \times \m{J}^0}$ denotes the power set of ${\m{J}^0 \times \m{J}^0}$.

Lemma 1 of \cite{chesher_ecta2017} applied to our case shows that the inequalities \eqref{eq:primal-Arstein} are equivalent to:
\begin{eqnarray*}
  \mmp \left[ \bQ_j \subseteq A ~|~ S_j=s \right]
  \leq
  \mmp \left[ Q_j \in  A ~|~ S_j=s \right] \;\; \;\forall A \in \bLambda^{\cup}_{j}(s).
\end{eqnarray*}

Next, consider any $s \in [c_j, c_j+\eps)$ for $\eps$ of Assumption \ref{aspt:dist_bQj}(i).
We have that
\begin{eqnarray*}
  \mmp \left[ \bQ_j \subseteq A ~|~ S_j=s \right]
  \leq
  \mmp \left[ Q_j \in  A ~|~ S_j=s \right] \;\; \;\forall A \in \bLambda^{\cup+}_{j},
\end{eqnarray*}
and taking limits on both sides as $s \downarrow c_j$ leads to
\begin{eqnarray}\label{eq:Arstein:right}
  \mmp \left[ \bQ_j \subseteq A ~|~ S_j=c_j^+ \right]
  \leq
  \mmp \left[ Q_j \in  A ~|~ S_j=c_j \right] \;\; \;\forall A \in \bLambda^{\cup+}_{j},
\end{eqnarray}
where we use the continuity of $\mmp \left[ Q_j \in  A ~|~ S_j=s \right]$ wrt $s$ (Assumption \ref{aspt:continuity})
and
the existence of  side limit  $\mmp \left[ \bQ_j \subseteq A ~|~ S_j=c_j^+ \right]$ (Assumption \ref{aspt:dist_bQj}(ii)).
Applying an analogous argument to the left of the cutoff $c_j$ leads to
\begin{eqnarray}\label{eq:Arstein:left}
  \mmp \left[ \bQ_j \subseteq A ~|~ S_j=c_j^- \right]
  \leq
  \mmp \left[ Q_j \in  A ~|~ S_j=c_j \right] \;\; \;\forall A \in \bLambda^{\cup-}_{j}.
\end{eqnarray}

If there is $A \in \bLambda^{\cup+}_{j} \cap \bLambda^{\cup-}_{j}$, then both \eqref{eq:Arstein:right} and 
\eqref{eq:Arstein:left} are true, which leads to 
\begin{eqnarray}\label{eq:Arstein:left:right}
	\max \left\{~
		\mmp\left[ \bQ_j \subseteq A | S_j= c_j^+ \right]
		~;~
		\mmp\left[ \bQ_j \subseteq A | S_j= c_j^- \right]
	~\right\}
	\leq
	\mmp \left[ Q_j \in  A ~|~ S_j=c_j \right].
\end{eqnarray}

In summary, we have an inequality for every $A \in \bLambda^{\cup+}_{j} \cup \bLambda^{\cup-}_{j}$.
There are three possibilities: $A \in \bLambda^{\cup+}_{j} \cap \bLambda^{\cup-}_{j}$ 
(Inequality \ref{eq:Arstein:left:right}),
$A \in \bLambda^{\cup+}_{j} \setminus \bLambda^{\cup-}_{j}$ (Inequality \ref{eq:Arstein:right}),
and
$A \in \bLambda^{\cup-}_{j} \setminus \bLambda^{\cup+}_{j}$ (Inequality \ref{eq:Arstein:left}).
\\*
$\square$

\subsection{Proof of Proposition \ref{result:bounds_RD}}
\label{proof:result:bounds_RD}

\indent

Define 
\[
\delta_{j,k}(s) = \frac{
	\mmp\left[ Q_j=(j,k)|S_j=s\right]
	}{
	\mmp\left[ \bQ_j \cap \{ (j,k) \} \neq \emptyset |S_j=s \right]
	},
\]
where we know $\delta_{j,k}(s)$ is well defined for $s$ in a neighborhood of $c_j$ because $(j,k)$ is a comparable pair (Definition \ref{def:compairs})
and because of Assumption \ref{aspt:sharp_bQj}.

By Assumptions \ref{aspt:continuity} and \ref{aspt:dist_bQj}, 
the side limits of $\delta_{j,k}(s)$ as $s \downarrow c_j$ and $s \uparrow c_j$ are well defined and equal to  
$\delta_{j,k}^+$ and $\delta_{j,k}^-$, respectively.

Take $g\in \m{G}$ of Assumption \ref{aspt:continuity}.
For $s \geq c_j$,
\begin{align*}
& \mme \left[ g(Y) \left| \bQ_j \cap \{(j,k)\} \neq \emptyset, S_j=s \right. \right]
\\*
& = \delta_{j,k}(s) ~ \mme \left[ g(Y) \left| Q_j=(j,k), \bQ_j \cap \{(j,k)\} \neq \emptyset, S_j=s  \right. \right]
\\
& \hspace{.5cm} + (1-\delta_{j,k}(s)) ~ \mme \left[ g(Y) \left| Q_j \neq (j,k), \bQ_j \cap \{(j,k)\} \neq \emptyset, S_j=s \right. \right]
\\
& = \delta_{j,k}(s) ~\mme \left[ g(Y(j)) \left| Q_j=(j,k), S_j=s \right. \right]
\\
& \hspace{.5cm} + (1-\delta_{j,k}(s)) ~ \mme \left[ g(Y) \left| Q_j \neq (j,k), \bQ_j \cap \{(j,k)\} \neq \emptyset, S_j=s  \right. \right],
\end{align*}
where we use the cutoff characterization and the fact that $\{Q_j = (j,k)\} \subseteq \{ \bQ_j \cap \{(j,k)\} \}$.
Taking the limit as $s \downarrow c_j$,
\begin{align*}
& \mme \left[ g(Y) \left| \bQ_j \cap \{(j,k)\} \neq \emptyset, S_j=c_j^+ \right. \right]
\\
& = \delta_{j,k}^+ ~\mme \left[ g(Y(j)) \left| Q_j=(j,k), S_j=c_j \right. \right]
\\
& \hspace{.5cm} + (1-\delta_{j,k}^+) ~ \mme \left[ g(Y) \left| Q_j \neq (j,k), \bQ_j \cap \{(j,k)\} \neq \emptyset, S_j=c_j^+  \right. \right],
\end{align*}
where again all limits are well defined by Assumptions \ref{aspt:continuity} and \ref{aspt:dist_bQj}. 
Repeating the derivation for $s < c_j$ and making $s \uparrow c_j$,
\begin{align*}
& \mme \left[ g(Y) \left| \bQ_j \cap \{(j,k)\} \neq \emptyset, S_j=c_j^- \right. \right]
\\
& = \delta_{j,k}^- ~\mme \left[ g(Y(k)) \left| Q_j=(j,k), S_j=c_j \right. \right]
\\
& \hspace{.5cm} + (1-\delta_{j,k}^-) ~ \mme \left[ g(Y) \left| Q_j \neq (j,k), \bQ_j \cap \{(j,k)\} \neq \emptyset, S_j=c_j^-  \right. \right].
\end{align*}

We know the expectations on the left-hand sides of the last two equations, but we do not know the $\delta$s or the expectations on the right-hand sides.
Above the cutoff, the goal is to partially identify  $\mme \left[ g(Y(j)) \left| Q_j=(j,k), S_j=c_j \right. \right]$ using
the distribution of $g(Y)$ conditional on $\bQ_j \cap \{(j,k)\} \neq \emptyset$ and $S_j=c_j^+$
plus knowledge of a strictly positive lower bound on $\delta_{j,k}^+$, i.e., $\lbar{\delta}_{j,k}^+$.
Likewise, below the cutoff, 
the goal is to partially identify  $\mme \left[ g(Y(k)) \left| Q_j=(j,k), S_j=c_j \right. \right]$ using
the distribution of $g(Y)$ conditional on $\bQ_j \cap \{(j,k)\} \neq \emptyset$ and $S_j=c_j^-$
plus $\lbar{\delta}_{j,k}^-$.
 This problem falls within the framework of \cite{horowitz1995}, specifically Corollary~4.1. In addition, Lemma~5(i) in \cite{KroftMourifieVayalinkal2024} provides a general truncation formula of the bounds in Corollary~4.1 of \cite{horowitz1995} which accommodate arbitrary outcome types---whether continuous, discrete, or mixed. Applying this result to the mixtures considered above completes the proof. In the binary and continuous cases, the bounds simplify respectively to the expressions given below. \\
\textbf{Binary Outcome.}
\begin{align*}
& \max \left\{
	1- \frac{1}{\lbar{\delta}^+_{j,k}}
	\mmp \left[ Y=0 \left| \bQ_j \cap \{(j,k)\} \neq \emptyset, S_j=c_j^+ \right. \right]
	~,~ 
	0 
	\right\} 
\\
&  
\hspace{1cm} 
\leq 
~ \mme\left[Y(j) \left| Q_j=(j,k), S_j=c_j \right. \right] ~ 
\leq  
\\ 
& \min \left\{ 
	\frac{1}{\lbar{\delta}^+_{j,k}} 
	\mmp \left[ Y=1 \left| \bQ_j \cap \{(j,k)\} \neq \emptyset, S_j=c_j^+ \right. \right]
	~,~ 
	1
	\right\},
\end{align*}
and
\begin{align*}
& 
\max \left\{
	1- \frac{1}{\lbar{\delta}^-_{j,k}}
	\mmp \left[ Y=0 \left| \bQ_j \cap \{(j,k)\} \neq \emptyset, S_j=c_j^- \right. \right]
	~,~ 
	0
\right\} 
\\
&  
\hspace{1cm}
\leq 
~ \mme\left[ Y(k) \left| Q_j=(j,k), S_j=c_j \right. \right] ~ 
\leq  
\\ 
&
\min \left\{
	\frac{1}{\lbar{\delta}^-_{j,k}}
	\mmp \left[ Y=1 \left| \bQ_j \cap \{(j,k)\} \neq \emptyset, S_j=c_j^- \right. \right]
	~,~ 
	1
	\right\}.
\end{align*}
\textbf{Continuous Outcome.}
\begin{align*}
 & \mme\left [g(Y)  \left|  \bQ_j \cap \{ (j,k) \} \neq \emptyset,  g(Y) < F_{j,k+}^{-1}(\lbar{\delta}_{j,k}^+), S_j=c_j^+ \right. \right] 
 \\
 &  
\hspace{1cm} \leq 
 ~ 
 \mme\left[g(Y(j)) \left|  Q_j=(j,k), S_j=c_j \right. \right] 
 ~
 \leq  
\\ 
 & \mme\left [g(Y) \left|  \bQ_j \cap \{ (j,k) \} \neq \emptyset,  
 g(Y) > F_{j,k+}^{-1}(1-\lbar{\delta}_{j,k}^+), S_j=c_j^+ \right. \right], 
 \end{align*}
and
\begin{align*}\label{eq:bounds_ASF_k}
 & \mme \left[ g(Y) \left|  \bQ_j \cap \{ (j,k) \} \neq \emptyset,  g(Y) < F_{j,k-}^{-1}(\lbar{\delta}_{j,k}^-), S_j=c_j^- \right. \right] 
 \\
 & \hspace{1cm} \leq ~ \mme\left[ g(Y(k)) \left| Q_j=(j,k), S_j=c_j \right. \right] ~ \leq  
\\ 
 & \mme \left[ g(Y) \left|  \bQ_j \cap \{ (j,k) \} \neq \emptyset,  
 g(Y) > F_{j,k-}^{-1}(1 - \lbar{\delta}_{j,k}^-), S_j=c_j^- \right. \right].
\end{align*}

$\square$

\renewcommand{\theequation}{C-\arabic{equation}} \setcounter{equation}{0} 
\renewcommand{\thetable}{C-\arabic{table}} \setcounter{table}{0}
\renewcommand{\thefigure}{C-\arabic{figure}} \setcounter{figure}{0}

\section{Empirical Appendix}
\label{sec:app:empirical}

\paragraph{Descriptive statistics.} Tables \ref{apptable:ds_programs} and \ref{apptable:ds_students} provide descriptive statistics about the programs used as case studies in Section \ref{sec:empir:results} and their applicants.

\paragraph{Average structural functions and the importance of preferences for graduation outcomes: comparing applicants across cutoffs in Figure \ref{fig:k1178_asf}.} This paragraph complements the discussion of Figure \ref{fig:k1178_asf} in Section \ref{sec:empir:results}. To assess the similarity of applicants across the different cutoff pairs $(j,k)$ considered in Figure \ref{fig:k1178_asf}, we begin by identifying the set of applicants in sample for estimation of bounds for the average structural function $\mme\left[ Y(k) \left| Q_j=(j,k), S_j=c_j \right. \right]$ at cutoff $c_j$ for each program pair $(j,k)$, and extracting their primary scores. For each pair $(j,k)$, we compute the average pairwise Euclidean distance between the primary scores vectors of any two applicants within that sample. This provides a measure of within-cutoff dispersion, denoted $d_{jk}$. We then calculate the average pairwise Euclidean distance between any two applicants in the analogous sample for distinct pairs $(j_1, k_1)$ and $(j_2, k_2)$, denoted $d_{j_1k_1, j_2k_2}$, which captures between-cutoff dispersion. Table \ref{tab:Yk_agvdist_1178} shows the average ratio $d_{j_1k, j_2k} / d_{j_1k}$ for $k = 1178$ (Bachillerato de Ingreso Comun at UChile) across the relevant combinations of $j_1$ and $j_2$. A ratio close to one suggests that applicants at the $(j_1, k)$ cutoff are, on average, as similar to applicants at the $(j_2, k)$ cutoff as they are to each other---supporting the idea that students across adjacent cutoffs are comparable in their primary scores.

\begin{table}[H]
	\caption{Descriptive statistics: Programs}\label{apptable:ds_programs}
 \centering
 	\vspace{-.4cm}
    \footnotesize
	\begin{tabular}{lcc|cc|cc}
\hline\hline
& \multicolumn{2}{c|}{All programs} & \multicolumn{2}{c|}{Medicine} & \multicolumn{2}{c}{Bachillerato} \\
& \multicolumn{2}{c|}{ } & \multicolumn{2}{c|}{at PUC} & \multicolumn{2}{c}{at UChile} \\
& Mean & S.dev. & Mean & S.dev.  & Mean & S.dev. \\
\hline
Total applicants (per year) & 474 & 351 & 1,094 & 138 & 2430 & 317 \\
Total admitted stud. (per year) & 66 & 48 & 90 & .90 & 385 & 39 \\
Year-to-year absolute change in cutoff & .85 & 18.2 & 2.5 & 11.5 & 2.0 & 6.8 \\
\hline
Cutoff & \multicolumn{2}{c|}{ } & 780 & 7.4 & 670 & 4.0 \\
Number of programs & \multicolumn{2}{c|}{1,191} & \multicolumn{2}{c|}{ }  \\
\hline\hline
\end{tabular}
\caption*{\footnotesize The first column of the table provides descriptive statistics on all programs involved in the centralized college assignment mechanism in Chile between 2004 and 2010. The second and third columns provide descriptive statistics on the medicine program at PUC Santiago and the Bachillerato de Ingreso Comun at UChile (two of our programs of interest in Section \ref{sec:empir:results}).
}
\end{table}

\vspace{-1cm}

\begin{table}[H]
	\caption{Descriptive statistics: Students}\label{apptable:ds_students}
 \centering
 \vspace{-.4cm}
 \footnotesize
	\begin{tabular}{lcc|cc|cc}
\hline\hline
& \multicolumn{2}{c|}{All students} & \multicolumn{2}{c|}{Medicine} & \multicolumn{2}{c}{Bachillerato} \\
& \multicolumn{2}{c|}{ } & \multicolumn{2}{c|}{at PUC} & \multicolumn{2}{c}{at UChile} \\
& Mean & S.dev. & Mean & S.dev. & Mean & S.dev. \\
\hline
Number of programs in ROL & 4.86 & 2.20 & 5.18 & 2.01 & 5.33 & 1.84 \\
ROL strictly shorter than permitted & .80 & .40 & .77 & .42  & .79 & .41 \\
Assigned (to any prog.) & .65 & .48 & .68 & .46 & .71 & .45 \\
Rank of assigned prog., cond. on assigned & 2.24 & 1.59 & 2.58 & 1.68 & 2.64 & 1.59 \\
Reapplies & .19 & .39 & .25 & .43 & .31 & .46 \\
Graduates from any program & .77 & .42 & .90 & .31 & .83 & .37  \\
Grad. from assigned prog., cond. assigned & .22 & .41 & .32 & .48 & .16 & .37 \\
\hline
Number of students & \multicolumn{2}{c|}{519,409} & \multicolumn{2}{c|}{6,438} & \multicolumn{2}{c}{14,571} \\
\hline\hline
\end{tabular}

\caption*{\footnotesize The first column of the table provides descriptive statistics on the population of participants to the centralized college assignment mechanism in Chile between 2004 and 2010. The second and third columns provide descriptive statistics on  the subpopulation of students who included, respectively, medicine at PUC Santiago and Bachillerato de Ingreso Comun at UChile (two of our programs of interest in Section \ref{sec:empir:results}) in their rank-ordered list (ROL).
}
\end{table}

\vspace{-1cm}

\begin{table}[H] 
\caption{Relative average distance between students across cutoffs $c_j$} \label{tab:Yk_agvdist_1178}
\begin{center}
\vspace{-.8cm}
\footnotesize

\begin{tabular}{llccccc}

\hline\hline
&&\multicolumn{5}{c}{$j_2$}\\[.5ex]
 &&Medicine&Medicine&Odontology&Kinesiology&Math \tabularnewline
 &&UChile&U.Sant. de Ch.&UChile&UChile&PUC \tabularnewline
\cline{3-7}
&Medicine, UChile&1&.98&.98&1&1.13 \tabularnewline
&Med., U.Sant. de Ch.&1.09&1&1.02&.98&1.13 \tabularnewline
$j_1$ & Odontology, UChile&1.05&.98&1&.97&1.11 \tabularnewline
& Kinesiology, UChile&1.16&1.03&1.06&1&1.13 \tabularnewline
& Math, PUC&1.1&.99&1.01&.95&1 \tabularnewline
\hline\hline
\end{tabular}

\caption*{\footnotesize The table reports the average ratio $d_{j_1k, j_2k} / d_{j_1k}$ for $k = 1178$ (Bachillerato de Ingreso Común at UChile), computed across relevant combinations of 
$j_1$ and $j_2$. Here, $d_{j_1k}$ is the average pairwise distance between any two applicants in sample for estimation of bounds for the average structural function $\mme\left[ Y(k) \left| Q_{j_1}=(j_1,k), S_{j_1}=c_{j_1} \right. \right]$ at cutoff $c_{j_1}$. Similarly, $d_{j_1k, j_2k}$ is the average Euclidean distance between the primary scores vectors of any two applicants in the analogous samples for $(j_1, k)$ and $(j_2, k)$. Ratios near one indicate that applicants at the $(j_1,k)$ cutoff are, on average, as similar to those at the $(j_2,k)$ cutoff as they are to each other—supporting comparability of student across these cutoffs in terms of primary scores.}
\end{center} 
\end{table}

\begin{table}[h] 
\caption{Local samples and estimated $\lbar{\delta}_{j,k}$s for $k=$ Bachillerato de Ingreso Comun at UChile} \label{tab:samplesizes_deltas_k1178}
\begin{center}
\begin{small}
\vspace{-.4cm}


\begin{tabular}{lcccccccc}
\hline \hline
& \multicolumn{2}{c}{(1)} & \multicolumn{2}{c}{(2)} & \multicolumn{2}{c}{(3)} & \multicolumn{2}{c}{(4)} \\ 
& \multicolumn{2}{c}{SPO} & \multicolumn{2}{c}{SPO} & \multicolumn{2}{c}{WPO} & \multicolumn{2}{c}{WPO} \\
& \multicolumn{2}{c}{UMAS} & \multicolumn{2}{c}{UMAS} & \multicolumn{2}{c}{UMAS} & \multicolumn{2}{c}{UMAS} \\ 
& \multicolumn{2}{c}{fields} & \multicolumn{2}{c}{} & \multicolumn{2}{c}{fields} & \multicolumn{2}{c}{ } \\ 
& $< c_j$ & $\geq c_j$ & $< c_j$ & $\geq c_j$ & $< c_j$ & $\geq c_j$ & $< c_j$ & $\geq c_j$ \\
\cline{2-9} \\
\multicolumn{5}{l}{\underline{Local first-best $j$: Medicine at U.Chile}}\tabularnewline
Obs. with $\bQ_j \cap \{(j,k)\} \neq \emptyset$&87&48&153&51&140&181&1001&370 \tabularnewline
Share listing $<$8 choices&.67&.87&.38&.82&.8&.96&.9&.97 \tabularnewline
$\lbar{\delta}_{j,k}$&1&1&1&1&.92&.83&.89&.28 \tabularnewline

\multicolumn{5}{l}{\underline{Local first-best $j$: Medicine at U.Santiago}} \tabularnewline
Obs. with $\bQ_j \cap \{(j,k)\} \neq \emptyset$&101&17&104&19&201&34&239&89 \tabularnewline
Share listing $<$8 choices&.78&.82&.75&.73&.89&.91&.89&.94 \tabularnewline
$\lbar{\delta}_{j,k}$&.87&1&.84&1&.43&1&.36&.51 \tabularnewline

\multicolumn{5}{l}{\underline{Local first-best $j$: Odontology at U.Chile}} \tabularnewline
Obs. with $\bQ_j \cap \{(j,k)\} \neq \emptyset$&49&41&121&49&111&111&936&250 \tabularnewline
Share listing $<$8 choices&.38&.63&.15&.53&.72&.86&.89&.9 \tabularnewline
$\lbar{\delta}_{j,k}$&.86&1&.75&1&.23&.68&.18&.29 \tabularnewline

\multicolumn{5}{l}{\underline{Local first-best $j$: Kinesiology at U.Chile}}\tabularnewline
Obs. with $\bQ_j \cap \{(j,k)\} \neq \emptyset$&78&20&99&21&307&30&544&94 \tabularnewline
Share listing $<$8 choices&.51&.75&.4&.71&.87&.83&.89&.93 \tabularnewline
$\lbar{\delta}_{j,k}$&.6&1&.47&1&.15&1&.08&.34 \tabularnewline

\multicolumn{5}{l}{\underline{Local first-best $j$: Math at PUC}} \tabularnewline
Obs. with $\bQ_j \cap \{(j,k)\} \neq \emptyset$&36&77&36&77&36&235&36&235 \tabularnewline
Share listing $<$8 choices&.72&.76&.72&.76&.72&.92&.72&.92 \tabularnewline
$\lbar{\delta}_{j,k}$&1&.8&1&.8&1&.15&1&.15 \tabularnewline
[.5em]
\hline\hline
\end{tabular}

\end{small}
\begin{tabular}{p{14.3cm}}
\footnotesize{The table describes the sample used in the estimation of the effect on outcomes of interest of assignment to Bachillerato de Ingreso Comun at UChile ($k$ in the table), under different sets of assumptions used to construct $\bQ_j$. Given each set of assumptions (shown at the top of each column), the table shows the number of students within a 30-point bandwidth on either side of the cutoff to program $j$ for whom $\bQ_j$ contains the pair $(j,k)$ and the share of these students who listed strictly fewer than eight choices in their ROLs. For each pair $(j,k)$ and each set of assumptions, the table also shows the estimated $\lbar{\delta}_{j,k}^+$ and $\lbar{\delta}_{j,k}^-$ used in the estimation of the bounds on the treatment effects. In the derivation of the bounds for the average structural function shown in Figure \ref{fig:k1178_asf}, only $\lbar{\delta}_{j,k}^-$ is used, as shown in Eq. \eqref{eq:bounds_ASF_k}. We still include $\lbar{\delta}_{j,k}^+$ for completeness and illustration purposes as the same logic applies to both $\lbar{\delta}_{j,k}^-$ and $\lbar{\delta}_{j,k}^+$.}
\end{tabular} 
\end{center} 
\end{table}

\begin{table}[h] 
\caption{Local samples and estimated $\lbar{\delta}_{j,k}$s for $j=$ medicine at PUC} \label{tab:samplesizes_deltas_j1258}
\begin{center}
\begin{small}
\vspace{-.4cm}


\begin{tabular}{lcccccccc}
\hline \hline
& \multicolumn{2}{c}{(1)} & \multicolumn{2}{c}{(2)} & \multicolumn{2}{c}{(3)} & \multicolumn{2}{c}{(4)} \\ 
& \multicolumn{2}{c}{SPO} & \multicolumn{2}{c}{SPO} & \multicolumn{2}{c}{WPO} & \multicolumn{2}{c}{WPO} \\
& \multicolumn{2}{c}{UMAS} & \multicolumn{2}{c}{UMAS} & \multicolumn{2}{c}{UMAS} & \multicolumn{2}{c}{UMAS} \\ 
& \multicolumn{2}{c}{fields} & \multicolumn{2}{c}{} & \multicolumn{2}{c}{fields} & \multicolumn{2}{c}{ } \\  
& $< c_j$ & $\geq c_j$ & $< c_j$ & $\geq c_j$ & $< c_j$ & $\geq c_j$ & $< c_j$ & $\geq c_j$ \\
\cline{2-9} \\
\multicolumn{5}{l}{\underline{Next-preferred $k$: medicine at UChile}} \tabularnewline
Obs. with $\bQ_j \cap \{(j,k)\} \neq \emptyset$&382&390&382&390&537&427&537&427 \tabularnewline
Share listing $<$8 choices&.88&.92&.88&.92&.91&.93&.91&.93 \tabularnewline
$\lbar{\delta}_{j,k}$&1&.92&1&.92&.67&.31&.67&.31 \tabularnewline

\multicolumn{5}{l}{\underline{Next-preferred $k$: medicine at U.Concepcion}} \tabularnewline
Obs. with $\bQ_j \cap \{(j,k)\} \neq \emptyset$&76&17&76&17&255&182&255&182 \tabularnewline
Share listing $<$8 choices&.57&.64&.57&.64&.87&.96&.87&.96 \tabularnewline
$\lbar{\delta}_{j,k}$&.71&1&.71&1&.21&.11&.21&.11 \tabularnewline

\multicolumn{5}{l}{\underline{Next-preferred $k$: medicine at U.Santiago}} \tabularnewline
Obs. with $\bQ_j \cap \{(j,k)\} \neq \emptyset$&61&2&61&2&123&38&123&38 \tabularnewline
Share listing $<$8 choices&.7&.5&.7&.5&.85&.97&.85&.97 \tabularnewline
$\lbar{\delta}_{j,k}$&.86&1&.86&1&.43&.51&.43&.51 \tabularnewline

\multicolumn{5}{l}{\underline{Next-preferred $k$: sciences at PUC}} \tabularnewline
Obs. with $\bQ_j \cap \{(j,k)\} \neq \emptyset$&70&9&70&11&72&58&77&147 \tabularnewline
Share listing $<$8 choices&.89&.66&.89&.54&.9&.94&.9&.96\tabularnewline
$\lbar{\delta}_{j,k}$&1&1&1&1&.97&.45&.9&.17 \tabularnewline

\multicolumn{5}{l}{\underline{Next-preferred $k$: engineering at PUC}} \tabularnewline
Obs. with $\bQ_j \cap \{(j,k)\} \neq \emptyset$&50&20&72&25&93&92&967&175 \tabularnewline
Share listing $<$8 choices&.74&.64&.51&.51&.86&.92&.96&.93 \tabularnewline
$\lbar{\delta}_{j,k}$&.83&.78&.58&.62&.45&.17&.04&.08 \tabularnewline
[.5em]
\hline \hline
\end{tabular}

\end{small}
\begin{tabular}{p{14.3cm}}
\footnotesize{The table describes the sample used in the estimation of the effect on outcomes of interest of assignment to medicine at PUC ($j$ in the table) relative to the effect of assignment to several next-preferred options $k$, under different sets of assumptions used to construct $\bQ_j$. Given each set of assumptions (shown at the top of each column), the table shows the number of students within a 30-point bandwidth around the cutoff to medicine at PUC for whom $\bQ_j$ contains the pair $(j,k)$ and the share of these students who listed strictly fewer than eight choices in their ROLs. For each pair $(j,k)$ and each set of assumptions, the table also shows the estimated $\lbar{\delta}_{j,k}^+$ and $\lbar{\delta}_{j,k}^-$ used in the estimation of the bounds on the treatment effects.}
\end{tabular} 
\end{center} 
\end{table}

\end{document}